%% file: paper.tex
\documentclass[times,twocolumn,final]{elsarticle}

\usepackage{medima}
\usepackage{framed,multirow}

\usepackage{amssymb}
\usepackage{latexsym}

\usepackage{url}
\usepackage{xcolor}
\usepackage{hyperref}
\usepackage{subfiles}
\usepackage[inline]{enumitem}
\usepackage{amsmath}
\usepackage{mathtools}
\usepackage{booktabs}
\usepackage{array}
\usepackage{newtxtext}
\usepackage{multirow}

\definecolor{newcolor}{rgb}{.8,.349,.1}

\newcommand{\RevText}[1]{#1}

\newcommand{\RevTextBold}[2]{#1}

\journal{Manuscript submitted to \textit{Medical Image Analysis}}

\begin{document}

\verso{Kaiwen Xu \textit{et~al.}}

\begin{frontmatter}


\title{Body Composition Assessment with Limited Field-of-view Computed
  Tomography: A Semantic Image Extension Perspective}



\author[1]{Kaiwen \snm{Xu}\corref{cor1}}
\cortext[cor1]{Corresponding author.}
\ead{kaiwen.xu@vanderbilt.edu}

\author[1]{Thomas \snm{Li}}
\author[2]{Mirza S.~\snm{Khan}}
\author[1]{Riqiang \snm{Gao}}
\author[2]{Sanja L.~\snm{Antic}}
\author[1]{Yuankai \snm{Huo}}
\author[2]{Kim L.~\snm{Sandler}}
\author[2]{Fabien \snm{Maldonado}}
\author[1,2]{Bennett A.~\snm{Landman}}



\address[1]{Vanderbilt University, 2301 Vanderbilt Place, Nashville,
  37235, United States}

\address[2]{Vanderbilt University Medical Center, 1211 Medical Center
  Drive, Nashville, 37232, United States}

\received{\#\#\#}
\finalform{\#\#\#}
\accepted{\#\#\#}
\availableonline{\#\#\#}
\communicated{\#\#\#}

\begin{abstract}
Field-of-view (FOV) tissue truncation beyond the lungs is common in
routine lung screening computed tomography (CT). This poses
limitations for opportunistic CT-based body composition (BC)
assessment as key anatomical structures are missing. Traditionally,
extending the FOV of CT is considered as a CT reconstruction problem
using limited data. However, this approach relies on the projection
domain data which might not be available in application. In this work,
we formulate the problem from the semantic image extension perspective
which only requires image data as inputs. The proposed two-stage
method identifies a new FOV border based on the estimated extent of
the complete body and imputes missing tissues in the truncated
region. The training samples are simulated using CT slices with
complete body in FOV, making the model development self-supervised. We
evaluate the validity of the proposed method in automatic BC
assessment using lung screening CT with limited FOV. The proposed
method effectively restores the missing tissues and reduces BC
assessment error introduced by FOV tissue truncation. In the BC
assessment for large-scale lung screening CT datasets, this
correction improves both the intra-subject consistency and the
correlation with anthropometric approximations. The developed method
is available at \url{https://github.com/MASILab/S-EFOV}.
\end{abstract}


\end{frontmatter}


\section{Introduction}
\subfile{sec-introduction}\label{sec:introduction}

\section{Method}
\subfile{sec-method}

\section{Experiments and Results}\label{sec:experiments-and-results}
\subfile{sec-experiment-result}

\section{Discussion}\label{sec:discussion}
\subfile{sec-discussion}

\section{Conclusion and Future Work}
\subfile{sec-conclusion}

\section*{Acknowledgments}
\subfile{sec-acknowledge}

\appendix

\section{Comparison of General-purpose Image Completion Methods for Stage-2}
\label{app:comp-image-completion}
\subfile{sec-appendix-comp-image-completion.tex}

\section{Identification of Body, Field-of-view, and Lung Masks}
\label{app:the-masks}
\subfile{sec-appendix-the-masks.tex}

\section{Generalizability on Conventional Chest CT beyond Lung Cancer Screening}
\label{app:generalizability-analysis}
\subfile{sec-appendix-generalizability-analysis.tex}

\bibliographystyle{model2-names.bst}\biboptions{authoryear}
\bibliography{refs}

\end{document}

%% file: sec-introduction.tex
\begin{figure*}[!t]
\centering
\includegraphics[scale=0.6]{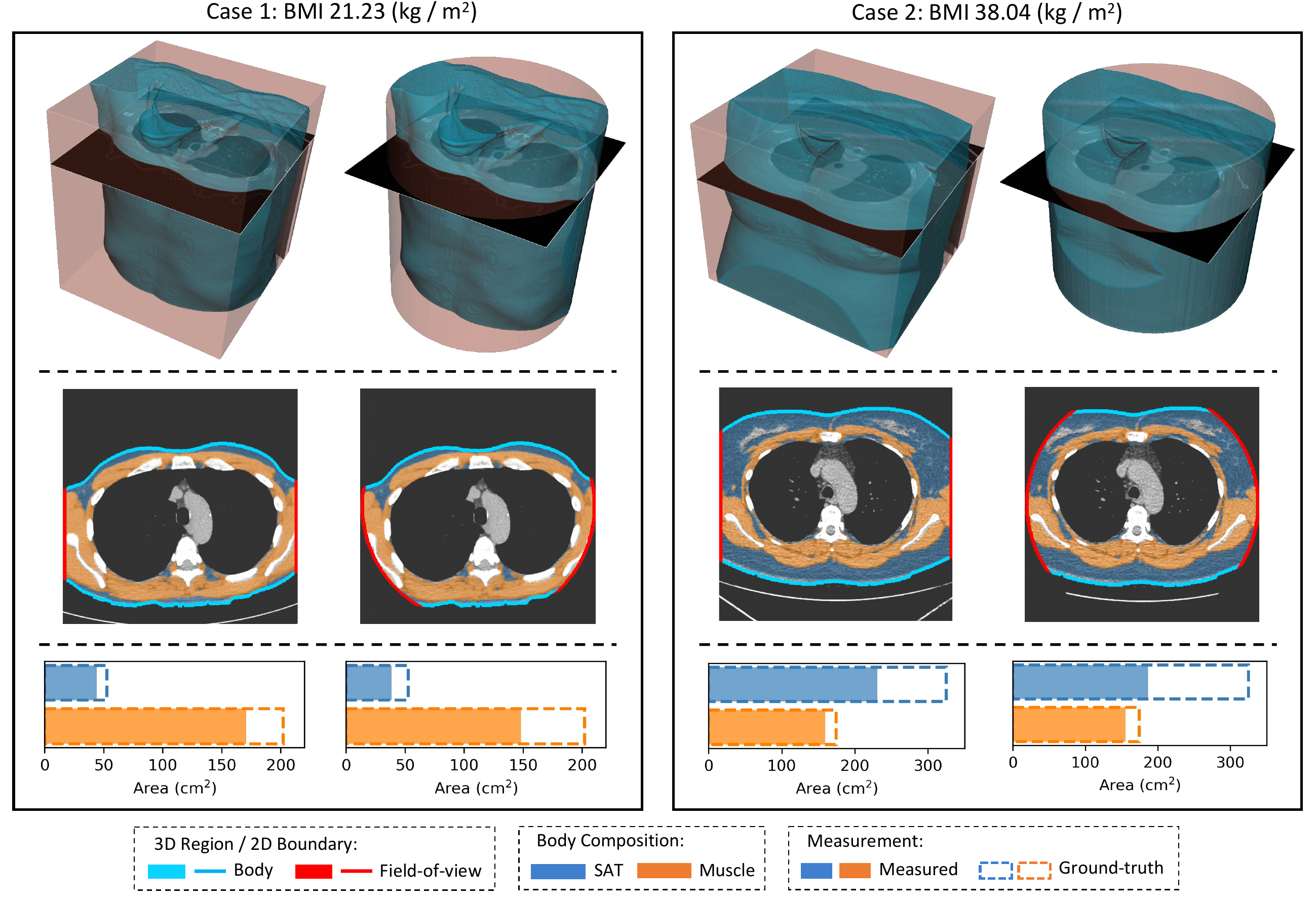}
\caption{\label{fig:problem} {\normalfont Typical FOV truncation in
    lung screening CT and its impact on body composition
    assessment. The presented examples are simulated using two lung
    screening CT scans with complete body at least at the level of the
    fifth thoracic vertebral body. As the same truncation pattern
    replicates across all cross-sectional slices for the same scan,
    the cubic or tube-like shape of the field-of-view region and the resulting
    artificial body surface can be clearly identified in 3D view (top
    row). The cross-sectional truncation effect is detailed by the
    blue segments and red segments indicating the true boundary of the
    human body and the artificial boundary of FOV respectively (second
    row). The resulting segmentation masks for SAT and muscle are
    overlaid with highlighted cross-sectional slices (second row). The
    offsets between assessment results based on truncated slices and
    complete slices are presented in the bar plots (third row).
    BMI=Body Mass Index. SAT=Subcutaneous Adipose Tissue.
} }
\end{figure*}

Computed tomography (CT) assessment of body composition (BC) has the
advantage of clear separation of adipose tissue, muscle, and organs
\citep{Thibault2012}. CT-based approaches are particularly beneficial
as CT examinations are often already available as a common imaging
study conducted for various clinical indications, allowing for
“opportunistic” assessment of BC which requires no additional
examination procedure \citep{Pickhardt2021, Pishgar2021, Pickhardt2022}.
Chest low dose computed tomography (LDCT) is the
standardized routine practice in lung cancer screening due to its high
sensitivity for nodule detection and malignancy risk evaluation
despite the lowered radiation exposure, making it an attractive
modality for opportunistic BC analysis for the lung cancer screening
population \citep{Krist2021}. Prior studies have demonstrated the
feasibility of using cross-sectional areas measured on single or
multiple cross-sectional slides in abdominal or thoracic CT as
surrogate markers for whole body compositions \RevTextBold{\citep{Shen2004,
    Mathur2020, Troschel2020, Best2022, Bridge2022}}{To address comment R3.C1. Added reference}.
However, the labor-intensive
nature of manual (or semi-automatic) annotation is a major roadblock
for both population scale evaluation and routine clinical
reporting. For this reason, artificial intelligence-based
approaches have been introduced for fully automatic BC assessment in
several recent studies \citep{Bridge2018, Lenchik2021, Magudia2021,
  Weston2019, Xu2022}.

As BC assessment is not among the primary clinical indications for
routine lung cancer screening CT examinations, certain imaging
limitations may exist. For instance, tissue truncation caused by
limited field-of-view (FOV) is a well-known issue for BC assessment
using thoracic CT \citep{Troschel2020}. In lung cancer screening, the
imaging acquisition protocol may even intentionally limit the FOV to
increase the imaging quality in the lung region
\citep{Gierada2009, Kazerooni2014, ACR2014}. In a prior study \citep{Xu2022}, the authors
introduced a morphology based cross-sectional FOV truncation severity
evaluation index and revealed that up to 96.1\% of scans in the CT arm
of National Lung Screening Trial (NLST) were associated with
significant tissue truncation caused by FOV limitation. Even though we
observed a lower rate (69.4\%) of severe FOV limitation in a recently
acquired in-house lung screening program, the issue was still frequent
enough to preclude consistent automatic BC assessment
application. Fig \ref{fig:problem} shows typical FOV
limitation-caused tissue truncation in lung screening LDCT and the
introduced shifts in BC assessment. Several studies have opted for
selective assessment of regions fully visible in FOV, e.g., using
pectoralis muscle or paraspinous muscle as surrogate for muscle
measurement \citep{Bak2019, Gazourian2020, Lenchik2021,
  Pishgar2021}. However, evidence suggests that the regional
evaluation can insufficiently represent whole-body assessment
\citep{Kim2016, Troschel2020}. In additional, significant portions of
available BC information in the CT images could be ignored following
this approach.

Extending the FOV to recover the missing tissues provides an
alternative solution for BC assessment with limited FOV
CT. Traditionally, the FOV extension of CT image was considered as an
image reconstruction problem with incomplete projection data \citep{Ogawa1984}.
\RevTextBold{
When the object exceeds
the effective data collection region, also known as the scan FOV (SFOV), of the CT scanner,
the object will be truncated in the reconstructed CT image, and significant
increase in image intensity near the FOV borders will appear at the truncated locations,
which is commonly referred as the "cupping" artifact \citep{Ruchala2002}.
}{To address R2.C3 and R3.C3.
address cupping artifact in introduction} Several
earlier works employed heuristic extrapolation methods to extend the
data in projection domain \citep{Ohnesorge2000, Hsieh2004,
  Sourbelle2005}. Recently, deep learning-based approaches have been
proposed to further improve the truncation correction, operating on
either the projection domain \citep{Ketola2021}, or a combination of
image and projection domains \citep{Fournie2019, Huang2021}.
\RevTextBold{However,
the projection data that are required by these methods}{To address R3.C2. Fix a typo}
might not be
available in many application scenarios. Typically, in retrospective
studies where the data acquisition is already completed, only the
reconstructed data in the image domain are stored and transferred.
\RevTextBold{
In addition, the FOV truncation in lung screening LDCT is mainly caused
by the reconstruction FOV (RFOV) and display FOV (DFOV), or a combination of both,
where the image intensity near the FOV borders can still be faithfully
reconstructed from the projection data collected in SFOV. This is mainly a result
of the intended restriction of the output CT FOV
where the adopted SFOVs in practice
are usually the same as those used in conventional full-sized  chest CT, e.g., 500 mm in diameter \citep{Gierada2009,
        Troschel2020}. As a result, the cupping
artifacts are rare in lung screening LDCT.
This simplifies the FOV extension for these CT images
as such it can be solved as an image completion problem.}{To address R2.C3 and R3.C3. The specialty of
lung screening LDCT in terms of cupping artifact and why it is relevant}

Image completion refers to the process of filling-in target regions
with contextual plausible contents based on the semantic information
provided by the remaining of the image
\citep{Iizuka2017}. Convolutional Neural Network (CNN) architectures
are widely used in modern image completion models with potential to
generate realistic imaging contents \citep{Iizuka2017, Isola2017,
  Li2020, Liu2018, Nazeri2019, Pathak2016, Yu2019}. The models are
typically developed in a self-supervised manner, where the input and
ground-truth data pairs are generated by applying centrally located
square \citep{Isola2017, Pathak2016}, or randomly generated
\citep{Isola2017, Li2020, Liu2018} corruption patterns on raw
images. In the medical imaging domain, the technique has shown its
capability of generating anatomically consistent structures, and has
been used, for example, to remove lesions or unwanted markers to
enhance downstream analysis including registration, segmentation, or
classification \citep{Armanious2020, Kang2021, Shen2021}. Compared to
the inpainting tasks, the extension of the original image boundary, or
outpainting, poses additional challenges as less information are
provided as boundary conditions. Though several studies have
demonstrated promising results in extending the image boundary of
natural images \citep{Krishnan2019, Wang2019}, it is still
an under-explored task to achieve anatomically consistent FOV
extension for medical images.

In this work, we sought to solve the CT FOV extension problem in the
image domain and formulated it as a semantic image extension task. A
two-stage procedure was designed to achieve fully automatic slice-wise
FOV extension of lung screening LDCT. In the first stage, the bounding
box covering the entire body region was predicted, which provided an
estimation for the appropriate extension ratio of the raw FOV. In the
second stage, the truncated anatomical structures in the region
outside of initial FOV region were automatically generated. To provide
training samples, we generated synthetic FOV truncation cases using CT
slices without tissue truncation. Unlike the randomly generated
corruption patterns commonly used in current literature, this
simulation was based on domain knowledge of FOV determination during
the CT acquisition procedure. To evaluate the validity of the
developed method in real application, we employed a prior developed
automatic BC assessment pipeline for lung cancer screening LDCT
\citep{Xu2022}. We evaluated the proposed semantic FOV extension
method on both synthetic cases and real-world lung cancer screening
LDCT with FOV truncation. The evaluation was based on human perceptual
studies conducted with trained clinical experts and assessments of the
methods' capability in correcting the BC measurement shifts caused by
FOV truncation. We evaluated the effectiveness of three different
general-purpose image completion methods under the proposed
framework. In addition, we characterized the generalizability and limitations of the
proposed method on chest CT scans acquired with a broader spectrum of
clinical indications beyond lung cancer screening.

%% file: sec-method.tex
\begin{figure*}[!t]
  \centering
  \includegraphics[scale=0.65]{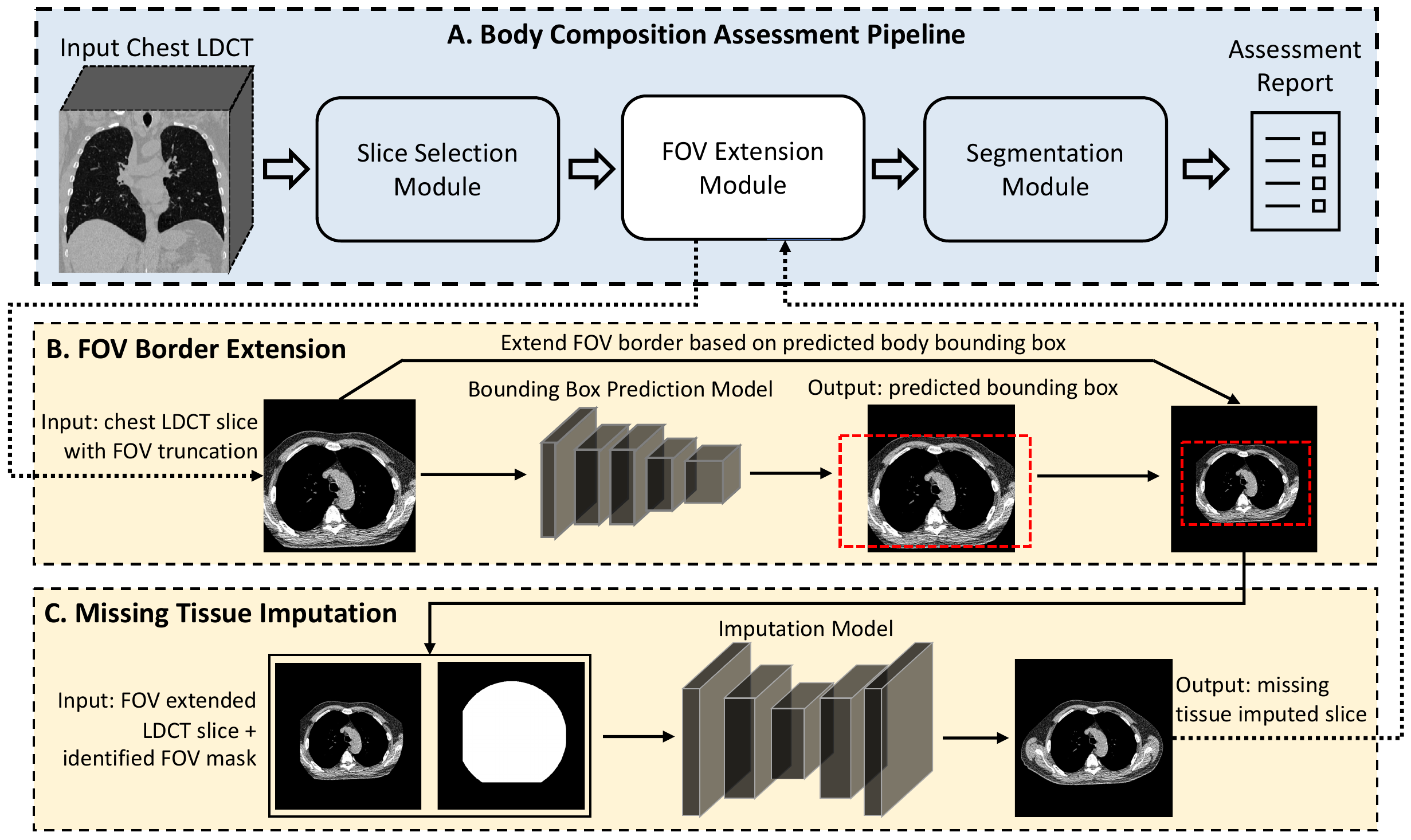}
  \caption{\label{fig:method-overall-workflow} {\normalfont
  \RevTextBold{The scheme of the proposed
      two-stage FOV extension method and its integration with body composition assessment pipeline. (A)
      The body composition assessment pipeline with the FOV extension module
      integrated as a preprocessing step.
      (B) Given a CT slice with FOV truncation as input, the stage-1 model
      extends the display FOV based on the predicted bounding box of
      the complete body. (C) The stage-2 model imputes missing
      tissues in the truncated region. LDCT=Low-dose CT. FOV=Field-of-view.}
  {To address R4.C3. Split the method figure into two figures: inference and data simulation + training.}
  }}
\end{figure*}

To extend the FOV of an image, there are two questions need to be
answered:
\begin{enumerate*}[label=(\arabic*)]
\item how to determine the new image border; and
\item how to determine the new contents in the extended region.
\end{enumerate*}
As opposed to the border extension task for natural images, where the
expected extension space may be arbitrary and usually need to be
specified manually \citep{Wang2019}, the actual anatomical boundary of
the human body can be roughly estimated even with partially visible
anatomy. With this consideration, we designed a two-stage framework
for semantic FOV extension by
\begin{enumerate*}[label=(\arabic*)]
\item extending the FOV border based on the estimated extent of the
  complete body; then
\item imputing missing tissue in the truncated region of the boundary
  extended images.
\end{enumerate*}
\RevTextBold{
Fig \ref{fig:method-overall-workflow} shows a combined overview of the method,
including the workflow of each stage and its
integration with the BC assessment pipeline.
}{Updated to address R4.C3}

\subsection{Two-stage Framework for Semantic FOV Extension}\label{sec:method-two-stage-framework}

\subsubsection{FOV Border Extension}\label{sec:method-fov-border-extension}

We formulated the task to identify the extent of the complete body as
a regression problem to estimate the axis-aligned minimum bounding box
(denoted “bounding box”) of the untruncated body region. The model $G$
took a slice $x$ with limited FOV-caused body tissue truncation as
input and outputted an estimation $G(x)$ for bounding box coordinates
of the complete body. For the training data, we used the FOV
truncation slice and ground-truth body bounding box data pairs
simulated using the slices with the complete body in FOV (detailed in
Section \ref{sec:synthetic-data}). To guide the model
training, we employed the generalized intersection over union (GIoU)
loss introduced in \cite{Rezatofighi2019}. The GIoU between two
arbitrary convex shape $A$ and $B$ is defined as \RevTextBold{
    \begin{equation}\label{eq:giou}
  \mathcal{G}(A, B) = \frac{\vert A \cap B \vert} {\vert A \cup B \vert} -
  \frac{\vert C \backslash \left( A \cup B \right) \vert}
  {\vert C \vert},
\end{equation}
}{Fixed a typo in eq.1} where $C$ is the smallest enclosing convex object of $A \cup B$.
$\vert \cdot \vert$ represent the number of elements in a set.
The first term, which follows the definition of conventional intersection
over union (IoU), assesses the degree of overlapping, and the second
term evaluates the normalized empty space between the two
regions. This combined representation provides an approximation for
IoU, while overcoming the difficulties of IoU as an objective for
model training.  With $(x, b)$ demoting the data pair of input CT
slice and ground-truth body bounding box coordinates, we
defined our GIoU objective as
\begin{equation}\label{eq:giou_term}
  \mathcal{L}_{GIoU}(G) = \mathbb{E}_{(x,b)}
  \left[\mathcal{G}\left( \mathcal{B}(b), \mathcal{B}(G(x)) \right)\right],
\end{equation}
where $\mathcal{B}(\cdot)$ represented the region defined by predicted
or ground-truth bounding-box coordinates. To further accelerate and
stabilize the training process, we also included the conventional mean
squared error (MSE) between the predicted and ground-truth bounding
box coordinates as the second objective:
\begin{equation}\label{eq:mse_term}
  \mathcal{L}_{MSE}(G) = \mathbb{E}_{(x,b)}
  \left[
    \Vert b - G(x) \Vert_2
    \right].
\end{equation}
The final loss function was given by a combination of the MSE term and
the GIoU term, which was in the form of
\begin{equation}\label{eq:combined_obj}
  \mathcal{L}_{total} = \mathcal{L}_{MSE} + \lambda\mathcal{L}_{GIoU}.
\end{equation}
\RevTextBold{Fig \ref{fig:method-training} (B) shows an overview of training
 of this module, and Fig \ref{fig:method-overall-workflow} (B) shows the
integration of the developed module with the overall workflow.}
{Update to address R4.C3}

With the predicted bounding box of the complete body, the FOV border
of the raw image was extended to fully cover the estimated extent of
the complete body. Since in most application cases the body region
locates approximately at the center of FOV, we simplified the FOV
border extension to symmetric padding which was controlled by an
estimated extension ratio
\begin{equation}\label{eq:extension-ratio}
  R = R_0 \cdot R_{est},
\end{equation}
where $R_{est}$ was the extension ratio using which the extended FOV
can exactly cover the
predicted bounding box.
\RevTextBold{As prediction errors for the body extent bounding box
always exist, $R_{est}$ alone may fail to cover the complete
actual body extent for a significant proportion of cases in application.
For this reason, we introduced the
empirically determined multiplier $R_0$ ($>1$)
such that the extended FOV can successfully cover
the body extent for most cases.}
{To address R3.C13. Why the additional multiplier is necessary.}
\RevTextBold{Based on the estimated extension ratio $R$,
the input image was symmetric padded, then resized to the dimension of input image.
The physical dimensions of image pixels were scaled by
factor $R$ correspondingly.}
{To address R2.C4. The determination of image and pixel size of
the FOV extended image}

\subsubsection{Image Completion}\label{sec:image-completion-methods}
\begin{figure*}[!t]
  \centering
  \includegraphics[scale=0.70]{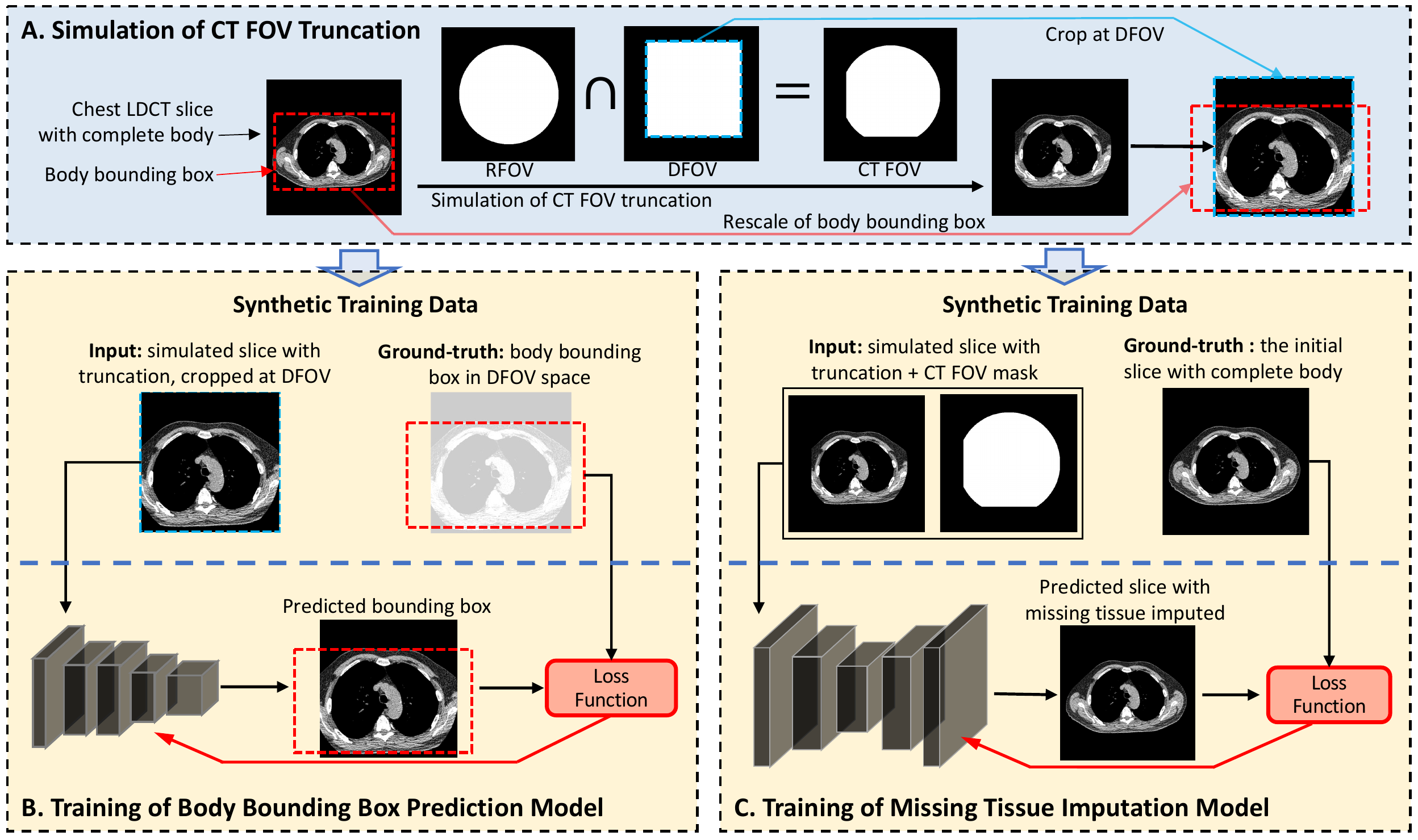}
  \caption{\label{fig:method-training} {\normalfont
  \RevTextBold{The simulation of synthetic training data and the self-supervised
  model training of the two proposed stages.
      (A) The training samples are generated using cross-sectional slices
      with complete body, by applying synthetic FOV as the
      intersection of RFOV and DFOV
      simulating the FOV determination procedure during CT
      acquisition.
      (B) The input data for the stage-1 model are the simulated slices with
      tissue truncation which are cropped at the DFOV. The bounding box of the complete body is
      pre-determined, shifted, and rescaled in corresponding with the
      cropping operation to serve as the ground-truth for stage-1
      model development. The model is developed using ResNet-18 as the
      backbone, and a loss function consisting of a MSE term and a
      Generalized IoU term.
      (C) The input data for the stage-2 model are the simulated slices and FOV patterns
      in the original image space. The initial slices with complete body in FOV are used as
      the ground-truth. We evaluate three published
      general-purpose image completion methods detailed in \ref{app:comp-image-completion}. FOV mask input may or may
      not be required depending on the image completion method.
        LDCT=Low-dose CT. FOV=Field-of-view. DFOV=Display FOV.
      RFOV=Reconstruction FOV.}
  {To address R4.C3. Split the method figure into two figures: inference and data simulation + training.}
  }}
\end{figure*}

The target of the second-stage model was to reconstruct the missing
tissues outside of FOV region. The model took the CT slice with
extended image border and optionally the FOV region mask as inputs,
and outputted a predicted image with missing tissue imputed. In the
training phase, the FOV region, input corrupted slice, and
ground-truth uncorrupted version were simulated using CT slice with
complete body in FOV (detailed in Section
\ref{sec:synthetic-data}). During the inference phase, the output
image generated by the first stage was directly forwarded to the
second-stage model.
\RevTextBold{ The FOV region can be given by the initial FOV
mask in the original image space with symmetric padding
and resizing based on the
same extension ratio defined in Eq (\ref{eq:extension-ratio}).
}{To address R2.C4}
\RevTextBold{Fig \ref{fig:method-training} (C) shows an overview of the training of this module.
Fig \ref{fig:method-overall-workflow} (C) demonstrated the integration
of this module with the overall workflow.
}{Updated to address R4.C3}

In our study, we evaluated three general-purpose solutions for
the image completion task: pix2pix \citep{Isola2017}, PConv-UNet \citep{Liu2018},
and RFR-Net \citep{Li2020}. \RevTextBold{The detailed evaluations of these methods
are given in \ref{app:comp-image-completion}.}
{All contents relevant to the comparison of the
three image completion solutions
are moved to appendix to address R4.C2.}
As RFR-Net outperformed the other two methods,
we used RFR-Net as the default method for the second stage
in the rest of this study.

\subsection{Synthetic Data Generation}
\label{sec:synthetic-data}
In image completion model development, it is a common practice to
synthesize the corrupted image using uncorrupted raw images by
applying a randomly generated free-form mask. The models are trained to
predict the corresponding uncorrupted version given the corrupted
version as input \citep{Li2020, Liu2018}. This makes the model
development self-supervised and easy to scale on a large
dataset. Inspired by this observation, we designed a synthetic data
generation procedure for the development and evaluation of the
proposed two-stage approach for FOV extension, which consisted of the
following steps:
\begin{enumerate*}[label=(\arabic*)]
\item identification of slices with the complete body in FOV;
\item simulation of FOV truncation patterns; and
\item paired data generation.
\end{enumerate*}
Herein, we assume the regions representing body and FOV in the CT
images are readily known. The solutions we adopted in this study to
automatically generate required regional masks are detailed in
\ref{app:the-masks}.
\RevTextBold{
An overview of the synthetic data generation and
data workflow for model development is demonstrated in
Fig \ref{fig:method-training}.
}{Updated to address R4.C3}

\subsubsection{FOV Truncation Severity Quantification}
To identify the FOV limitation-caused tissue truncation and assess the
truncation severity, we adopt the Tissue Cropping Index (TCI) which
was initially introduced in \cite{Xu2022}. TCI evaluated the
truncation severity for the given CT slice by the proportion of
artificial body boundary caused by FOV truncation in all detected body
boundaries. Given the body region mask $\mathcal{M}_{body}$ and FOV
region mask $\mathcal{M}_{FOV}$, TCI was defined as
\begin{equation}\label{eq:TCI}
  \mathcal{T} \left( \mathcal{M}_{body}, \mathcal{M}_{FOV} \right) =
  \frac{\vert \mathcal{E}(\mathcal{M}_{body}) \cap
    \mathcal{E}(\mathcal{M}_{FOV}) \vert}
  { \vert \mathcal{E} (\mathcal{M}_{body}) \vert },
\end{equation}
where $\mathcal{E}(\cdot)$ represented the set of boundary pixels of a
2D binary mask. The TCI value ranged from zero to one, with a
non-zero value indicating the existence of body tissue truncation and
a larger value indicating more severe tissue truncation. In our
synthetic data generation, we used a TCI value of zero to filter out
slices with a complete body in FOV. At the scan level, we defined the
scan TCI as the averaged slice-wise TCI across T5, T8, and T10 levels.
The TCI value can give an approximated stratification for both
slice-wise and scan-wise truncation severity. We
empirically setup a four-level system:
\begin{enumerate*}[label=(\arabic*)]
\item trace level, $\text{TCI} \in \left( 0, 0.15 \right]$;
\item mild level, $\text{TCI} \in \left( 0.15, 0.3 \right]$;
\item moderate level, $\text{TCI} \in \left( 0.3, 0.5 \right]$; and
\item severe level, $\text{TCI} > 0.5$.
\end{enumerate*}

\subsubsection{FOV Truncation Pattern Simulation}\label{sec:truncation-simulation}
\begin{figure}[!t]
  \centering
  \includegraphics[scale=0.27]{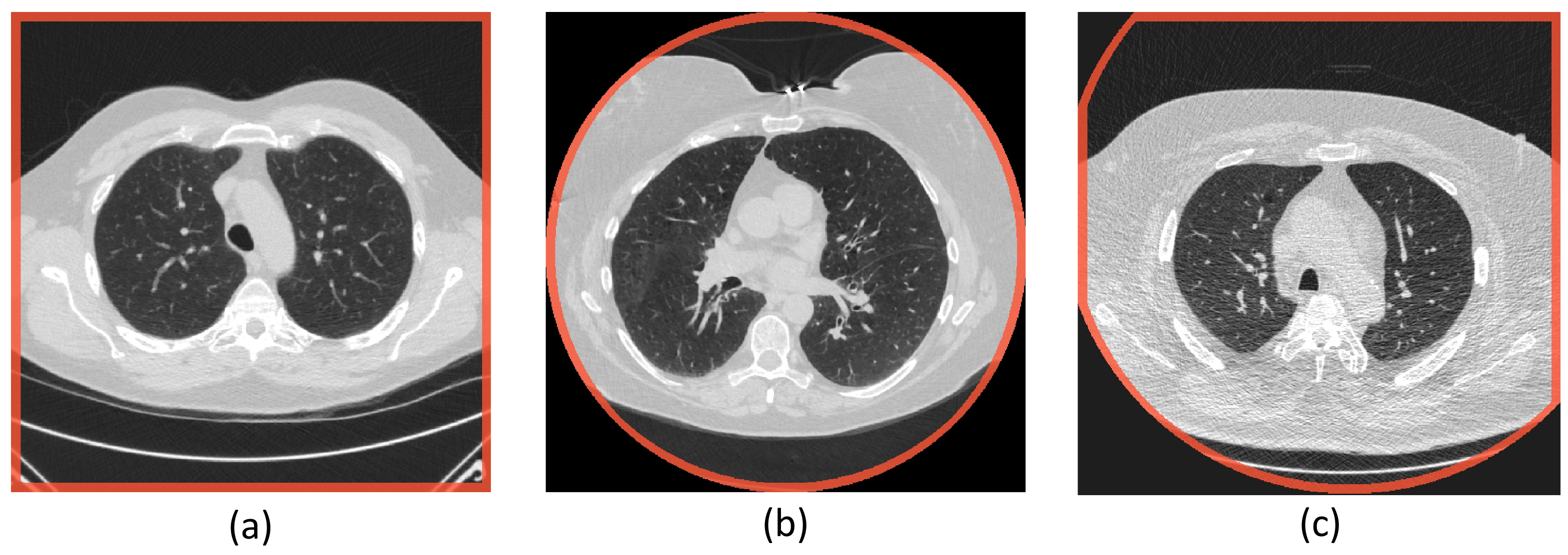}
  \caption{\label{fig:truncation-type} {\normalfont Typical FOV
      patterns in lung screening CT. The images are shown using HU
      window [-1200, 300]. The FOV borders are highlighted in red.
      The FOV patterns are the results of the intersection
      between a circular reconstruction field-of-view and a
      square display field-of-view. HU=Hounsfield Unit. FOV=Field-of-view.} }
\end{figure}

\begin{figure}[!t]
  \centering
  \includegraphics[scale=0.39]{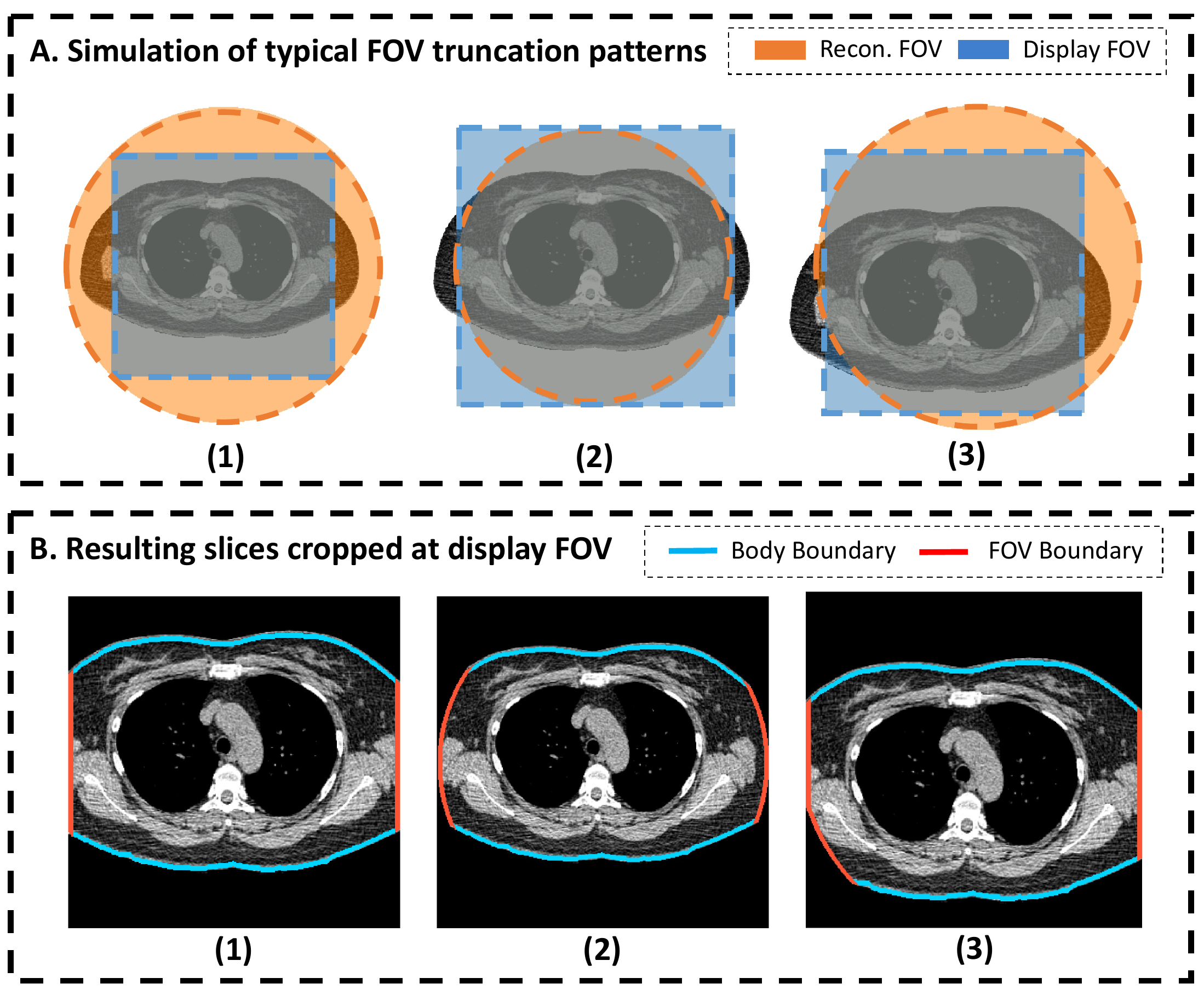}
  \caption{\label{fig:simulation} {\normalfont Simulation of typical
      FOV truncation patterns using a CT slice with complete body in
      FOV. The patterns are determined by a circular
      RFOV and a squared DFOV (A.1, A.2, and A.3). Artificial
      tissue truncation is generated by imputing the pixels outside of
      CT FOV with background intensity. The slices are further cropped
      at the DFOV, simulating the DFOV selection during CT acquisition
      procedure (B.1, B.2, and B.3). FOV=Field-of-view.
      RFOV=Reconstruction Field-of-view. DFOV=Display Field-of-view.} }
\end{figure}

The following three spatial region concepts are closely relevant to
the determination of CT FOV:
\begin{itemize}
\item {\bf{Scan Field-of-view (SFOV)}}. The SFOV is the region from
  which the projection data are collected during CT acquisition
  \citep{Seeram2015}. The size of SFOV is determined by the scanner
  limitation and can be adjusted based on specific application. In lung cancer
  screening, this parameter is usually set to 500 mm. The SFOV
  determines the maximum spatial region that the image can be
  reconstructed.
\item {\bf{Reconstruction Field-of-view (RFOV)}}. The RFOV, or
  reconstruction circle, is the circular region in which the image
  data is reconstructed from the projection domain. RFOV can be equal
  or smaller than SFOV. In general, reducing RFOV can improve the
  quality of the reconstructed image \citep{Salimova2022}, and is a
  commonly used strategy in lung cancer screening to improve the image
  quality in the lung regions \citep{Gierada2009, Kazerooni2014, ACR2014}.
  \RevTextBold{The role of RFOV in the determination of CT FOV
  is visually demonstrated in Fig \ref{fig:truncation-type} and
  Fig \ref{fig:simulation} (A, the yellow region).}{To address R2.C2}
\item {\bf{Display Field-of-view (DFOV)}}. After the reconstruction, a
  squared region needs to be specified, to which the data will be
  cropped or padded to form the final output image. We follow the same
  notation as in Chapter 3 of \cite{Seeram2015}, and term this
  squared region as display field-of-view (DFOV). The DFOV is selected
  to partially or entirely cover the RFOV, which provides a way to
  further adjust the anatomical region to be displayed. \RevTextBold{The role of DFOV in
  the determination of CT FOV
  is visually demonstrated in Fig \ref{fig:truncation-type} and
  Fig \ref{fig:simulation} (A, the blue region).}{To address R2.C2}
\end{itemize}

The final CT FOV is determined by the intersection of RFOV and
DFOV (Fig \ref{fig:truncation-type} and Fig \ref{fig:simulation}).
\RevTextBold{However, when the object exceeds the SFOV,
cupping artifacts can appear near those CT FOV borders that are overlapping
or close to the borders of SFOV where the truncation exist.
In our study, we only considered
the FOV truncation without the cupping artifacts, in consideration of
its extremely low occurrence observed in the application in lung screening LDCT.
A detailed discussion in relevant limitations in terms of generalizability are given in Section \ref{sec:discussion}.
Two chest CT scans with cupping artifacts identified
from a conventional chest CT dataset are given in Table \ref{tab:failure-mode}.
}{To address R2.C3}

It is common that parts of the squared DFOV extrude the boundary of
the circular RFOV, resulting in output image regions without available
reconstructed data (Fig \ref{fig:truncation-type}).
In the reconstructed images, these “invalid”
regions are imputed with a pre-defined value, which is controlled by the
"Pixel Padding Value (0028, 0120)" under DICOM standard (DICOM PS3.3 2016c: Information
Object Definitions).

Depending on the relative size and location between RFOV and DFOV, the
FOV truncation can present in three distinguishable
patterns. Fig \ref{fig:truncation-type} shows the typical
examples of the three major truncation patterns in lung screening
CT. In  Fig \ref{fig:truncation-type} (a), the selected DFOV is
fully inside the region of RFOV. This generates a slice where all
pixels are with valid reconstruction value. The artificial body
boundaries are located at the edge of the image. In
Fig \ref{fig:truncation-type} (b), the DFOV is selected to
exactly match the extent of the RFOV, which leads to a slice with all
valid pixels located inside the centered reconstruction circular. The
artificial body boundaries are at the edge of the circular region and
inside the image region. Fig \ref{fig:truncation-type} (c)
represents a middle status between Fig \ref{fig:truncation-type} (a)
and Fig \ref{fig:truncation-type} (b), with DFOV
selected smaller than the extent of the RFOV, but not fully covered in
RFOV. This gives a truncation pattern where artificial body
boundaries exist on both the image borders and the internal region as arc
segments of the reconstruction circular.

We designed a random procedure to generate synthetic FOV truncation
patterns simulating the FOV truncation in CT acquisition described
above (Fig \ref{fig:method-training}, A).
First, we specified the probabilities of generating each of the
three truncation patterns, with $P_a$, $P_b$, and $P_c$ corresponding
to the probability to generate pattern represented by
Fig \ref{fig:truncation-type} (a),
Fig \ref{fig:truncation-type} (b), and
Fig \ref{fig:truncation-type} (c), respectively. A pseudo
circular RFOV was generated at the center of input image space, with
diameter determined by a ratio $R_{RFOV} (\le 1)$ relative to the full
image dimension. For type Fig \ref{fig:truncation-type} (a),
the DFOV was automatically determined as the largest square region fit
into the RFOV, with location centered at the image center and side
length as $1 / \sqrt{2}$ of the diameter of the RFOV. For
Fig \ref{fig:truncation-type} (b), the DFOV was the bounding
box of RFOV, with location centered at the image center and side
length same as the diameter of RFOV. For type
Fig \ref{fig:truncation-type} (c), a squared DFOV region was
selected inside the extent of the RFOV with side length specified by a
ratio $R_{DFOV} (\le 1)$ relative to the RFOV diameter, and with
center location defined by displacement $(X_{DFOV}, Y_{DFOV})$
relative to the image center. $R_{RFOV}$, $R_{DFOV}$, $X_{DFOV}$, and
$Y_{DFOV}$ were randomly generated inside a pre-defined range, with
specific constraints to confine with geometric limitations. 
\subsubsection{Synthetic Data Pairs}
Combining the simulated FOV truncation pattern and CT slice with
complete body in FOV, we generated the paired synthetic data used in
model development and evaluation (Fig \ref{fig:method-training}, B and C). To further increase the
generalizability of the trained model, we applied random scaling,
rotation, and translation on the CT slice. Corresponding operations
were applied on associated binary masks simultaneous. The following two groups
of data were derived based on this process.

{\bf{Corrupted-uncorrupted pairs and FOV masks}}. The FOV region was
defined as the intersection of the simulated RFOV and DFOV. The
artificial truncation was generated by imputing the regions outside
this FOV by a predefined value as indication for invalid pixels. This
truncated slice and the corresponding untruncated version formed the
corrupted-uncorrupted image pair. Combining with the FOV region
mask, these data were designed for the image complete model
development (Fig \ref{fig:method-training}, C).

{\bf{Ground-truth body bounding box for cropped slice}}. The processed
CT slices with artificial truncation were further cropped at the DFOV,
simulating the same process during CT acquisition. The bounding box of
the body region was defined as pixel coordinates in image space and
first identified in the complete image space before the DFOV cropping
operation. After the cropping, these coordinates were shifted and
rescaled to the cropped image space, usually resulting in a bounding
box extrudes the image border. The cropped slice and ground-truth
body bounding box in its space formed the training data pair for the
bounding box prediction model (Fig \ref{fig:method-training}, B).

\subsection{Evaluation}
The evaluation of synthetic image results is challenging as the
commonly used intensity-based similarity metrics, e.g., $L_1$, $L_2$,
PSNR, and SSIM, can lead to a significant preference for blurry images
\citep{Liu2018}. For this reason, the human perceptual study is
commonly used as the gold standard to evaluate the output image
quality of image generative models \citep{Chuquicusma2018, Isola2017,
  Liu2018, Schlegl2019, Tang2021a}. In addition, anatomical
consistency of the synthetic contents is critical for medical imaging
applications. We proposed to evaluate the anatomical consistency of
FOV extension results using a previously developed BC assessment tool
\citep{Xu2022}, with the assumption that a biologically consistent
image completion algorithm should generate CT slices that can be
properly processed by the pre-trained segmentation model and reduce
the measurement offsets caused by FOV truncation. We further
integrated the developed FOV extension module into this BC assessment
tool and evaluated the application validity for CT-based BC assessment
under the context of routine lung screening.

\subsubsection{Body Composition Assessment} \label{sec:bc-assessment}
{\bf{Multi-level BC assessment using thorax CT}}. Using the
cross-sectional BC areas measured on axial slices selected at certain
landmark as surrogate for whole-body BC evaluation is a
well-established approach for CT-based BC assessment
\citep{Fintelmann2018, Troschel2019, Mathur2020}. In this study, we
followed the multi-level approach for thoracic CT introduced in
\cite{Best2022}, where axial slices at the level of the fifth,
eighth, and 10th thoracic vertebral bodies (T5, T8, and T10) were
selected for evaluation. The primary outputs included the
accumulated \RevTextBold{subcutaneous adipose tissue (SAT)}{Define abbreviations when first appear}
and muscle areas ($\text{cm}^2$) measured at three
levels. The measurements can be further divided by the height squared
to form the muscle and SAT indexes ($\text{cm}^2/
\text{m}^2$) for a normalized description of the body composition
profile regardless of the size of the patient.

{\bf{Implementation of fully automatic pipeline}}. A deep learning
pipeline was introduced in \cite{Bridge2022} to achieve fully
automation of the above method. However, the pipeline was developed on
chest CT scans of lung cancer patients prior to lobectomy, the
protocol of which could be significantly different from routine lung
screening CT. In \cite{Xu2022}, we implemented a fully automatic
multi-level BC assessment pipeline specifically for lung cancer
screening LDCT scans with a similar two-stage framework. The pipeline
consisted of a slice selection module based on a 3D regression model,
which identified the levels of T5, T8, and T10 vertebral bodies from a
CT volume, and a BC segmentation module implemented using 2D Nested
U-Net \citep{Zhou2018}, which delineated the cross-sectional areas
corresponding to each BC component. Once the two-stage semantic FOV
extension pipeline is developed, it can be integrated as an additional
processing module after the slice selection module and before the BC
segmentation module of the original BC assessment pipeline
(Fig \ref{fig:method-overall-workflow}, A). Thus, the final evaluation
outputs are the BC indexes evaluated on FOV extended slices, providing
a correction for measurement offsets caused by FOV tissue truncation.

\subsubsection{Evaluation on Synthetic Paired Data}\label{sec:evaluation-synthetic-data}
We evaluated the performance of the developed model on the synthetic
paired data generated using untruncated CT slices of subjects withheld
from the training phase. Only cases generated by T5, T8, or T10 slices
and with complete lung region in FOV were included to best represent
the application scenario. \RevTextBold{In additional to a direct evaluation of pixel-wise difference
in Hounsfield Unit (HU) based on Root Mean Square Error (RMSE),
the following two evaluations were adopted:}
{Include pixel-wise RMSE as part of resolution for R2.C1}

{\bf{Visual Turing test}}. To evaluate the quality of synthetic
contents, we designed a visual Truing test inspired by
\cite{Chuquicusma2018} and \cite{Schlegl2019}. For a random subsample of
synthetic cases with moderate or severe truncation ($\text{TCI} >
0.3$), half of the cases were generated by the trained pipeline, while
the other half were corresponding untruncated images of the selected
samples. The order of the samples was randomly shuffled before
presented to two clinical experts to independently classify each case
into real or synthetic category. We also provided the readers with
synthetic FOV pattern used in each sample. The mean accuracy and
inter-rater consistency were recorded.

{\bf{Correction of BC measurement shift}}. The pretrained BC
segmentation model was applied to
untruncated, synthetic truncated, and reconstructed slices,
with the segmentations on untruncated slices as ground-truth.
We quantitatively
evaluated the model’s capability to correct the BC assessment
shifts caused by FOV truncation.
\RevTextBold{First, the Dice Similarity Coefficients (DSC) were used to
assess the improvements in agreement with the ground-truth segmentations.}
{To address R1.C1}
Then, the performance was assessed by the reduction in the
shift of BC measurements.
\RevTextBold{In addition to measured area ($\text{cm}^2$) of SAT and muscle,
we included the mean attenuation (HU) of each of the considered body composition in the evaluation,
in consideration of their potential implementation in applications \citep{Pickhardt2022}.}
{We added the mean tissue attenuation in HU as part of the resolution to address R2.C1.}

\subsubsection{Evaluation on Lung Screening CT Volumes with Limited FOV}
\label{sec:evaluation-real}
We evaluated the effectiveness of the FOV extension method on real
lung screening CT data with systematic FOV truncation. The evaluation
was conducted using the automatic multi-level BC assessment pipeline
with FOV extension module integrated
(Section \ref{sec:bc-assessment}). 
\RevTextBold{In application, the FOV extension module was only applied for
 slices with detected FOV tissue truncation (TCI $>$ 0).}{
    Add details regarding how the non-cropping slices are handled, which is necessary
to understand Table \ref{tab:correlation-anthropometric} and
    Table \ref{tab:intra-subject-consistency} updated
to address R3.C11 and R4.C1}
As the ground-truth BC
data were not available due to FOV truncation, the evaluation was
based on human perceptual study and two indirect quantitative
assessments:
\begin{enumerate*}[label=(\arabic*)]
\item intra-subject consistency; and
\item correlation with anthropometric approximations.
\end{enumerate*}

{\bf{Expert review for application validity}}. We designed a human
perceptual study to evaluate the application validity of the pipeline
output in BC assessment. The evaluation was based on a combined review
for the quality of reconstructed images and BC segmentations. We
formulated a quality score system with nine digital numbers from 1
(exceptional) to 9 (poor). This quality score was further stratified
into
\begin{enumerate*}[label=(\arabic*)]
\item {\bf{Succeed}} -- quality score $\le$ 3, for cases only with trivial
  defect or without any noticeable defect;
\item {\bf{Acceptable}} -- quality score between 4 and 6, for cases
  with certain defects but still can be included for downstream
  analysis; and
\item {\bf{Failed}} -- quality score $\ge$ 7, for cases that should be
  excluded from downstream analysis due major defects
\end{enumerate*}
Two trained clinical experts were asked to independently review each
case in a selected cohort. The quality scores were recorded together
with comments for the identified primary quality issue in each case.

{\bf{Intra-subject consistency}}. Multiple screens (e.g.,
annually) for the same subject are usually conducted in lung cancer
screening. Even though the BC profiles for the same individual may
change over time, the overall correlation between measurements on the
same subject should be stronger than the correlation between
measurements on different individuals. However, this intra-subject
consistency can be significantly reduced by the FOV limitation-caused
tissue truncation. Under this assumption, the benefit of the proposed
FOV extension method can be assessed by the improved overall
correlation in BC measurements between longitudinal screens.

{\bf{Correlation with anthropometric approximations}}. An
anthropometric approximation for whole-body fat mass (FM) and fat-free
mass (FFM) computed from weight and height was given in
\cite{Kuch2001}. The method was developed by fitting a non-linear
relationship with Bioelectrical Impedance Analyses results as
ground-truth. FFM was expressed as
\begin{equation}\label{eq:anthropometric-v2}
  \text{FFM (kg)} =
  \begin{cases*}
    5.1 \times H^{1.14} \times W^{0.41} & \text{~for men},\\
    5.34 \times H^{1.47} \times W^{0.33} & \text{~for women},
  \end{cases*}
\end{equation}
with $H$ and $W$ representing the height (m) and weight (kg) of the
subject, respectively. FM was computed by subtracting FFM from the
overall mass, i.e., $\text{FM (kg) = \text{weight (kg)}} - \text{FFM
  (kg)}$. FM and FFM indexes (kg/$\text{m}^2$) were defined by the
estimated FM and FFM normalized by height (m) square. The correlation
between measured BC indexes and anthropometric approximations were
usually used to evaluate the validity of CT-based BC assessment
\citep{McDonald2014, Pishgar2021} (muscle index vs. FFM index, and SAT
index vs. FM index). In our evaluation, we reported the improvement in
these correlations as evidence in support of the effectiveness of the
proposed method.

%% file: sec-experiment-result.tex
In this section, we introduce the data preparation, model development,
and evaluation of the proposed semantic FOV extension method.
Experiments and analyses were conducted in Python\texttrademark ~3.7.4,
PyTorch\texttrademark ~1.9.0, CUDA\texttrademark ~11.3, and R 4.1.2. The
pretrained pipeline is available in the form of docker container and
can be accessed by following the instructions at
\url{https://github.com/MASILab/S-EFOV}.

\subsection{Lung Screening CT Dataset}
In this study, we included two lung cancer screening CT datasets.

{\bf{CT arm of National Lung Screening Trial (NLST)}}. NLST
\citep{Schaapveld2011} is the largest randomized controlled trial to
evaluate the effectiveness of LDCT in lung cancer screening. 53,454
eligible participants were enrolled in the program from August 2002
through April 2004. 26,722 were randomly assigned to the CT
arm. Longitudinal data are available, with up to three annual screens
for those who continuously enrolled and have not been diagnosed with
lung cancer during previous screens. The anthropometric measurements,
including height and weight, were self-reported at the time of
enrollment right before the first screen. In this study, we randomly
sampled 1,280 subjects from the CT arm of NLST, with 3,586 available
LDCT scans in total.

{\bf{Vanderbilt Lung Screening Program (VLSP)}}. VLSP
(\url{https://www.vumc.org/radiology/lung}) is an on-going LDCT-based
lung cancer screening program conducted at Vanderbilt University
Medical Center. In this study, we used the VLSP data to develop our
two-stage pipeline as more untruncated cases are available. This
included 1,490 CT scans of 887 subjects enrolled since 2013. All data
were de-identified and acquired under internal review board
supervision (IRB\#181279).

The demographic and imaging protocol statistics of these two study
cohorts are summarized in Table \ref{tab:cohort}.
\RevTextBold{The smaller DFOV in NLST could be explained by the
strict requirement on the FOV size in the NLST imaging protocol,
where including unnecessary amount of additional body tissue beyond the lung was even considered
as a type of imaging quality issue \citep{Gierada2009, Schaapveld2011}.
Although this restriction might not be strictly enforced by later
lung screening programs equipped with more advanced scanner platforms than NLST (e.g.,
    minimum requirement for 16-slice MDCT),
optimization of the FOV to the lung field for each patient is still recommended \citep{Kazerooni2014, ACR2014}.
}{To address R3.C6.}

\begin{table}[!t]
  \caption{\label{tab:cohort} {\normalfont Characteristics of study
      cohorts. SD=Standard Deviation.} } \centering
  \footnotesize
  \begin{tabular}{lrr}\toprule
    \textbf{Characteristics} & \textbf{VLSP} & \textbf{NLST}\\\midrule
    \textbf{Demographic} & & \\
    \hspace{3mm} No. of subject &   887 &   1280 \\
    \hspace{3mm} No. of female (\%) & 399 (45.0) & 527 (41.2) \\
    \hspace{3mm} Age at baseline (y) $\pm$ SD & 64.0 $\pm$ 5.6 & 61.4 $\pm$ 4.9 \\
    \hspace{3mm} BMI at baseline (kg / m$^2$) $\pm$ SD & 28.3 $\pm$ 6.0 & 27.7 $\pm$ 4.8 \\\midrule
    \textbf{Imaging} & &\\
    \hspace{3mm} No. of scans   &  1490 &   3586 \\
    \hspace{3mm} Effective mAs $\pm$ SD & 45.6 $\pm$ 30.3 & 36.9 $\pm$ 7.9 \\
    \hspace{3mm} kVp $\pm$ SD & 119.0 $\pm$ 4.3 & 121.0 $\pm$ 4.4 \\
    \hspace{3mm} Display FOV (cm) $\pm$ SD & 36.9 $\pm$ 3.6 & 33.4 $\pm$ 3.4\\\bottomrule
  \end{tabular}
\end{table}

\subsection{Data Preparation}\label{sec:data-preparation}
\subsubsection{Image Quality Review}\label{sec:image-quality-review}
We reviewed all selected lung screening LDCT scans and excluded cases
with severe imaging artifacts. The type
of imaging artifacts included: incomplete imaging data,
imaging data corruption, beam hardening, cupping artifact,
severe imaging noise, and non-standard body
positioning. For all
qualified images, the FOV mask, lung mask, and body mask were
identified using the procedures described in \ref{app:the-masks}.
We further reviewed the generated region masks and excluded those cases with
significant defects. This combined review process filtered out
5 (0.3\%) scans of the VLSP cohort and 74 (2.1\%) scans of the NLST
cohort in total.
\RevTextBold{We identified one VLSP scan and two NLST scans that are
associated with cupping artifacts. This accounted for 0.06\% in all
included LDCT scans.
}{As evidence to support the resolution for comment R2.C3}
For all included scans, the levels of T5, T8, and T10 were estimated using
the vertebral level identification module developed in
\cite{Xu2022}. The per-slice TCI values were calculated based on Eq
(\ref{eq:TCI}).

\subsubsection{Candidate Slice Identification}
We used the untruncated slices in the VLSP dataset to generate the
synthetic data. As the intended application was focused on the BC
assessment using the T5, T8, and T10 axial slices, we first defined an
inclusion range to cover those slices with anatomy close to these
locations. Briefly, we defined a linear body part regression (BPR)
score for each slice based on the relative location to T5 and T10
levels, with T10 level of BPR score 0 and T5 level of BPR score
1. Then, all slices with BPR scores between -0.2 and 1.2 were marked
as in-range slices. Among the in-range slices, we further filtered out
those with zero TCI value, which indicated no tissue truncation in
these slices. This process filtered out 89,992 slices from 1,018 CT
volumes across 669 unique subjects.
Among these subjects, we random
sampled 549 subjects to form a training cohort, with 71,319 candidate
slices in total. Within the remaining subjects, we identified those with
at least one T5, T8, or T10 slice in the candidate slice set, and
split them into a 60 subject validation cohort, and a 60 subject
testing cohort. For the validation and testing cohort, we only
considered the T5, T8, and T10 slices, which led to 148 slices for
validation and 145 slices for testing.

\begin{figure*}[!t]
  \centering
  \includegraphics[scale=0.5]{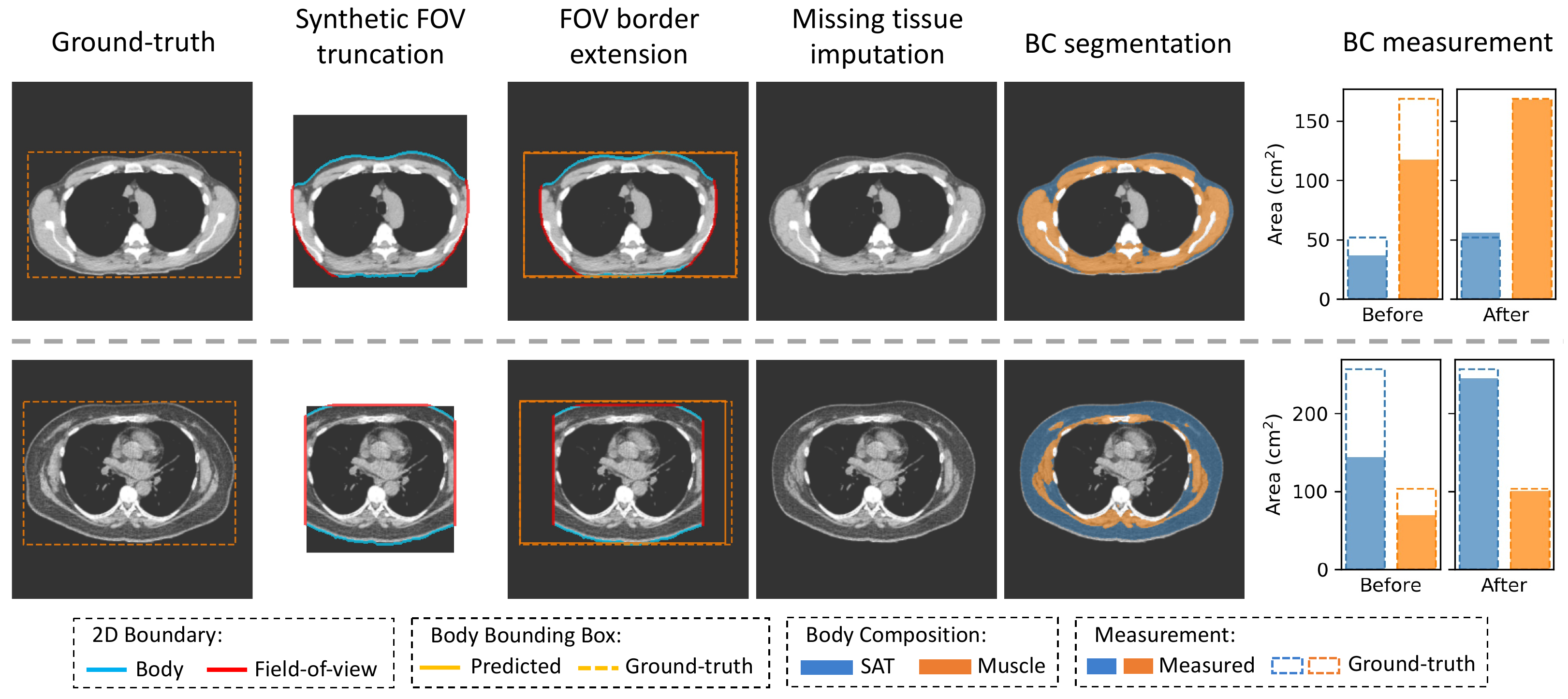}
  \caption{\label{fig:result-internal} {\normalfont Typical results of
      the developed FOV extension method on synthetic FOV truncation
      samples. The two ground-truth slices are selected at the fifth
      and eighth thoracic vertebral body levels from two patients with
      dramatically different body composition profiles. Synthetic FOV
      masks are applied to the raw slices, generating slices with
      tissue truncation. The predicted complete body bounding box is
      compared against the ground-truth bounding box. The anatomical
      consistency of the generated structures is evaluated by a prior
      developed BC assessment tool. The correction for BC assessment
      offset is demonstrated using the bar plots that compare the
      measurements on truncated or reconstructed slices against
      measurements on paired untruncated slices. FOV=Field-of-view.
      BC=Body Composition. SAT=Subcutaneous Adipose Tissue.} }
\end{figure*}

\begin{figure*}[!t]
  \centering
  \includegraphics[scale=0.67]{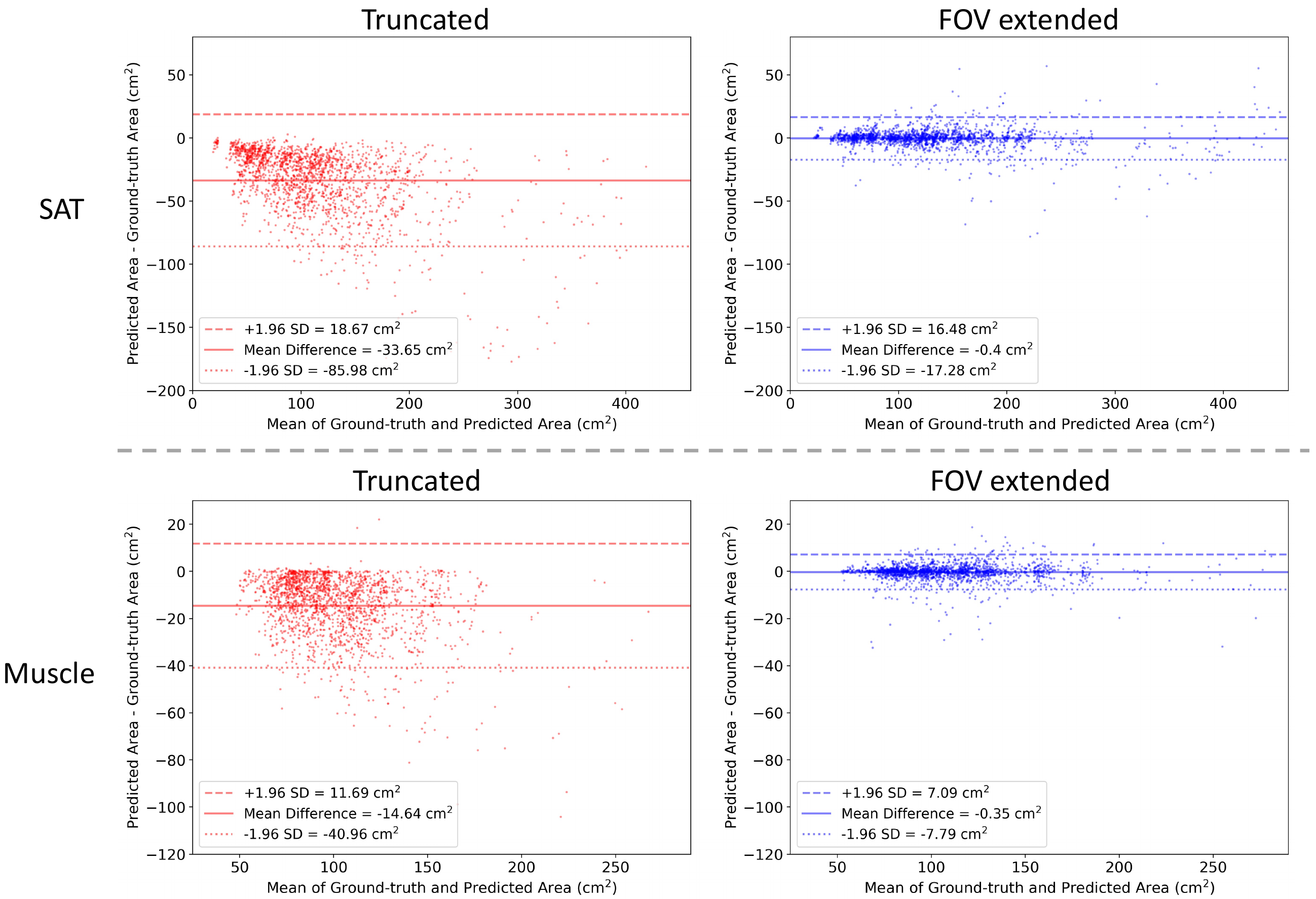}
  \caption{\label{fig:bland-altman} {\normalfont Correction of FOV
      truncation caused BC assessment offset by the
      proposed FOV extension method. The Bland-Altman plots compare
      the slice-wise body BC assessment performed on truncated or
      reconstructed slices with measurement performed on paired
      untruncated slices. The data are collected on 1,940 synthetic
      FOV truncation slices with mild to severe truncation severity
      level. SD=Standard Deviation. FOV=Field-of-view. BC=Body Composition.} }
\end{figure*}

\subsubsection{Synthetic Data Pairs}
\RevTextBold{For preprocessing of the CT slices,
the extraneous information outside the identified body mask
    (Section \ref{sec:image-quality-review} and \ref{app:the-masks}),
e.g., the scan tables and clothes,
 were removed by replacing the intensity of
pixels with HU intensity of air.
}{To address R3.C8}
 Then, the intensity window $\left[
  -150, 150 \right]$ (HU) was applied to highlight relevant
tissues. The resulting slices, together with pre-identified body
region masks, were further resized to $256\times 256$ before the
synthetic data generation procedure. We adopt the following
configuration for the synthetic data generation procedure introduced
in Section \ref{sec:synthetic-data}. The probability to
generate three types of truncation patterns $P_a$, $P_b$, and $P_c$,
were set to 0.5, 0.3, and 0.2, respectively, with emphasis on the
first two types of truncation patterns.  $R_{RFOV}$ was uniformly
sampled in range [0.6, 0.9], which determined the size of RFOV. For
the first two types of truncation patterns, the DFOV was automatically
identified once RFOV was given. For the third type pattern generation,
the $R_{DFOV}$ was uniformly sampled in range [0.7, 1.0], which
determined the size of DFOV. The offsets on two dimensions relative to
the image center, $X_{DFOV}$ and $Y_{DFOV}$, were sampled in $\left[
  -D/2, D/2 \right]$ , where $D = 256 \cdot (1 - R_{DFOV})\cdot
R_{RFOV}$ was the dimension difference between the RFOV diameter and
DFOV side length. This guarantees the generated DFOV contained inside
the extent of RFOV. On the CT slice augmentation, the random scale
ratio was sampled between 0.7 and 1.0. The maximum rotation degree was
set as 15\textdegree. The maximum translation in ratio of the image
dimension was 0.1 in the anterior-posterior direction and 0.2 in the
transverse direction. As a result of this random augmentation of raw CT
slice, the body region may extrude the original image border in some
cases, resulting in inaccurate body bounding box assessment. We
excluded these cases for the training of the body bounding box
estimation model, while keeping them for the image completion model
development.

In our implementation, the randomized synthetic sample generation
procedure was integrated into the model training process and applied
to each training CT slice when it was loaded. In contrast, the paired
data used in validation and testing are pre-generated and specifically
configured. Briefly, we generated 1,000 samples for each slice using
the same randomized synthetic data generation procedure for training
data. Cases with body regions that extruded the image boundary were
excluded. To best represent the application situation in lung cancer
screening, we excluded cases with incomplete lung regions. To balance
the data regarding different truncation severity, we set the limit of
maximum five cases for each severity levels defined in Section
\ref{sec:synthetic-data} for each slice. This resulted in 2,600
samples for validation and 2,657 samples for testing.

\subsection{Pipeline Development}
\label{sec:results-pipeline-development}
The development of the models in the proposed two-stage pipeline was
based on the same sets of training and validation data. The models
were first trained on the training set, and the best epoch was
selected as the one with the best performance on the validation
set. For all models, the input slices were normalized from $\left[
  -150, 150 \right]$ to range $\left[ -1, 1 \right]$, with non-FOV
region imputed with value 0.

{\bf{FOV border extension}}. We implemented the body bounding box
prediction model using ResNet-18 pre-trained on ImageNet as backbone,
with the last layer replaced with a fully connected layer with four
output channels representing the pixel-space coordinates of the
bounding box. We empirically set the weight $\lambda$ to 1500 to
balance the MSE term and GIoU term in loss function (Eq \ref{eq:combined_obj}). The model was
trained with batch-size of 20, optimized using Adam optimizer with
weight decay $1\times 10^{-4}$. The learning rate was set as $2 \times
10^{-3}$ . The model was trained for 200 epochs in total. The trained
model achieved the performance of $0.976 \pm 0.015$ in IoU on testing
samples. With the predicted bounding box, the extension ratio $R$ defined
as in Eq (\ref{eq:extension-ratio}) was determined to extend the image
border symmetrically to cover the estimated body region.
\RevTextBold{
Using the estimated $R_{est}$ alone (set $R=R_{est}$ in Eq \ref{eq:extension-ratio})
generated extended FOV covering complete body region only for 78.3\% of cases in the test set,
indicating the necessity of the extra extension $R_0$.
In our evaluation, a
5\% extra extension ($R_0=1.05$) was able to consistently produce
extended FOV border that covers the complete body region, with
98.4\% success rate on the test set.
}{To address R3.C13. Why the extra extension is necessary}

{\bf{Image completion}}. For image completion stage, we evaluated the
three published methods mentioned in Section \ref{sec:image-completion-methods}.
\RevTextBold{The detailed training configurations are given in
\ref{app:comp-image-completion}.}{All contents relevant to the comparison of the
three image completion solutions
are moved to appendix to address R4.C2.}

\subsection{Evaluations and Results}
\subsubsection{Evaluation on Synthetic Paired Data}
\label{sec:evaluation-synthetic-data-results}
We evaluated the developed FOV extension models on pre-generated
synthetic paired data following the methods described in Section
\ref{sec:evaluation-synthetic-data}.

\begin{figure}[!t]
  \centering
  \includegraphics[scale=0.67]{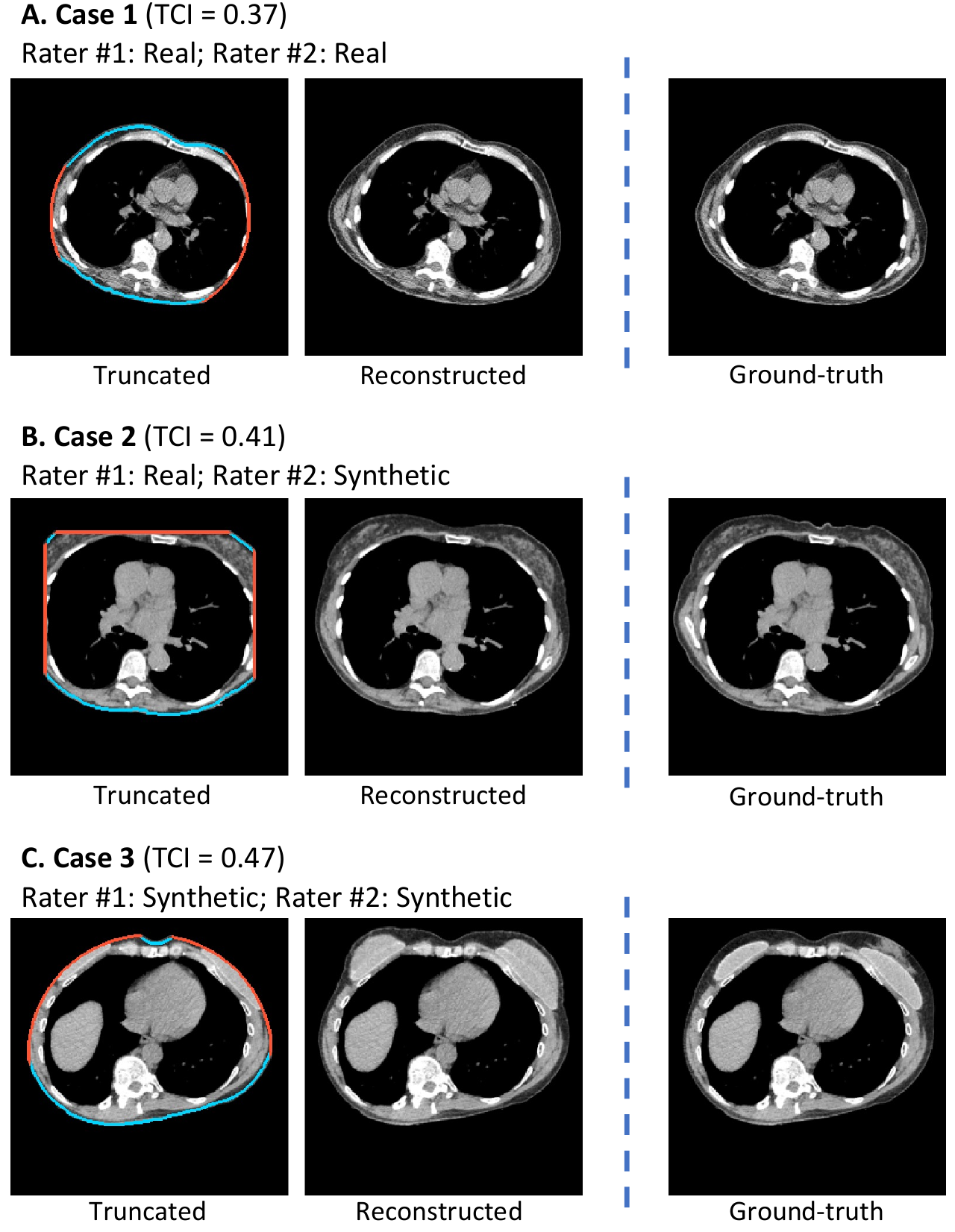}
  \caption{\label{fig:visual-turing-test-example} {\normalfont
  \RevTextBold{Example cases in visual Turing test to assess the effectiveness of the
  image completion stage (Section \ref{sec:evaluation-synthetic-data-results}). Two clinical experts were asked to
  identify synthetic cases independently from a mixed
  set of equally numbered synthetic and real cases.
  All three demonstrated cases are selected from the
  synthetic group, where the reconstructed (synthetic) slices were
  provided along with the truncation patterns
  to the raters instead of the real slices.
  For {\bf{Case 1}}, both raters misclassified it as real.
  For {\bf{Case 2}}, one rater classified it as real while the other rater
      classified it as synthetic. For {\bf{Case 3}}, both raters successfully
  identified the case as synthetic. TCI=Tissue Truncation Index.
  }{Added this figure as part of
  resolution for comment R1.C2 and R3.C12.}
  } }
\end{figure}

{\bf{Visual Turing test}}. We randomly sampled 100 synthetic samples
with TCI $> 0.3$ from the withheld testing dataset (Section
\ref{sec:data-preparation}). We prepared the data following the
practice of visual Turing test (Section
\ref{sec:evaluation-synthetic-data}). Two trained clinical experts
independently classified each case into fake or real category. The
mean accuracy of the two raters was 0.71, and inter-rater consensus
was 0.68.
\RevTextBold{On 35 out of 50 synthetic cases, at least one rater identified
the case properly. These cases were associated with slightly higher TCI value (indicating
more severe truncation)
comparing to the rest of the cases (0.47 $\pm$ 0.11 vs. 0.42 $\pm$ 0.08).
}{To address R1.C2. easiest / most difficult to distinguish by the observers.}
\RevTextBold{Fig \ref{fig:visual-turing-test-example} demostrate
the results of three example
cases selected from the synthetic group.}{Added to address comment R1.C2 and R3.C12.}

{\bf{Correction of BC measurement shift}}. Fig
\ref{fig:result-internal} shows the results on two samples for
qualitative evaluation. Both samples were generated using slices
without tissue truncation (zero TCI value). The BC assessment results
on untruncated slice were considered as the ground-truth
measurements. The measurements on truncated slice and FOV extended
slice were compared against the ground-truth to evaluate the
effectiveness of the correction. In Fig \ref{fig:bland-altman},
we use Bland-Altman plot to evaluate the capability of the method to
systematically correct the under-estimation of area ($\text{cm}^2$)
in the BC assessment
caused by FOV truncation. 
\RevTextBold{Table \ref{tab:bc-correction} shows the
effectiveness of the image completion method in restoring missing body tissues in the truncated regions,
which was assessed by pixel-wise RMSE, DSC, and BC measurements including
area ($\text{cm}^2$) and attenuation (HU) of SAT and muscle.}{To address R1.C1 and R2.C1.
Added the evaluations based on pixel-wise RMSE and BC attenuation.}
 The metrics to characterize the difference between
the ground-truth assessment and assessment without correction are
included to provide a reference.

\setlength{\tabcolsep}{5pt}
\begin{table*}[!t]
  \caption{\label{tab:bc-correction} {\normalfont \RevTextBold{Evaluation of
  the effectiveness of the image completion method in restoring
  missing body tissues in the truncated regions.
  The experiments were conducted on synthetic
  truncation pairs generated by the procedure described
  in Section \ref{sec:synthetic-data}, by evaluating the
  improvement comparing reconstructed slices to truncated slices in the agreement with untruncated slices.
  The evaluation was based on
  pixel-wise RMSE and DSC, as well as the
  BC measurements including area ($\text{cm}^2$) and
  attenuation (HU) of SAT and muscle, using the untruncated
  slice and the segmentation and measurements on it as ground-truth.
   The evaluation
  was stratified by truncation severity levels as defined in
  Section \ref{sec:synthetic-data}. The results are displayed
  in mean and SD for pixel-wise RMSE and DSC, and in
  RMSE and 95\% CI for BC measurements.
  Statistical significance in difference was assessed by Mann Whitney U Test for pixel-wise RMSE and DSC, and
  Wilcoxon Signed-rank Test on bootstrap samples for BC measurements. $P < 0.001$ for all comparisons between
  truncated and reconstructed. RMSE=Root Mean Square Error.
  DSC=Dice Similarity Coefficients. BC=Body Composition. SAT=Subcutaneous Adipose Tissue. HU=Hounsfield Unit.
  SD=Standard Deviation. CI=Confidence Interval.
  }
  {Updated this table to address comments R1.C1, R2.C1, and R4.C2.
  Added evaluation based on pixel-wise RMSE, DSC, and tissue attenuation.
  Moved the comparison
  between the three models to the appendix section to reduce the complexity of the main text.
  }
  } } \centering \footnotesize

   \begin{tabular}{lccccc}\toprule
   & \multicolumn{5}{c}{{\bf{Truncation Severity Level}}} \\\cmidrule(lr){2-6}
   \bf{Metric} & \bf{Trace ($N=715$)} & \bf{Mild ($N=715$)} & \bf{Moderate ($N=713$)} & \bf{Severe ($N=514$)} & \bf{Overall ($N=2657$)} \\\midrule
   \it{Pixel-wise RMSE (HU), Mean $\pm$ SD} & & & & & \\
       \hspace{3mm} \it{Truncated} & 5.33 $\pm$ 3.22 & 11.54 $\pm$ 5.23 & 17.18 $\pm$ 5.73 & 24.11 $\pm$ 7.83 & 13.81 $\pm$ 8.66 \\
       \hspace{3mm} \it{Reconstructed} & 2.18 $\pm$ 1.20 & 4.80 $\pm$ 2.09 & 7.55 $\pm$ 2.77 & 11.49 $\pm$ 3.52 & 6.12 $\pm$ 4.09 \\
   \it{Dice Similarity Coefficient, Mean $\pm$ SD} & & & & &\\
       \hspace{3mm} \it{Truncated}     & 0.98 $\pm$ 0.02 & 0.94 $\pm$ 0.03 & 0.88 $\pm$ 0.05 & 0.79 $\pm$ 0.07 & 0.90 $\pm$ 0.08 \\
       \hspace{3mm} \it{Reconstructed} & 0.99 $\pm$ 0.01 & 0.98 $\pm$ 0.02 & 0.96 $\pm$ 0.03 & 0.93 $\pm$ 0.04 & 0.97 $\pm$ 0.03 \\
   \it{SAT Area ($\text{cm}^2$), RMSE (95\% CI)} & & & & & \\
       \hspace{3mm} \it{Truncated}     & 7.88 (7.36, 8.57) & 20.48 (19.26, 22.34) & 39.09 (36.88, 42.09) & 65.41 (62.31, 68.82) & 36.95 (35.46, 38.45) \\
       \hspace{3mm} \it{Reconstructed} & 1.69 (1.45, 2.10) & 3.85 (3.37, 4.51)    & 8.23 (6.91, 10.32)   & 12.91 (11.59, 14.70) & 7.42 (6.79, 8.26) \\
   \it{SAT Attenuation (HU), RMSE (95\% CI)} & & & & & \\
       \hspace{3mm} \it{Truncated}     & 0.57 (0.53, 0.63) & 1.37 (1.28, 1.50) & 2.71 (2.49, 3.09) & 4.33 (4.06, 4.66) & 2.49 (2.36, 2.64) \\
       \hspace{3mm} \it{Reconstructed} & 0.34 (0.31, 0.40) & 0.80 (0.72, 0.92) & 1.26 (1.15, 1.45) & 2.04 (1.86, 2.27) & 1.20 (1.13, 1.29) \\
   \it{Muscle Area ($\text{cm}^2$), RMSE (95\% CI)} & & & & & \\
       \hspace{3mm} \it{Truncated}     & 3.59 (3.25, 3.96) & 10.81 (10.08, 11.77) & 18.35 (17.46, 19.42) & 29.39 (27.72, 31.41) & 17.08 (16.38, 17.87) \\
       \hspace{3mm} \it{Reconstructed} & 0.69 (0.61, 0.77) & 2.18 (1.86, 2.73)    & 3.54 (3.14, 4.25)    & 5.56 (4.89, 6.50)    & 3.28 (2.99, 3.63) \\
   \it{Muscle Attenuation (HU), RMSE (95\% CI)} & & & & & \\
       \hspace{3mm} \it{Truncated}     & 0.39 (0.33, 0.50) & 1.26 (1.12, 1.45) & 2.41 (2.24, 2.62) & 3.10 (2.87, 3.38) & 1.97 (1.87, 2.08) \\
       \hspace{3mm} \it{Reconstructed} & 0.20 (0.19, 0.23) & 0.61 (0.55, 0.67) & 1.00 (0.91, 1.14) & 1.48 (1.34, 1.67) & 0.89 (0.84, 0.97) \\\bottomrule
   \end{tabular}
\end{table*}
\setlength{\tabcolsep}{6pt}

\subsubsection{Evaluation on Real FOV Truncation Scans}
\label{sec:evaluation-real-fov-scans}
In addition to the evaluation of synthetic FOV truncation samples, we
evaluated our pipeline on real lung cancer screening LDCT scans
following the methods described in Section
\ref{sec:evaluation-real}.
\RevTextBold{Table \ref{tab:truncation-statistics} shows the
statistics of the scan-wise truncation severity levels and the anthropometric
characteristics of each level in the two included the lung screening LDCT datasets.
NLST scans were associated with more severe truncation compared to VLSP scans, which
consistent with the even restricted DFOV size in NLST compared to VLSP (Table \ref{tab:cohort}).
Scans with more severe truncation were associated with higher weight, BMI, and FM index in both of the datasets.
}{As part of resolution for R4.C1.}
\RevTextBold{For LDCT scans in both VLSP and NLST,}{To address R3.C11. Add VLSP results.}
we obtained the BC
assessment results using both the original version of BC assessment
pipeline developed in \cite{Xu2022} and the enhanced version with the
FOV extension module integrated to correct the measurement offsets
caused by FOV truncation.

\begin{table*}[!t]
  \caption{\label{tab:truncation-statistics} {\normalfont
  \RevTextBold{
  Anthropometric characteristics of scan truncation severity levels.
      The included scans are those filtered out by image quality review and
      with complete height and weight data. BMI=Body Mass Index. FM=Fat Mass. FFM=Fat-free Mass.
  }{To address R4.C1.}
  } } \centering
  \footnotesize
  \begin{tabular}{llcccccc}\toprule
    \textbf{Cohort} & \bf{Severity Level} & \bf{No. of Scans (\%)} & \textbf{Height (m)} & \textbf{Weight (kg)} & \textbf{BMI (kg/m$^2$)} &
    \textbf{FM Index (kg/m$^2$)} & \textbf{FFM Index (kg/m$^2$)} \\\midrule
    \multirow{6}{*}{\textbf{VLSP}}
    & \it{None} & 78 (6.3\%) & 1.70 $\pm$ 0.10 & 70.3 $\pm$ 15.7 & 24.1 $\pm$ 4.3 & 6.8 $\pm$ 3.2 & 17.4 $\pm$ 1.7 \\
    & \it{Trace} & 690 (55.4\%) & 1.72 $\pm$ 0.10 & 79.4 $\pm$ 16.8 & 26.6 $\pm$ 4.6 & 8.4 $\pm$ 3.5 & 18.2 $\pm$ 1.7\\
  & \it{Mild} & 355 (28.5\%) & 1.72 $\pm$ 0.11 & 90.1 $\pm$ 17.9 & 30.5 $\pm$ 5.1 & 11.6 $\pm$ 4.1 & 18.9 $\pm$ 1.8\\
  & \it{Moderate} & 107 (8.6\%) & 1.69 $\pm$ 0.10 & 98.5 $\pm$ 25.2 & 34.4 $\pm$ 7.5 & 15.1 $\pm$ 5.9 & 19.2 $\pm$ 2.3\\
  & \it{Severe} & 16 (1.3\%) & 1.67 $\pm$ 0.07 & 104.2 $\pm$ 20.6 & 37.2 $\pm$ 5.9 & 18.1 $\pm$ 5.0 & 19.1 $\pm$ 1.5\\\cmidrule(lr){2-8}
  & \it{Overall} & 1,246 & 1.72 $\pm$ 0.10 & 83.8 $\pm$ 19.5 & 28.3 $\pm$ 5.8 & 9.9 $\pm$ 4.6 & 18.4 $\pm$ 1.9\\\midrule
    \multirow{6}{*}{\textbf{NLST}}
    & \it{None} & 18 (0.5\%) & 1.70 $\pm$ 0.09 & 66.1 $\pm$ 17.9 & 22.6 $\pm$ 4.1 & 5.65 $\pm$ 2.5 & 17.0 $\pm$ 1.9\\
    & \it{Trace} & 638 (18.2\%) & 1.75 $\pm$ 0.10 & 79.4 $\pm$ 17.2 & 25.7 $\pm$ 4.3 & 7.4 $\pm$ 3.1 & 18.3 $\pm$ 1.6\\
  & \it{Mild} & 1,190 (34.0\%) & 1.75 $\pm$ 0.09 & 82.9 $\pm$ 16.4 & 27.0 $\pm$ 4.1 & 8.4 $\pm$ 3.1 & 18.6 $\pm$ 1.6\\
  & \it{Moderate} & 1,214 (34.7\%) & 1.71 $\pm$ 0.10 & 83.1 $\pm$ 18.3 & 28.4 $\pm$ 4.7 & 10.1 $\pm$ 3.5 & 18.3 $\pm$ 1.9\\
  & \it{Severe} & 442 (12.6\%) & 1.68 $\pm$ 0.10 & 89.0 $\pm$ 20.4 & 31.2 $\pm$ 5.5 & 12.7 $\pm$ 4.2 & 18.6 $\pm$ 1.9\\\cmidrule(lr){2-8}
  & \it{Overall} & 3,502 & 1.73 $\pm$ 0.10 & 83.0 $\pm$ 18.0 & 27.8 $\pm$ 4.8 & 9.3 $\pm$ 3.7 & 18.4 $\pm$ 1.8\\
  \bottomrule
  \end{tabular}
\end{table*}

\begin{figure}[!t]
  \centering
  \includegraphics[scale=0.34]{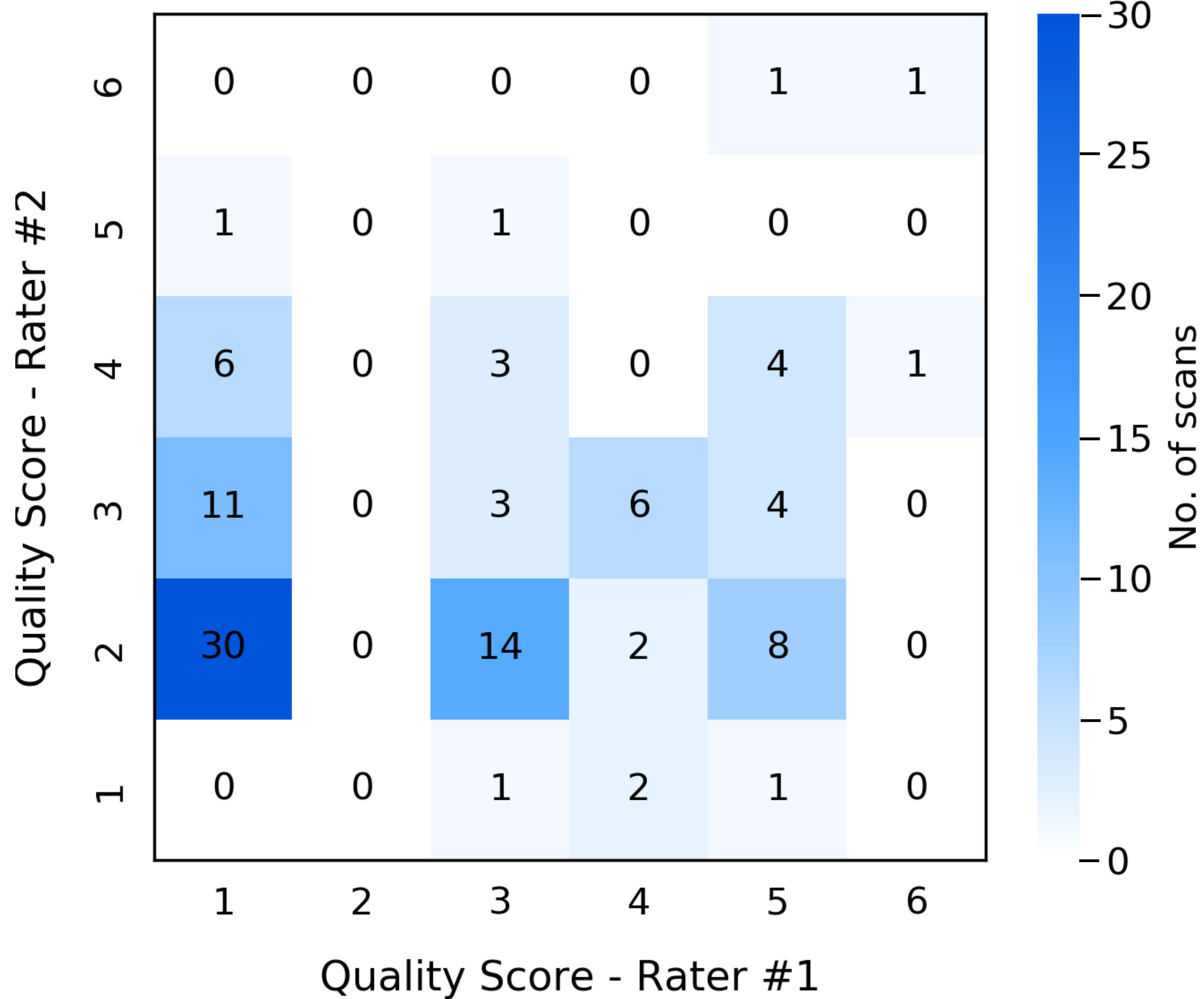}
  \caption{\label{fig:nlst-qa} {\normalfont
  \RevTextBold{Distribution of quality scores given by two expert raters on the
  FOV extension quality of 100 random NLST scans with moderate to severe tissue truncation.
  The quality score system was defined from 1 (exceptional) to 9 (poor).
  As none of the sampled scans was assigned with score higher than 6 (Failed),
      we reduce the range as from 1 to 6 in the confusion matrix plot.}
  {To address comment R1.C2. Add a table/fig for expert review results.}
  } }
\end{figure}

\begin{figure*}[!t]
  \centering
  \includegraphics[scale=1.10]{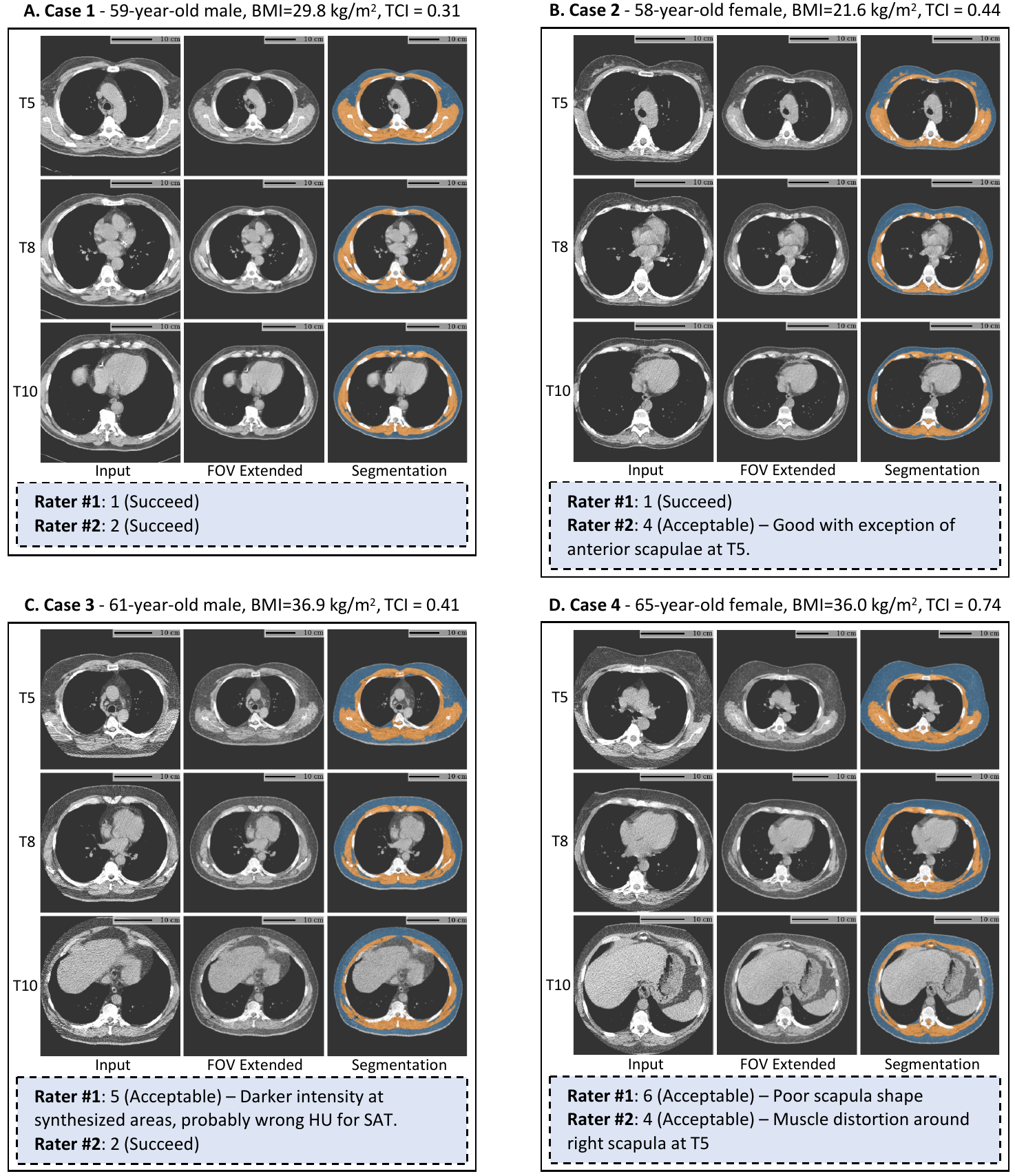}
  \caption{\label{fig:result-external} {\normalfont
  \RevTextBold{Results of BC assessment with FOV extension (image grid) and expert reviews (blue box) on
      lung screening LDCT with limited FOV.
  The pipeline identifies the axial slices
  corresponding to the T5, T8, and T10 vertebral bodies from 3D
  chest CT volume (first column). The FOV of each slice is
  extended with missing body tissue imputed (second column). The measurements of BC
  are based on the segmentation masks predicted on the
  FOV extended images (third column). The four selected cases are among the
  100 NLST scans with moderate to severe truncation (TCI $>$ 0.3)
      included for the expert quality review conducted with
  two trained experts (Section \ref{sec:evaluation-real-fov-scans}).
   All CT slices are displayed using window [-150,
        150] HU. BC=Body Composition. FOV=Field-of-view. BMI=Body Mass Index. TCI=Tissue Truncation Index.}
  {Updated this figure to address comment R1.C2. Expand to four cases and add the rating scores
  to the image so the reader can understand what kind of results were typically
      represented by the quality review categories, togather with brief case descriptions.}
  } }
\end{figure*}

{\bf{Expert review for application validity}}.
We randomly sampled 100 NLST scans with moderate to severe
tissue truncation (TCI $> 0.3$) and presented the FOV extended images
and final segmentation masks to two trained clinical experts who
independently reviewed and rated the quality of the results for
downstream application. None of the cases were rated as {\bf{Failed}}
(quality score $\ge 7$). 39 cases were labeled as {\bf{Acceptable}}, with
quality score between 4 and 6, by at least one rater and with at least
one clear description for identified issue. Even with identifiable
defects, these results were considered as valid for downstream
analysis by both raters. 
\RevTextBold{Those labeled as {\bf{Acceptable}} were associated with
slightly higher TCI value (more severe truncation) comparing to those considered {\bf{Succeed}}
(0.46 $\pm$ 0.10 vs. 0.43 $\pm$ 0.10).
BMI distributions were similar between those labeled as {\bf{Acceptable}} (28.7 $\pm$ 5.8)
and those labeled as {\bf{Succeed}} (28.6 $\pm$ 4.7).}{To address R1.C2. Add additional characterizations for the
Failed/Acceptable stratification}
The identified defect include:
\begin{enumerate*}[label=(\arabic*)]
\item unrealistic scapula shape, 24 cases;
\item unrealistic intensity in subcutaneous region, 4 cases;
\item body extrude the image border, 2 cases; and
\item unrealistic breast implant shape in female, 1 case.
\end{enumerate*}
\RevTextBold{However, the inter-rater consistency was poor, with Intraclass Correlation
Coefficient between the two raters being 0.15. By categorizing the ratings into
{\bf{Failed}} (score $\le 3$) and {\bf{Acceptable}} (score $> 3$),
    the Cohen's Kappa Coefficient was 0.09 (none to slight agreement) between the two raters.
}{To address comment R4.C4}
\RevTextBold{The distribution of rating scores given by the
two raters is presented in Fig \ref{fig:nlst-qa}.
Fig \ref{fig:result-external} shows the
pipeline results, quality scores, and review comments
on four cases included for this quality review.
}
{To address comment R1.C2.}

{\bf{Intra-subject consistency}}.
\RevTextBold{For NLST, we identified all longitudinal
pairs, e.g., the baseline screen and second follow-up screen of the
same subject, which resulted in 3,110 pairs. As
the time distance between two consecutive screens for the same subject was
approximately fixed at one year based on NLST protocol, we put the
identified pairs into two categories: {\it{1-year-pair}} (2,081
pairs) and {\it{2-year-pair}} (1,029 pairs).
In consistency with NLST, we identified those longitudinal pairs in
VLSP with time distance between 0.5 and 1.5 years and categorized these pairs
as {\it{1-year-pair}} (505 pairs), while longitudinal pairs with time distance
between 1.5 and 2.5 years were identified and categorized as {\it{2-year-pair}} (191 pairs).
The longitudinal pairs were further stratified into different truncation
severity level, where the pair-wise severity level was defined as the
maximum severity level of the two paired scans.
The correlations between the measurement results on two
longitudinal scans were assessed using Spearman’s rank correlation
coefficients with and with FOV extension.
Statistical significance in difference between the correlations with
and with FOV extension were
assessed by the method of \cite{silver2004testing}, which compared two
dependent correlations with non-overlapping variables.
The results are summarized in Table
\ref{tab:intra-subject-consistency}.}
{To address R3.C11 and R4.C1. Update experiments on intra-subject consistency.}

{\bf{Correlation with anthropometric approximation}}. 
\RevTextBold{
We obtained the anthropometric approximations of FFM, and FM indexes (kg/m2)
based on the formulas described in Section \ref{sec:evaluation-real}.
For VLSP, the required height and weight data were obtained
before each of LDCT screen, with 1,246
scans with available corresponding anthropometric metrics.
For NLST, the anthropometric data were obtained
at enrollment right before the baseline screens.
For this reason, the effectiveness of the
approximation in NLST was expected to be strongest for the baseline screens and
decreased for the follow-up screens. Thus, we
categorized the NLST scans based on screen years: 1) {\it{Screen-0}} (1,232
scans); {\it{Screen-1}} (1,158 scans); and {\it{Screen-2}} (1,112 scans).
{\it{Screen-(0, 1, 2)}} represent the baseline screen, first
follow-up screen, and second follow-up screen, respectively.
We further categorized the scans by truncation severity levels.
We assessed the correlation between measured SAT index and FM index and
the correlation between muscle index and FFM index using Spearman’s
rank correlation coefficients with and with FOV extension.
Statistical significance in difference between the correlations with
and with FOV extension were
assessed by the method of \cite{hittner2003monte}, which compared two
dependent correlations with overlapping variables.
The results are summarized in Table \ref{tab:correlation-anthropometric}.
}{To address R3.C11 and R4.C1. Update the anthropometric correlation tests}

\begin{table*}[!t]
\caption{\label{tab:intra-subject-consistency} {\normalfont 
\RevTextBold{The comparisons of intra-subject consistency between
BC measurements on longitudinal scan pairs with and without FOV extension.
The scan-wise BC measurements were defined as the summation of BC areas measured on the cross-sectional
slices at T5, T8, and T10. For NLST, the intra-subject pairs were formed by the possible
combinations of two screens for each subject, categorized into {\it{1-year-pair}} and {\it{2-year-pair}} based on
the years of trial the two scans were conducted.
For VLSP, pairs with time distance between 0.5 and 1.5 years were categorized as {\it{1-year-pair}}, while
pairs with time distance between 1.5 and 2.5 years were categorized as {\it{2-year-pair}} in consistency with NLST.
The pairs are further categorized into different truncation severity levels defined
as the maximum of scan-wise truncation severity levels of the two paired scans.
Spearman’s rank correlation coefficients with 95\% confidence interval are reported.
Results on groups with too few cases (e.g., less than six pairs) are not included.
As FOV extension is not needed for pairs without truncation, only results on raw images are displayed when it is possible.
Statistical significance are evaluated by method of \citep{silver2004testing} for comparison between two non-overlapping
dependent correlations. The tests are two-sided, with p-values reported.
BC=Body Composition. FOV=Field-of-view. SAT=Subcutaneous Adipose Tissue.
}{To address R3.C11 and R4.C1. 1) Add results on VLSP; 2) provide "reference" measurements.}
}} \centering \footnotesize
\begin{tabular}{llcccccc}\toprule
& & \multicolumn{3}{c}{{\bf{SAT area (cm$^2$)}}} & \multicolumn{3}{c}{{\bf{Muscle area (cm$^2$)}}}\\\cmidrule(lr){3-5}\cmidrule(lr){6-8}
\bf{Cohort} & \bf{Group (no. of pairs)} & Without Correction & FOV Extended & p-value & Without Correction & FOV Extended & p-value\\\midrule
\multirow{18}{*}{\textbf{VLSP}}& \it{No Truncation} & & & & & & \\
&\hspace{3mm} \it{1-year-pair (N=13)} & 0.967 (0.845, 1.000) & - & - & 0.918 (0.659, 0.994) & - & - \\
&\hspace{3mm} \it{2-year-pair (N=6)} & 0.886 (0.200, 1.000) & - & - & 0.771 (0.000, 1.000) & - & - \\
&\it{Trace Truncation} & & & & & & \\
&\hspace{3mm} \it{1-year-pair (N=233)} & 0.946 (0.923, 0.961) & 0.941 (0.915, 0.958) & $<$.001 & 0.940 (0.914, 0.958) & 0.946 (0.923, 0.961) & $<$.001 \\
&\hspace{3mm} \it{2-year-pair (N=79)} & 0.894 (0.809, 0.947) & 0.893 (0.803, 0.951) & .89 & 0.955 (0.921, 0.971) &0.943 (0.895, 0.965) & .69\\
&\it{Mild Truncation} & & & & & & \\
&\hspace{3mm} \it{1-year-pair (N=181)} & 0.929 (0.896, 0.949) & 0.933 (0.899, 0.954) & .64 & 0.936 (0.908, 0.956) & 0.946 (0.922, 0.963) & .001\\
&\hspace{3mm} \it{2-year-pair (N=72)} & 0.886 (0.797, 0.941) & 0.890 (0.805, 0.944) & .69 & 0.911 (0.814, 0.962) & 0.925 (0.860, 0.961) & .49 \\
&\it{Moderate Truncation} & & & & & & \\
&\hspace{3mm} \it{1-year-pair (N=69)} & 0.810 (0.701, 0.881) & 0.908 (0.840, 0.947) & $<$.001 & 0.865 (0.751, 0.931) & 0.944 (0.901, 0.964) & $<$.001 \\
&\hspace{3mm} \it{2-year-pair (N=31)} & 0.762 (0.502, 0.898) & 0.906 (0.749, 0.967) & $<$.001 & 0.849 (0.663, 0.945) & 0.904 (0.806, 0.944) & .50 \\
&\it{Severe Truncation} & & & & & & \\
&\hspace{3mm} \it{1-year-pair (N=9)} & 0.400 (-0.381, 0.890) & 0.617 (-0.242, 1.000) & .02 & 0.850 (0.352, 1.000) & 0.983 (0.817, 1.000) & .16\\
&\hspace{3mm} \it{2-year-pair (N=3)} & - & - & - & - & - & -\\\cmidrule(lr){2-8}
&\it{Overall} & & & & & & \\
&\hspace{3mm} \it{1-year-pair (N=505)} & 0.950 (0.937, 0.959) & 0.961 (0.949, 0.969) & $<$.001 & 0.930 (0.910, 0.946) & 0.952 (0.940, 0.961) & $<$.001\\
&\hspace{3mm} \it{2-year-pair (N=191)} & 0.927 (0.895, 0.947) & 0.941 (0.915, 0.959) & $<$.001 & 0.927 (0.889, 0.953) & 0.938 (0.912, 0.957) & .42\\\midrule
\multirow{18}{*}{\textbf{NLST}} & \it{No Truncation} & & & & & &\\
&\hspace{3mm} \it{1-year-pair (N=4)} & - & - & - & - & - & -\\
&\hspace{3mm} \it{2-year-pair (N=1)} & - & - & - & - & - & -\\
& \it{Trace Truncation} & & & & & &\\
&\hspace{3mm} \it{1-year-pair (N=207)} & 0.952 (0.932, 0.965) & 0.946 (0.923, 0.960) & $<$.001 & 0.934 (0.897, 0.959) & 0.927 (0.880, 0.955) & .028 \\
&\hspace{3mm} \it{2-year-pair (N=105)} & 0.955 (0.920, 0.972) & 0.951 (0.914, 0.970) & $<$.001 & 0.934 (0.886, 0.962) & 0.937 (0.898, 0.958) & .84 \\
& \it{Mild Truncation} & & & & & &\\
&\hspace{3mm} \it{1-year-pair (N=645)} & 0.936 (0.922, 0.949) & 0.936 (0.921, 0.948) & .96 & 0.942 (0.930, 0.952) & 0.936 (0.922, 0.947) & .21 \\
&\hspace{3mm} \it{2-year-pair (N=329)} & 0.885 (0.849, 0.911) & 0.886 (0.852, 0.915) & .50 & 0.912 (0.885, 0.933) & 0.909 (0.880, 0.932) & .76 \\
& \it{Moderate Truncation} & & & & & &\\
&\hspace{3mm} \it{1-year-pair (N=853)} & 0.901 (0.877, 0.918) & 0.926 (0.903, 0.942) & $<$.001 & 0.943 (0.935, 0.950) & 0.952 (0.943, 0.959) & $<$.001 \\
&\hspace{3mm} \it{2-year-pair (N=388)} & 0.875 (0.836, 0.904) & 0.898 (0.859, 0.927) & .026 & 0.941 (0.928, 0.950) & 0.953 (0.942, 0.960) & $<$.001 \\
&\it{Severe Truncation} & & & & & &\\
&\hspace{3mm} \it{1-year-pair (N=372)} & 0.797 (0.751, 0.837) & 0.916 (0.892, 0.934) & $<$.001 & 0.861 (0.818, 0.891) & 0.919 (0.893, 0.939) & $<$.001 \\
&\hspace{3mm} \it{2-year-pair (N=206)} & 0.782 (0.707, 0.841) & 0.894 (0.854, 0.921) & $<$.001 & 0.833 (0.775, 0.875) & 0.917 (0.877, 0.942) & $<$.001 \\\cmidrule(lr){2-8}
&\it{Overall} & & & & & &\\
&\hspace{3mm} \it{1-year-pair (N=2,081)} & 0.912 (0.901, 0.922) & 0.950 (0.942, 0.957) & $<$.001 & 0.944 (0.938, 0.949) & 0.951 (0.946, 0.956) & $<$.001\\
&\hspace{3mm} \it{2-year-pair (N=1,029)} & 0.880 (0.859, 0.898) & 0.925 (0.909, 0.938) & $<$.001 & 0.933 (0.924, 0.940) & 0.948 (0.940, 0.955) & $<$.001\\
\bottomrule
\end{tabular}
\end{table*}

\setlength{\tabcolsep}{5.5pt}
\begin{table*}[!t]
\caption{\label{tab:correlation-anthropometric} {\normalfont
\RevTextBold{
The comparisons of correlation between the height-square
normalized BC area indexes and anthropometric approximated BC
indexes with and without FOV extension. SAT index (cm$^2$/m$^2$) was compared against FM
index (kg/m$^2$), and Muscle index (cm$^2$/m$^2$) was compared
against FFM index (kg/m$^2$). For VLSP, the anthropometric approximations
were derived using weight and height obtained before each of the scan.
For NLST, the approximations were derived from the height and weight information
obtained at the baseline. Thus, we further stratified the NLST scans into years of
the screen in addition to the truncation severity level.
Spearman’s rank correlation
coefficients with 95\% confidence interval are reported.
Statistical significance are evaluated by the method of \citep{hittner2003monte}
for comparison between two overlapping dependent correlations.
The tests are two-sided, with p-values reported. BC=Body Composition. FOV=Field-of-view. SAT=Subcutaneous Adipose
    Tissue. FM=Fat Mass. FFM=Fat-free Mass.
}{To address R3.C11 and R4.C1. 1) Add VLSP results; 2) provide the reference measurements}
}} \centering \footnotesize
\begin{tabular}{llcccccc}\toprule
  & & \multicolumn{3}{c}{{\bf{SAT index (cm$^2$/m$^2$) vs. FM index (kg/m$^2$)}}} &
    \multicolumn{3}{c}{{\bf{Muscle index (cm$^2$/m$^2$) vs. FFM index (kg/m$^2$)}}}\\\cmidrule(lr){3-5}\cmidrule(lr){6-8}
    \bf{Cohort} & \bf{Severity Level (no. of scans)} & Without Correction & FOV Extended & p-value & Without Correction & FOV Extended & p-value\\\midrule
    \multirow{5}{*}{\bf{VLSP}}
    & \it{None (N=78)} & 0.780 (0.657, 0.862) & - & - & 0.692 (0.545, 0.801) & - & -\\
    & \it{Trace (N=690)} & 0.798 (0.765, 0.826) & 0.801 (0.769, 0.829) & .27 & 0.658 (0.612, 0.703) & 0.681 (0.638, 0.721) & $<$.001\\
    & \it{Mild (N=355)} & 0.782 (0.734, 0.822) & 0.789 (0.743, 0.828) & .02 & 0.605 (0.532, 0.671) & 0.659 (0.596, 0.717) & $<$.001\\
    & \it{Moderate to Severe (N=123)} & 0.653 (0.535, 0.750) & 0.673 (0.556, 0.767) & .36 & 0.362 (0.172, 0.528) & 0.579 (0.415, 0.709) & $<$.001\\\cmidrule(lr){2-8}
    & \it{Overall (N=1,246)} & 0.837 (0.815, 0.855) & 0.845 (0.823, 0.862) & $<$.001 & 0.606 (0.567, 0.646) & 0.668 (0.634, 0.699) & $<$.001\\\midrule
    \multirow{20}{*}{\bf{NLST}}
    & \it{None} & & & & & &\\
    & \hspace{3mm} \it{Screen-0} (N=9) & 0.762 (-0.078, 1.000) & - & - & 0.905 (0.315, 1.000) & - & - \\
    & \hspace{3mm} \it{Screen-1} (N=5) & - & - & - & - & - & -\\
    & \hspace{3mm} \it{Screen-2} (N=4) & - & - & - & - & - & -\\
    & \it{Trace} & & & & & &\\
    & \hspace{3mm} \it{Screen-0} (N=226) & 0.737 (0.653, 0.806) & 0.735 (0.649, 0.805) & .08 & 0.698 (0.611, 0.768) & 0.713 (0.632, 0.778) & .004 \\
    & \hspace{3mm} \it{Screen-1} (N=211) & 0.712 (0.611, 0.792) & 0.717 (0.618, 0.795) & .15 & 0.708 (0.613, 0.785) & 0.725 (0.635, 0.796) & .01\\
    & \hspace{3mm} \it{Screen-2} (N=201) & 0.702 (0.616, 0.776) & 0.700 (0.615, 0.773) & .41 & 0.704 (0.612, 0.780) & 0.724 (0.633, 0.794) & .19\\
    & \it{Mild} & & & & & & \\
    & \hspace{3mm} \it{Screen-0} (N=421) & 0.760 (0.707, 0.812) & 0.762 (0.708, 0.811) & .48 & 0.705 (0.645, 0.754) & 0.714 (0.655, 0.766) & .65\\
    & \hspace{3mm} \it{Screen-1} (N=379) & 0.727 (0.663, 0.778) & 0.730 (0.670, 0.781) & .84 & 0.716 (0.656, 0.766) & 0.721 (0.661, 0.772) & .49\\
    & \hspace{3mm} \it{Screen-2} (N=390) & 0.711 (0.648, 0.764) & 0.710 (0.647, 0.767) & .34 & 0.707 (0.645, 0.761) & 0.698 (0.634, 0.754) & .98\\
    & \it{Moderate to Severe} & & & & & & \\
    & \hspace{3mm} \it{Screen-0} (N=577) & 0.762 (0.723, 0.795) & 0.818 (0.785, 0.845) & $<$.001 & 0.709 (0.662, 0.748) & 0.746 (0.705, 0.779) & $<$.001 \\
    & \hspace{3mm} \it{Screen-1} (N=562) & 0.711 (0.665, 0.750) & 0.758 (0.718, 0.791) & $<$.001 & 0.705 (0.656, 0.748) & 0.741 (0.696, 0.780) & $<$.001\\
    & \hspace{3mm} \it{Screen-2} (N=517) & 0.718 (0.668, 0.764) & 0.745 (0.696, 0.786) & $<$.001 & 0.668 (0.617, 0.715) & 0.712 (0.669, 0.752) & $<$.001 \\\cmidrule(lr){2-8}
    & \it{Overall} & & & & & & \\
    & \hspace{3mm} \it{Screen-0} (N=1,232) & 0.778 (0.750, 0.803) & 0.818 (0.795, 0.839) & $<$.001 & 0.714 (0.687, 0.741) & 0.752 (0.727, 0.776) & $<$.001 \\
    & \hspace{3mm} \it{Screen-1} (N=1,158) & 0.759 (0.729, 0.786) & 0.802 (0.777, 0.825) & $<$.001 & 0.699 (0.667, 0.729) & 0.745 (0.718, 0.770) & $<$.001 \\
    & \hspace{3mm} \it{Screen-2} (N=1,112) & 0.744 (0.712, 0.773) & 0.782 (0.753, 0.807) & $<$.001 & 0.694 (0.659, 0.726) & 0.731 (0.700, 0.757) & $<$.001 \\
  \bottomrule
\end{tabular}
\end{table*}
\setlength{\tabcolsep}{6pt}

%% file: sec-discussion.tex
\noindent
{\bf{Effectiveness of semantic FOV extension}}.

\vspace{2mm}
In this work, we
proposed a two-stage framework for semantic FOV extension of lung
screening LDCT scans with limited FOV. For the first stage, our
results indicated the trained model can successfully identify the
bounding box of the complete body region given CT slice with limited
FOV (third column of Fig \ref{fig:result-internal}).
\RevTextBold{With an
empirical extension ratio multiplier ($R_0=1.05$), the model can
 reliably extend the FOV border as such it can cover the complete
body region (Section \ref{sec:results-pipeline-development}).
}{To address R3.C13. Rephrased this sentence.} For the second stage, the proposed training strategy
produced models that could effectively predict the missing tissues in
truncated regions (Fig \ref{fig:result-internal}, Fig
\ref{fig:result-external}, and Fig \ref{fig:image-completion-comparison}).
\RevTextBold{The proposed training strategy was effective for all three considered
general-purpose image completion methods.
A detailed comparison of the performance of these methods
is given in \ref{app:comp-image-completion}.}
{Address comment R4.C2. Move the comparison of the image completion methods to the appendix.}

In the visual Turing test, the mean
accuracy of the two experts to discriminate synthetic image from the
real image was only 0.71, even on the most difficult cases (TCI $>$
0.3) and with hints of the potential synthetic region of the
image.
\RevTextBold{The effectiveness of the image completion was
 confirmed by the reduction in
pixel-wise RMSE (Table \ref{tab:bc-correction}).
The anatomical consistency of the predicted contents was further confirmed by the
BC assessment result, including the improved agreement in BC segmentation
with original slice (Fig \ref{fig:result-internal},Table
\ref{tab:bc-correction} and Fig \ref{fig:result-external}) and the correction for the BC
assessment offset caused by FOV truncation (Table
\ref{tab:bc-correction} and Fig \ref{fig:bland-altman}).
}{As part of the resolution to address R1.C1 and R2.C1.}

Combining these observations, the proposed method successfully extended the FOV
border and generated anatomical consistent contents in truncated
regions.

\vspace{2mm}

\noindent
{\bf{Application validity for CT-based BC assessment in lung cancer
    screening}}.

\vspace{2mm}
To evaluate the application validity of the proposed
method, we integrated the trained semantic FOV extension pipeline into
a previously developed BC assessment pipeline as one additional
processing module. The expert review of the application validity indicated that
the results were reliable even on scans with relative severe FOV
tissue truncation (TCI $>$ 0.3). In certain cases that were associated
with partially or entirely missing scapula, the reconstructed scapula
bone structures could be distorted, which accounted for the primary
cause of defect identified in expert review
(Section \ref{sec:evaluation-real-fov-scans},
Fig \ref{fig:result-external} Case-2 and Fig \ref{fig:result-external} Case-4).
However, the BC analysis results
on these cases were still considered acceptable as most missing
anatomical structures of BC components were recovered anatomically
consistently (Fig \ref{fig:result-external}).

\RevTextBold{The FOV
extension correction significantly improved the overall intra-subject
correlation for both SAT and muscle measurements in both of the two included datasets (Table
\ref{tab:intra-subject-consistency}).
The improvement was consistent in both 1-year-pairs and 2-year-pairs.
Lowered consistency were observed in 2-year-pairs, which could be explained by
the longitudinal change in BC.
The evaluation stratified by truncation severity level revealed that the overall
improvement was mainly contributed by the significant improvement
for those pairs with one or two scans with moderate to severe truncation.
Among the pairs that consisted of scans with only trace truncation severity,
slight decreases in longitudinal consistency of
certain comparisons were observed after the FOV extension.
This might be caused by the potential measurement error introduced by the decreased
resolution due to the extension.

The overall correlations of SAT and muscle indexes with anthropometric approximated
FFM and FM indexes were also improved significantly in both of the included datasets (Table
\ref{tab:correlation-anthropometric}).
The stratified evaluation indicated
the improvement was more significant for those scans with moderate to severe truncation.
In NLST, the overall improvement was consistent for all three screen years. The
correlation decreased with the increase in time distance between the scan year and baseline.
This could be explained by the longitudinal change in BC, while the anthropometric approximations
in NLST were obtained at baseline.
In both NLST and VLSP, it is common for each of the severity groups alone to have lowered
correlations compared to the overall dataset, even for those with none and trace truncation.
This could be explained by the narrowed body composition distribution
in each strata compared to the overall dataset (Table \ref{tab:truncation-statistics}),
where the variations in measurements themself
could obscure the overall trend between the two measurements.
Differences between VLSP and NLST were observed, where VLSP was associated with higher
correlation between SAT index and FM index and lower correlation between Muscle index and FFM index.
This can possibly be explained by the demographic difference between the two datasets (Table \ref{tab:cohort}).
Nevertheless, with the FOV correction, we find stronger correlations with anthropometric approximations in both
VLSP (SAT index vs. FM index: 0.85; Muscle index vs. FFM index: 0.67) and NLST
(Screen-0. SAT index vs. FM index: 0.82; Muscle index vs. FFM index: 0.75) comparing to the same correlations
recently reported in \citep{Pishgar2021} (SAT index vs. FM index: 0.80; Muscle index vs. FFM index: 0.62) on a subset of
Multi-Ethnic Study of Atherosclerosis, where
the BC measurements were derived by a semi-automatic regional assessment approach using routine chest CT.
However, the effectiveness of this comparison may subject to the potential
demographic difference between the cohorts included in
these two studies.

These results indicated that the developed semantic FOV extension method improved
the overall BC measurement quality and demonstrated
the application validity of the method in opportunistic BC assessment using lung screening LDCT.
}{To address R3.C11 and R4.C1. Update result on the two
application validity experiments. The discussion is extended.}

\vspace{3mm}
\RevTextBold{

\noindent
{\bf{Limitations}}

\vspace{2mm}
{\bf{Generalizability to limited FOV CT scans acquired with other clinical indications.}}
The semantic FOV extension method presented in this study was developed and
tested on lung screening LDCT.
These chest CT scans were acquired with specific imaging protocols, e.g., non-contrast, low dose, and
    optimized FOV, for a specific target
population - older asymptomatic current and former heavy smokers, and for a specific clinical indication -
early detection of lung cancer.
Thus, the proposed method may not be well generalizable to CT scans acquired with
different imaging protocols, target population, or clinical indications. Most noticeably, the cupping
artifact, which is considered as common in CT scans with limited
FOV \citep{Ohnesorge2000, Hsieh2004, Sourbelle2005, Ketola2021, Huang2021}, was not addressed
in this study.
Although this decision was intentional based on the extremely low
occurrence of this issue observed in both of the two included lung
screening LDCT datasets (Section \ref{sec:data-preparation}),
we recognize this as a limitation in terms of generalizability of the presented method.
Specifically, the developed method is only applicable when
the FOV truncations are caused by RFOV and DFOV, or a combination of both, and
the truncations in projection data caused by SFOV do not have significant impact on the
reconstructed image intensity inside the CT FOV.
In addition, the inability of the developed pipeline to process scans with common imaging artifacts,
e.g., beam hardening artifact, severe imaging noise, and non-standard body positioning
(Section \ref{sec:data-preparation}), may pose challenges for the application of the method in a more heterogeneous scenario.
As the primary focus of this present study was on the
application in lung screening LDCT, we left the development for a
thorough solution to address these issues for future studies.

To further characterize the generalizability and potential limitations when applying to conventional chest CT scans,
we evaluated our developed method on a third dataset, which consisted of chest CT scans acquired with a broader
spectrum of clinical indications in daily clinical practice.
The details of this evaluation are presented in
\ref{app:generalizability-analysis}.

\vspace{2mm}
{\bf{Reference measurements for application validity assessments.}}
    In the application validity assessments on
    real lung screening LDCT data (Section \ref{sec:evaluation-real}),
    we employed the anthropometric approximations
    for FM and FFM indexes initially developed in \cite{Kuch2001} as references to assess
    the FOV extension method's effectiveness in
    correcting the BC measurement error introduced by FOV tissue truncation.
    Though the same method has been used in later study \cite{Pishgar2021} to provide references
    in assessing the validity of CT-based SAT and muscle measurements, the
    anthropometric assessments of BC are known to subject to lowered sensitivity \citep{Thibault2012}.
    Other validated whole body BC measurements,
    e.g., those obtained by Bioelectrical Impedance Analysis or Dual-energy X-ray
    Absorptiometry, can provide validated
    references to assess the improvement in BC measurement accuracy \citep{Thibault2012}.
    In addition, using the raw projection data of the scans truncated by limited RFOV and DFOV to
    recover the missing tissues would provide the ultimate references to assess the
    effectiveness of the FOV extension method in real application
    (Section \ref{sec:truncation-simulation}). However, due to limitations of the data source
    we were unable to perform these analyses.
}{Add limitations to address R2.C3, R3.C7, R3.C15, and R4.C1.
1) the cupping artifact and other generalizability issue; 2) inability of the pipeline to process
common imaging artifacts; 3) the lack of reference body composition
measurement for truncated cases}

%% file: sec-conclusion.tex
In this paper, we proposed a two-stage framework for slice-wise
semantic FOV extension for lung screening LDCT scans with body tissue
truncation caused by limited FOV. In the first stage, given a CT slice
with incomplete body region, we predicted the extent of the
complete body in the form of axis-aligned minimum bounding box. Based
on this estimation, the original FOV border was extended to cover the
estimated complete body region. The second stage was formulated as an
image completion problem, where the model was trained to impute the
missing body tissues in extended space. To generate paired data for
both model development and evaluation, we designed a synthetic sample
generation procedure simulating the FOV determination mechanism during
CT acquisition. To evaluate the anatomical consistency of the
generated body tissues, we utilized the pre-trained models in a
previously developed deep learning BC assessment pipeline for lung cancer
screening LDCT. In addition, we integrated the developed semantic FOV
extension method into the BC assessment pipeline and evaluated the
effectiveness of the method to correct BC assessment shift in the real
application.

We developed the semantic FOV extension pipeline on a large lung
cancer screening cohort. Evaluation results on synthetic samples
indicated the developed pipeline can effectively identify the extent
of the complete body and generate anatomically consistent tissues in
the truncated regions. The pipeline also consistently corrected the BC
measurement shifts caused by FOV truncation for CT slices with various
truncation severity. \RevTextBold{To evaluate the validity on real
FOV truncated data, we applied
the BC assessment pipeline with semantic FOV extension module
on the FOV truncation scans in VLSP and a subsample of the CT arm of NLST.
We observed
improvements in overall intra-subject consistency and overall correlation with
anthropometric approximations for FFM and FM in both datasets.
}{
To address R3.C11. Add VLSP into these test.
}

The proposed method demonstrates the possibility for extending the FOV
of CT image from the semantic image extension perspective.
Instead of
utilizing the CT projection data as mainly focused by current
literature, our method seeks to solve the problem by learning a deep
representation for complete body structures from a large dataset with complete
body structures in FOV. For this reason, the developed method
only requires data in the image domain as input, making it possible to
extend the limited FOV in applications where the CT projection data are
not available.
\RevTextBold{
However, the methodology is only applicable to certain types of limited FOV where
the truncations are caused by RFOV and DFOV, or a combination of both, and
the cupping artifacts that are associated with truncations caused by SFOV do not manifest.
}{To address R2.C3. Restrict the contribution to specific
types of CT FOV truncation}

From the clinical point-of-view, our work provides a solution for the
prevalent issue of limited FOV in opportunistic CT-based BC
assessment. Compared to the currently widely adopted solution of
regional assessment, our method allows for the whole axial slice
evaluation which is better correlated with full body evaluation and
makes use of all visible anatomical information in the FOV. As such,
BC assessment with our method has the potential to be more
realistically correlated with clinical outcomes. A rigorous clinical
comparison of the two approaches may be a valuable future study
direction. In addition, the fully automatic BC assessment for lung
screening LDCT has the potential to extend the value of the LDCT-based lung
cancer screening, especially with the semantic FOV extension
correction proposed in this work.

%% file: sec-acknowledge.tex
This research is supported or partly supported by the following
awards: NSF CAREER 1452485; R01 EB017230; R01 CA253923; U01 CA196405
to Massion; Grant UL1 RR024975-01of the National Center for Research
Resources and now Grant 2 UL1 TR000445-06 at the National Center for
Advancing Translational Sciences; Martineau Innovation Fund Grant
through the Vanderbilt-Ingram Cancer Center Thoracic Working Group;
NCI Early Detection Research Network 2U01CA152662 to PPM; and IBM PhD
Fellowship.

The authors thank the National Cancer Institute for access to
the data collected by the National Lung Screening Trial.

%% file: sec-appendix-comp-image-completion.tex
\RevText{This appendix section was added to address comment R4.C2, which collected all contents
relevant the comparison of the three image completion methods}

Following the strategy described in Section \ref{sec:image-completion-methods},
we evaluated the following three general-purpose methods for image completion:
\begin{itemize}
\item {\bf{pix2pix}} \citep{Isola2017}. The method employed a
  conditional generative adversarial network (cGAN) to provide a
  solution for general-purpose image-to-image translation problems
  where image completion was among the example applications. The cGAN
  model learned a transformation from observed image space to target
  image space. The objective was to optimize the generator to produce
  predictions that cannot be distinguished by an adversarially trained
  discriminator, which is known as the adversarial loss. In addition
  to the adversarial loss, a traditional $L_1$ loss was also included to
  encourage the generator to produce a fake image that is near to the
  ground-truth. The final objective function was a combination of
  these two terms. The FOV region mask was not required for either the
  training or inference phase for pix2pix.
\item {\bf{PConv-UNet}} \citep{Liu2018}. The method was based on the
  concept of partial convolution (PConv) which was a replacement for
  the normal convolution to handle the existence of invalid region in
  input images. In partial convolution operation, a mechanism was
  designed such that the outputs were only conditioned on valid pixels
  specified by a binary mask. The mask was updated through each layer
  in the forward pass, leading to a gradually shrinking invalid
  region. After passing through a sufficient number of stacked layers,
  the invalid region would eventually vanish. The method used a
  UNet-like structure similar to pix2pix as the generator, replacing
  all normal convolutional layers with partial convolution layers. The
  loss function contained multiple terms targeting both per-pixel
  reconstruction accuracy and semantic composition. This included
  \begin{enumerate*}[label=(\arabic*)]
  \item $L_1$ loss;
  \item perceptual loss and style loss based on features extracted by
    a pretrained VGG-16 model; and
  \item total variation loss.
  \end{enumerate*}
  A valid region mask was required during both the training and
  inference phase.
\item {\bf{RFR-Net}} \citep{Li2020}. Instead of imputing all missing
  regions in a single pass, the Recurrent Feature Reasoning (RFR)
  network took a recurrent strategy to first infer the contents near
  the valid region and then use the information as clues for further
  inference iteratively. In each iteration, the newly updated area was
  identified as the difference between the initial mask and the
  updated mask after several partial convolution layers. The missing
  features in this area were filled with a feature reasoning module
  with a UNet-like structure. The features generated in each iteration
  were merged to form a final feature map to generate the
  reconstructed image. In addition, an attention mechanism was
  introduced to improve the semantic consistency of the generated
  contents, especially with image contents at a far distance. The loss
  function was similar to PConv-UNet, including
  \begin{enumerate*}[label=(\arabic*)]
  \item $L_1$ loss; and
  \item perceptual loss and style loss based on features extracted by
    a pretrained VGG-16 model.
  \end{enumerate*}
  A valid region mask was also required for both training and
  inference stages.
\end{itemize}

In training, we set the
batch-size to 20 and total training epoch to 80 for all three methods.
For PConv-UNet and
RFR-Net, we set the first 40 epoch in initial training mode and the
last 40 epochs in finetune mode. For the convenience of reproducing
and comparison, we followed the original literatures \citep{Isola2017,
  Li2020, Liu2018} for the configuration of other training settings,
which included the number of layers in models, the balancing weights
of loss function terms, the optimization configurations, and learning
rates.

\begin{figure*}[!t]
  \centering
  \includegraphics[scale=0.45]{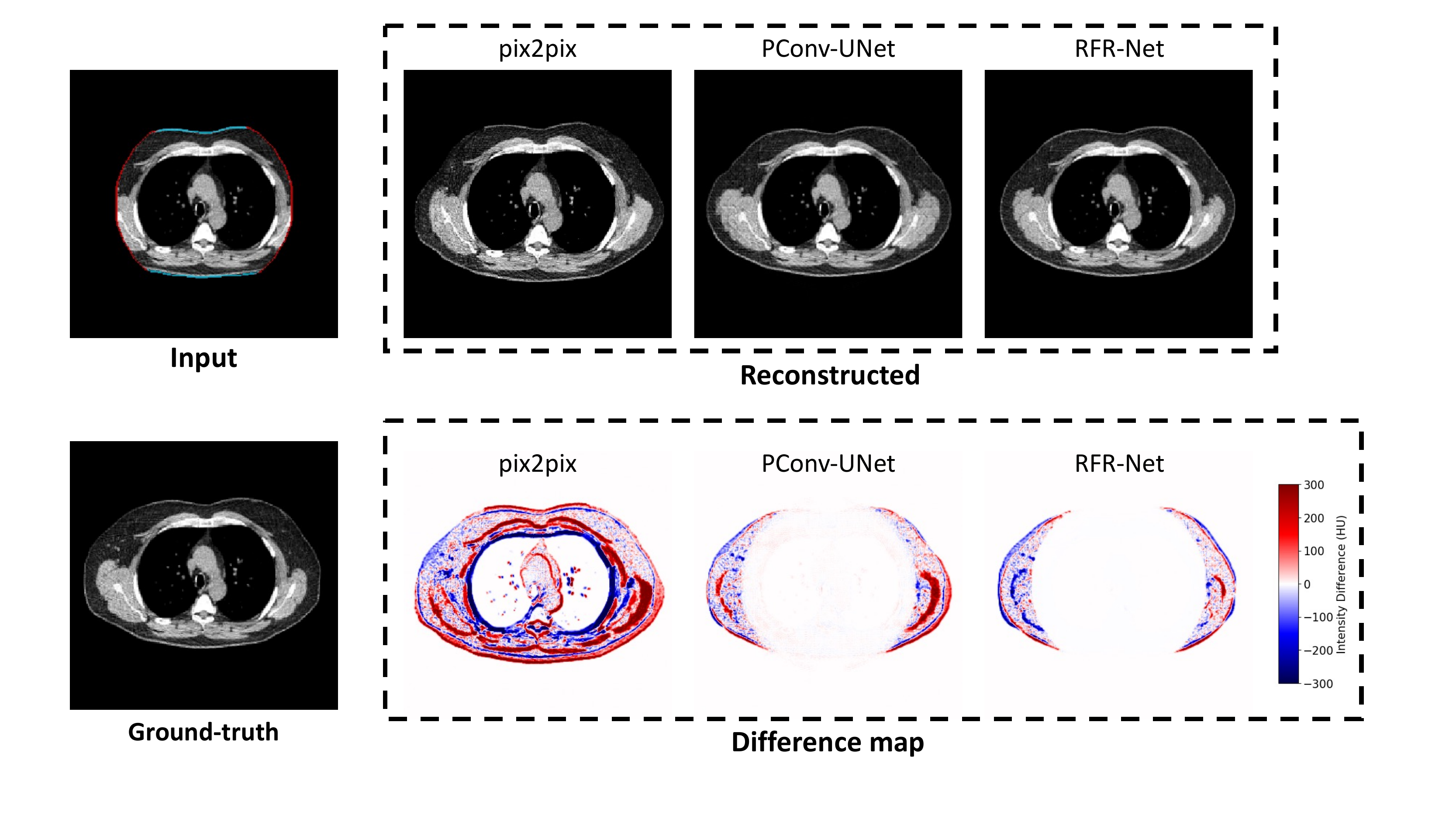}
  \caption{\label{fig:image-completion-comparison} {\normalfont
      Comparison between three image completion methods for the
      missing tissue imputation in truncated region. The reconstructed
      slices show the direct outputs of each method. The difference
      maps are generated by subtracting the reconstructed slices with
      ground-truth slice.} }
\end{figure*}

\begin{table*}[!t]
  \caption{\label{tab:image-completion-comparison} {\normalfont
  \RevTextBold{
  Quantitative comparison of the three general-purpose image completion models.
  The evaluation was based on
  pixel-wise RMSE and DSC, as well as the
  BC measurements including area ($\text{cm}^2$) and
  attenuation (HU) of SAT and muscle, using the untruncated
  slice and the segmentation and measurements on it as ground-truth.
  RMSE=Root Mean Square Error. DSC=Dice Similarity Coefficients.
  BC=Body Composition. SAT=Subcutaneous Adipose Tissue. HU=Hounsfield Unit.
  SD=Standard Deviation. CI=Confidence Interval.
  }{Updated to address R1.C1 and R2.C1. Include additional evaluation metrics.}
  }} \centering \footnotesize
   \begin{tabular}{lccc}\toprule
   & \multicolumn{3}{c}{{\bf{Image completion method}}} \\\cmidrule(lr){2-4}
   \bf{Metric} & \bf{pix2pix} & \bf{PConv-UNet} & \bf{RFR-Net} \\\midrule
   Pixel-wise RMSE (HU), Mean $\pm$ SD & 7.48 $\pm$ 4.48 & 6.52 $\pm$ 4.05 & 6.12 $\pm$ 4.09 \\
   Dice Similarity Coefficient, Mean $\pm$ SD & 0.96 $\pm$ 0.04 & 0.96 $\pm$ 0.04 & 0.97 $\pm$ 0.03 \\
   SAT Area ($\text{cm}^2$), RMSE (95\% CI) & 9.83 (9.11, 10.70) & 8.79 (8.14, 9.66) & 7.42 (6.79, 8.26) \\
   SAT Attenuation (HU), RMSE (95\% CI) & 1.91 (1.80, 2.04) & 1.48 (1.40, 1.60) & 1.20 (1.13, 1.29) \\
   Muscle Area ($\text{cm}^2$), RMSE (95\% CI) & 3.95 (3.67, 4.29) & 3.58 (3.32, 3.91) & 3.28 (2.99, 3.63) \\
   Muscle Attenuation (HU), RMSE (95\% CI) & 1.16 (1.10, 1.24) & 1.02 (0.96, 1.09) & 0.89 (0.84, 0.97) \\\bottomrule
   \end{tabular}
\end{table*}

We characterized the difference
between three image completion solutions. Fig
\ref{fig:image-completion-comparison} shows the results on the same
synthetic sample by the three methods, where the generated
reconstruction images are compared with the ground-truth image. The
residue maps are shown as the heatmaps. The quantitative comparison of the
three image completion methods is given in Table \ref{tab:image-completion-comparison}.
 As shown in Fig
\ref{fig:image-completion-comparison} and Table
\ref{tab:image-completion-comparison}, all three models were able to impute
anatomically
plausible contents in the truncated regions and correct the BC
measurement offset. However, there was a size change in the
reconstructed image generated by the pix2pix model (Fig
\ref{fig:image-completion-comparison}). This could be caused by the emphasis of the
loss function toward the semantic consistency by the adversarial term
instead of the reconstruction accuracy. The pix2pix model was
originally optimized for image-to-image translation of natural images
in which size change is irrelevant \citep{Isola2017}. However, this
property was undesired for medical image applications like BC
assessment. This caveat was further confirmed by the inferior performance in
BC measurement correction comparing to two other methods (Table
\ref{tab:image-completion-comparison}). For PConv-UNet and RFR-Net, both methods
maintained the size of the subject. On qualitative evaluation, RFR-Net
generated anatomical structures closer to the ground-truth, which was
a possible advantage of its recurrent inference strategy (Fig
\ref{fig:image-completion-comparison}). This observation was further
confirmed by the quantitative results in Table \ref{tab:image-completion-comparison}
as RFR-Net outperformed the other two methods consistently for all considered metrics.

%% file: sec-appendix-the-masks.tex
The development and evaluation of the proposed method replied on
multiple binary masks specifying certain regions in thoracic CT
volume. This included the FOV mask, lung mask, and body mask. Here, we
describe the solutions we employed in our study to automatically
generated these masks.

{\bf{FOV mask}}. As introduced in Section
\ref{sec:truncation-simulation}, once the cross-sectional FOV pattern
is determined, it is replicated through all cross-sectional slices in
a CT volume. The pixels inside this 3D FOV region are considered
valid, with intensity value representing the HU of the physical
material at the represented spatial location, while the pixels in the
non-FOV region are without any physical correspondence and need to be
imputed with a predefined value to form the final square shape
image. In the DICOM standard, this value is termed Pixel Padding
Value, which, by its design, should be outside of the range of normal
HU for the ease of identification of the non-FOV regions in
application. However, in real application, depending on the scanner
manufactures, the value may fall in the normal range of HU or without
specification in the header data structure. For this reason, we
designed the following algorithm to retrieve the FOV mask following a
data-driven approach.

With the assumption that the imputation value is constant across all
axial slices of the same CT scan, the intensity variation along the
vertical direction of the axial plane is zero in the non-FOV
cross-sectional region. On the contrary, this
variation is non-zero for the FOV cross-sectional region due to the
intensity difference of different materials and intrinsic noise during
data acquisition. Based on this observation, we obtained the FOV masks
by identifying the cross-sectional region with non-zero vertical
intensity variations. This trick reliably retrieved the FOV masks for
most of the lung screening CT scans in our study cohort. In a small
amount of CT scans, mainly in the NLST, the lower bound of
intensity window was set as -1000 (HU of air), which lead to zero
vertical intensity variation regions even inside the FOV region. We
mitigated the problem by retrieving the convex hull of the identified
non-zero vertical intensity variation region.

\RevTextBold{
When padding and resizing the CT slices with FOV truncation and corresponding FOV masks
as described in Section
\ref{sec:method-two-stage-framework},
care must be taken to avoid
the interpolation between the FOV and non-FOV regions, as the
operators may lead to unrealistic pixel values near the FOV borders.
If such interpolation is unavoidable, a certain amount of recession of FOV regions
and corresponding intensity correction are necessary to eliminate the
artificial pixel values. The amount of this recession can be estimated by
applying the same operator on the floating-point binary FOV masks,
and identifying the pixels with value between zero and one.
}{Add this note along with the contents added for addressing R2.C4}

{\bf{Lung mask}}. The lung masks were generated using the segmentation
model developed in \citep{Hofmanninger2020}, with pretrained model
available at (\url{https://github.com/JoHof/lungmask}). The model was
based on 2D UNet and performed prediction of lung region on individual
slices. The model was developed on a large and diverse cohort with
wide range of variation in FOV.

{\bf{Body mask}}. We used the morphology-based 3D body mask
identification tool initially developed in \citep{Tang2021} for body
region identification in abdominal and whole-body CT. The method
converted the input image into a binary mask using HU threshold of
-500. Then, the largest connected region was identified, which was
followed by slice-wise hole filling operation to impute the lung
regions.

%% file: sec-appendix-generalizability-analysis.tex
\begin{table*}[!t]
\caption{\label{tab:failure-mode} {\normalfont Representative cases
    for identified failure mode in LiVU dataset. All images are
    displayed using window [-150, 150] HU.
    FOV=Field-of-view. BC=Body Composition. SAT=Subcutaneous Adipose Tissue.
} }
\centering
\footnotesize
\begin{tabular}{l|m{30mm}|ccc}\toprule
  {\bf{Failure mode}} & \multicolumn{1}{c|}{\bf{Explanation}} & {\bf{Input}} & {\bf{FOV extended}} & {\bf{BC segmentation}} \\\midrule

  {\bf{Body positioning}}
  & Non-standard body positioning with arms in FOV, resulting in addition area measured comparing to normal positioning.
  & \raisebox{-0.45\height}{\includegraphics[scale=0.9]{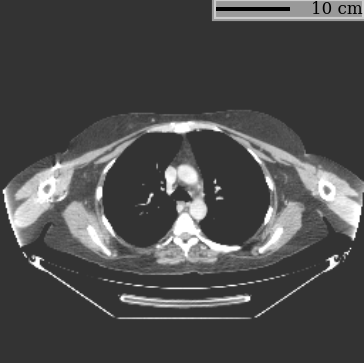}}
  & \raisebox{-0.45\height}{\includegraphics[scale=0.9]{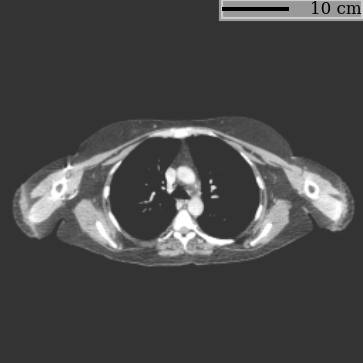}}
  & \raisebox{-0.45\height}{\includegraphics[scale=0.9]{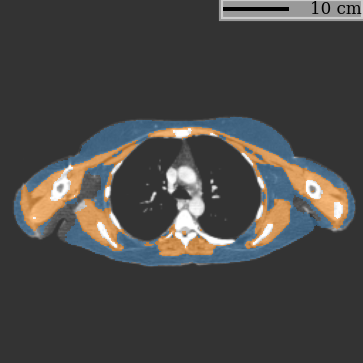}}\\\midrule

  {\bf{Cupping artifact}} & The increased intensity at the FOV border
  causing shift from standard HU. The method cannot mitigate this
  artifact, resulting in inaccurate BC segmentation.
  & \raisebox{-0.45\height}{\includegraphics[scale=0.9]{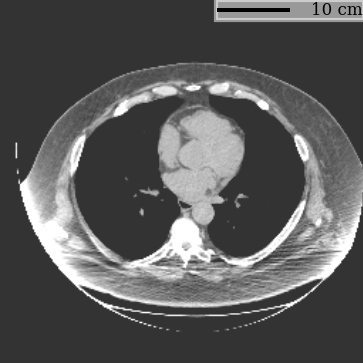}}
  & \raisebox{-0.45\height}{\includegraphics[scale=0.9]{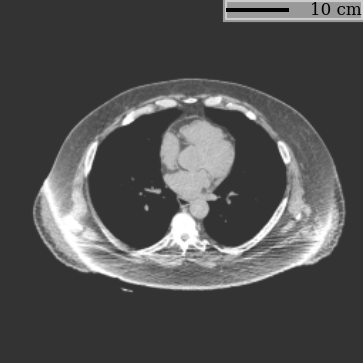}}
  & \raisebox{-0.45\height}{\includegraphics[scale=0.9]{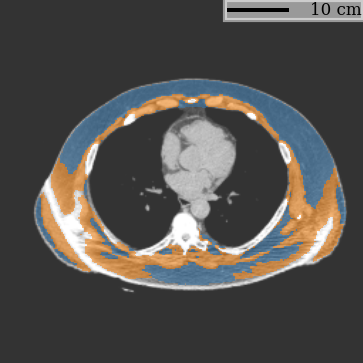}}\\\midrule

  {\bf{Intravenous contrast}}
  & Typical observed failure pattern associated with contrasted scans. Parts of muscle and SAT are missing in the segmentation mask.
  & \raisebox{-0.45\height}{\includegraphics[scale=0.9]{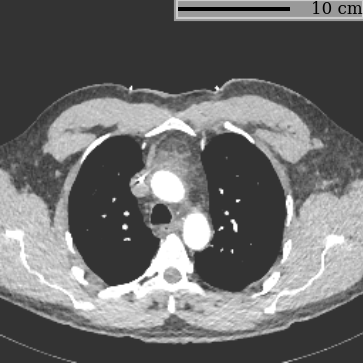}}
  & \raisebox{-0.45\height}{\includegraphics[scale=0.9]{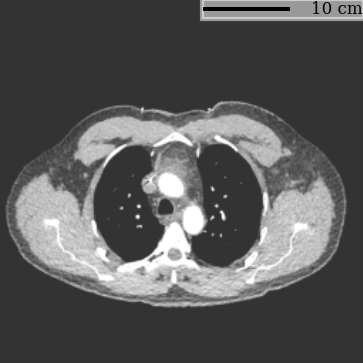}}
  & \raisebox{-0.45\height}{\includegraphics[scale=0.9]{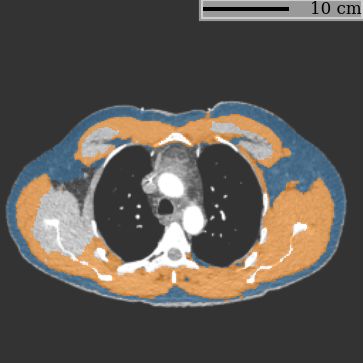}}\\\midrule

  {\bf{Pleural effusion}}
  & The segmentation module failed to distinguish the high intensity region in the lung (pleural effusion) from the muscle tissue.
  & \raisebox{-0.45\height}{\includegraphics[scale=0.9]{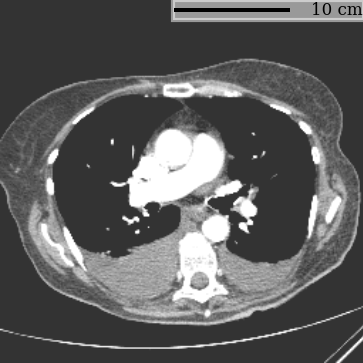}}
  & \raisebox{-0.45\height}{\includegraphics[scale=0.9]{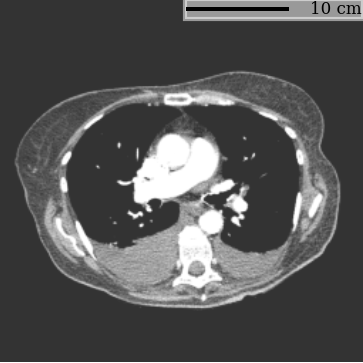}}
  & \raisebox{-0.45\height}{\includegraphics[scale=0.9]{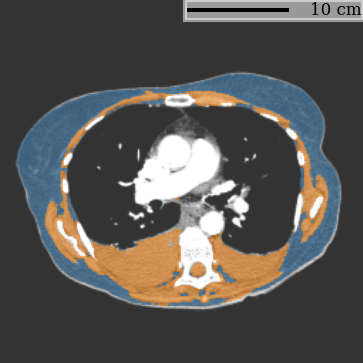}}\\\midrule

  {\bf{Beam hardening artifact }}
  & Severe imaging noise and beam hardening artifact associated with metal implant, resulting in inaccurate BC segmentation.
  & \raisebox{-0.45\height}{\includegraphics[scale=0.9]{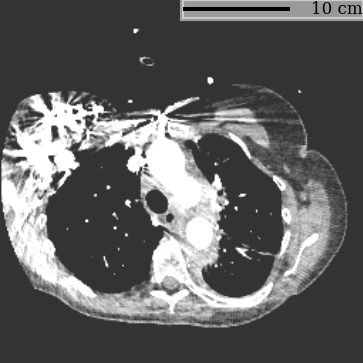}}
  & \raisebox{-0.45\height}{\includegraphics[scale=0.9]{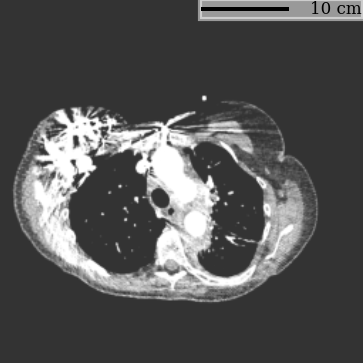}}
  & \raisebox{-0.45\height}{\includegraphics[scale=0.9]{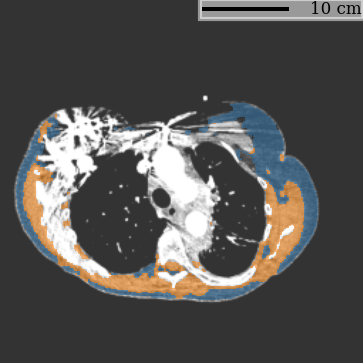}}\\
  \bottomrule
\end{tabular}
\end{table*}

\RevText{Extended based on the previous "Generalizability Analysis" section and changed it to
  a appendix,
as part of solution to address comments
R2.C3, R3.C5, R3.C9, R3.C10, and R3.C14. The data in this section were updated as
  a second review of the analyses identified multiple repeated records in the initial analysis}

\RevTextBold{The primary target of this present study was to
develop a solution for the systematic FOV truncation problem for lung screening LDCT,
which was a prohibitive factor for the application of opportunistic
BC assessment (Section \ref{sec:introduction}).
In Section \ref{sec:experiments-and-results},
we demonstrated the effectiveness and application validity of the developed FOV extension method.
However, it is unknown if the developed method can be generalized to clinically acquired chest CT
beyond lung cancer screening.
In this additional evaluation, we sought to characterize
  the generalizability and potential limitations of
the developed FOV extension method, as well as the enhanced
  BC assessment pipeline, for opportunistic BC assessment using routine diagnostic chest CT images.
As the primary interest was to characterize on how many and on which types of cases the method would fail,
  the evaluation was mainly based on qualitative visual assessment.
}{To address R3.C5, R3.C10 and R3.C14 - why evaluate on the
ImageVU, and the motivation for the choice of evaluation method}

\RevTextBold{The Longitudinal ImageVU (LiVU) cohort}
{Renamed the dataset to avoid conflict with ImageVU itself and in consistency with our other on-going projects}
was a retrospective chest CT study cohort initially designed for
longitudinal evaluation of incidental lung nodule, which was sampled from the ImageVU system
  {(\url{https://victr.vumc.org/our-programs/})}, a large
medical imaging databank that collects the routine diagnostic imaging
studies conducted at Vanderbilt University Medical Center (VUMC).
\RevTextBold{The inclusion criteria of LiVU selected those ImageVU subjects
with three consecutive chest CT studies
in five years, without any additional selection criteria.
As such, the associated chest CT scans were acquired with a broad spectrum of clinical indications.
We randomly sampled 953 subjects from the LiVU cohort,
and selected one scan per subject to form an evaluation dataset.
}{To address R3.C9 - the selection criteria for
 non LDCT cases from Vanderbilt}
\RevTextBold{The scans were performed between 2000 and 2019, with most (94.1\%) of the
  scans performed after 2014.}
{To address R3.C9 - the years of the scans}
All data were de-identified under internal review board supervision
(IRB\#181279). 
\RevTextBold{
57 scans with keyword "DERIVED" or "SECONDARY" included in the
DICOM tag "Image Type Attribute (0008, 0009)" were excluded, as these
  scans were not acquired directly from the CT scanners and were processed,
  e.g., affine transformed, before saving as DICOM data.
}{To address R3.C9 - are you describing that external scans were excluded}
\RevTextBold{We identified 15
intravenous (IV) contrast cases with significantly lowered intensity in
SAT and skeletal muscle comparing to the standard HU scale. These 15 CT scans were acquired from a same
dual-source CT scanner (Siemens\textregistered ~SOMATOM
Force\texttrademark) with ultra low-dose technique
implementation. We excluded these scans as the technique is known to
impact quantitative imaging analyses \citep{Wang2015} and lack of generalizability \citep{Vonder2021}.}
{To address R3.C9 - How were the 16 IV contrast cases excluded}
\RevTextBold{This led to a dataset consists of 881 scans acquired from 12
 model types of 5 scanner manufacturers.}{To address R3.C9 - why not have a real world sample of chest CTs
from multiple scanner types and manufactures}
Table \ref{tab:image-protocal-imagevu} shows the characterization of the
imaging protocol of this dataset.

\begin{table}[!t]
\caption{\label{tab:image-protocal-imagevu} {\normalfont Image
    protocol information of sampled LiVU dataset. SD=Standard Deviation.} } \centering
\footnotesize
\begin{tabular}{lr}\toprule
  {\bf{Parameter}} & {\bf{Value}}\\\midrule
  Effective mAs $\pm$ SD & 131.9 $\pm$ 81.1\\
  kVp $\pm$ SD & 108.6 $\pm$ 16.9\\
  Display FOV (cm) $\pm$ SD & 37.7 $\pm$ 4.9\\
  No. case with IV contrast (\%) & 591 (63.8)\\\bottomrule
\end{tabular}
\end{table}

The developed multi-level BC assessment pipeline with FOV
extension correction was applied on this dataset.
  We reviewed the quality of
generated results which include both the FOV extended images, and the
BC segmentation masks at the three levels.
 151 (17.1\%) cases were
identified with certain types of defects. The failure rate was higher
in the contrasted cases, with 19.5\% (115 out of 591) compared to 12.4\%
(36 out of 290) in non-contrast cases.
\RevTextBold{We identified 19 (2.2\%) FOV truncation cases associated with cupping artifacts,
where the developed pipeline failed to process.}
{To address R2.C3. Explicitly mention the number of cases with
cupping artifacts in the conventional CT dataset}
The segmentation module failed on 64 (7.3\%) pleural effusion cases, as the
module failed to distinguish the high intensity region in the lung from the muscle
tissue.  We identified 52 (5.9\%) cases with
non-standard body positioning. In addition, 20 (2.3\%) cases were failed due
to severe image noise or beam hardening artifact caused by metal
implant.
 Typical examples of each failure mode are given in Table
\ref{tab:failure-mode}.

In conclusion, we observed a
significant higher failure rate in the LiVU dataset. This could be explained by
the deviations in both of the scan protocols and patient characteristics
comparing to the lung screening LDCT datasets.
\RevTextBold{For instance, the occurrence of cupping artifact in scans with
FOV truncation was significantly more frequent
in the LiVU dataset than in the two included lung screening LDCT datasets (Section \ref{sec:data-preparation}).}
{To address R2.C3}
In addition, a significant portion of the scans in LiVU dataset were with IV contrast,
while IV contrast is usually not included in the lung cancer
screening protocols \citep{Gierada2009, Kazerooni2014}.
Pleural effusion was more prevalent in the
routine clinical scenarios that requiring a chest CT study comparing
to the asymptomatic screening context. The standard body positioning for lung screening LDCT,
with both arms above head, may not be required or difficult to implement in certain
clinical indications.
All these factors contribute to the challenges for the generalization of the developed method on routine clinical
chest CT scans.

%% file: paper.bbl
\begin{thebibliography}{59}
\expandafter\ifx\csname natexlab\endcsname\relax\def\natexlab#1{#1}\fi
\providecommand{\url}[1]{\texttt{#1}}
\providecommand{\href}[2]{#2}
\providecommand{\path}[1]{#1}
\providecommand{\DOIprefix}{doi:}
\providecommand{\ArXivprefix}{arXiv:}
\providecommand{\URLprefix}{URL: }
\providecommand{\Pubmedprefix}{pmid:}
\providecommand{\doi}[1]{\href{http://dx.doi.org/#1}{\path{#1}}}
\providecommand{\Pubmed}[1]{\href{pmid:#1}{\path{#1}}}
\providecommand{\bibinfo}[2]{#2}
\ifx\xfnm\relax \def\xfnm[#1]{\unskip,\space#1}\fi
\bibitem[{{American College of Radiology}(2022)}]{ACR2014}
\bibinfo{author}{{American College of Radiology}}, \bibinfo{year}{2022}.
\newblock \bibinfo{title}{Lung cancer screening center designation (revised
  11-9-2022)}.
\newblock \bibinfo{howpublished}{Available online:
  \url{https://accreditationsupport.acr.org/support/solutions/11000003422}}.
\newblock \bibinfo{note}{Accessed: 2023-01-20}.
\bibitem[{Armanious et~al.(2020)Armanious, Kumar, Abdulatif, Hepp, Gatidis and
  Yang}]{Armanious2020}
\bibinfo{author}{Armanious, K.}, \bibinfo{author}{Kumar, V.},
  \bibinfo{author}{Abdulatif, S.}, \bibinfo{author}{Hepp, T.},
  \bibinfo{author}{Gatidis, S.}, \bibinfo{author}{Yang, B.},
  \bibinfo{year}{2020}.
\newblock \bibinfo{title}{{ipA-MedGAN: Inpainting of Arbitrary Regions in
  Medical Imaging}}, in: \bibinfo{booktitle}{2020 IEEE International Conference
  on Image Processing (ICIP)}, \bibinfo{publisher}{IEEE}. pp.
  \bibinfo{pages}{3005--3009}.
\newblock \DOIprefix\doi{10.1109/ICIP40778.2020.9191207}.
\bibitem[{Bak et~al.(2019)Bak, Kwon, Han and Kim}]{Bak2019}
\bibinfo{author}{Bak, S.H.}, \bibinfo{author}{Kwon, S.O.},
  \bibinfo{author}{Han, S.S.}, \bibinfo{author}{Kim, W.J.},
  \bibinfo{year}{2019}.
\newblock \bibinfo{title}{{Computed tomography-derived area and density of
  pectoralis muscle associated disease severity and longitudinal changes in
  chronic obstructive pulmonary disease: a case control study}}.
\newblock \bibinfo{journal}{Respiratory Research} \bibinfo{volume}{20},
  \bibinfo{pages}{226}.
\newblock \DOIprefix\doi{10.1186/s12931-019-1191-y}.
\bibitem[{Best et~al.(2022)Best, Mercaldo, Bryan, Marquardt, Wrobel, Bridge,
  Troschel, Javidan, Chung, Muniappan, Bhalla, Meyers, Ferguson, Gaissert and
  Fintelmann}]{Best2022}
\bibinfo{author}{Best, T.D.}, \bibinfo{author}{Mercaldo, S.F.},
  \bibinfo{author}{Bryan, D.S.}, \bibinfo{author}{Marquardt, J.P.},
  \bibinfo{author}{Wrobel, M.M.}, \bibinfo{author}{Bridge, C.P.},
  \bibinfo{author}{Troschel, F.M.}, \bibinfo{author}{Javidan, C.},
  \bibinfo{author}{Chung, J.H.}, \bibinfo{author}{Muniappan, A.},
  \bibinfo{author}{Bhalla, S.}, \bibinfo{author}{Meyers, B.F.},
  \bibinfo{author}{Ferguson, M.K.}, \bibinfo{author}{Gaissert, H.A.},
  \bibinfo{author}{Fintelmann, F.J.}, \bibinfo{year}{2022}.
\newblock \bibinfo{title}{{Multilevel Body Composition Analysis on Chest
  Computed Tomography Predicts Hospital Length of Stay and Complications After
  Lobectomy for Lung Cancer}}.
\newblock \bibinfo{journal}{Annals of Surgery} \bibinfo{volume}{275},
  \bibinfo{pages}{e708--e715}.
\newblock \DOIprefix\doi{10.1097/SLA.0000000000004040}.
\bibitem[{Bridge et~al.(2022)Bridge, Best, Wrobel, Marquardt, Magudia, Javidan,
  Chung, Kalpathy-Cramer, Andriole and Fintelmann}]{Bridge2022}
\bibinfo{author}{Bridge, C.P.}, \bibinfo{author}{Best, T.D.},
  \bibinfo{author}{Wrobel, M.M.}, \bibinfo{author}{Marquardt, J.P.},
  \bibinfo{author}{Magudia, K.}, \bibinfo{author}{Javidan, C.},
  \bibinfo{author}{Chung, J.H.}, \bibinfo{author}{Kalpathy-Cramer, J.},
  \bibinfo{author}{Andriole, K.P.}, \bibinfo{author}{Fintelmann, F.J.},
  \bibinfo{year}{2022}.
\newblock \bibinfo{title}{{A Fully Automated Deep Learning Pipeline for
  Multi–Vertebral Level Quantification and Characterization of Muscle and
  Adipose Tissue on Chest CT Scans}}.
\newblock \bibinfo{journal}{Radiology: Artificial Intelligence}
  \bibinfo{volume}{4}.
\newblock \DOIprefix\doi{10.1148/ryai.210080}.
\bibitem[{Bridge et~al.(2018)Bridge, Rosenthal, Wright, Kotecha, Fintelmann,
  Troschel, Miskin, Desai, Wrobel, Babic, Khalaf, Brais, Welch, Zellers,
  Tenenholtz, Michalski, Wolpin and Andriole}]{Bridge2018}
\bibinfo{author}{Bridge, C.P.}, \bibinfo{author}{Rosenthal, M.},
  \bibinfo{author}{Wright, B.}, \bibinfo{author}{Kotecha, G.},
  \bibinfo{author}{Fintelmann, F.}, \bibinfo{author}{Troschel, F.},
  \bibinfo{author}{Miskin, N.}, \bibinfo{author}{Desai, K.},
  \bibinfo{author}{Wrobel, W.}, \bibinfo{author}{Babic, A.},
  \bibinfo{author}{Khalaf, N.}, \bibinfo{author}{Brais, L.},
  \bibinfo{author}{Welch, M.}, \bibinfo{author}{Zellers, C.},
  \bibinfo{author}{Tenenholtz, N.}, \bibinfo{author}{Michalski, M.},
  \bibinfo{author}{Wolpin, B.}, \bibinfo{author}{Andriole, K.},
  \bibinfo{year}{2018}.
\newblock \bibinfo{title}{{Fully-Automated Analysis of Body Composition from CT
  in Cancer Patients Using Convolutional Neural Networks BT - OR 2.0
  Context-Aware Operating Theaters, Computer Assisted Robotic Endoscopy,
  Clinical Image-Based Procedures, and Skin Image Analysis}}, in:
  \bibinfo{booktitle}{Context-Aware Operating Theaters, Computer Assisted
  Robotic Endoscopy, Clinical Image-Based Procedures, and Skin Image Analysis.
  CARE CLIP OR 2.0 ISIC 2018}, \bibinfo{publisher}{Springer International
  Publishing}, \bibinfo{address}{Cham}. pp. \bibinfo{pages}{204--213}.
\bibitem[{Chuquicusma et~al.(2018)Chuquicusma, Hussein, Burt and
  Bagci}]{Chuquicusma2018}
\bibinfo{author}{Chuquicusma, M.J.M.}, \bibinfo{author}{Hussein, S.},
  \bibinfo{author}{Burt, J.}, \bibinfo{author}{Bagci, U.},
  \bibinfo{year}{2018}.
\newblock \bibinfo{title}{{How to fool radiologists with generative adversarial
  networks? A visual turing test for lung cancer diagnosis}}, in:
  \bibinfo{booktitle}{2018 IEEE 15th International Symposium on Biomedical
  Imaging (ISBI 2018)}, \bibinfo{publisher}{IEEE}. pp.
  \bibinfo{pages}{240--244}.
\newblock \DOIprefix\doi{10.1109/ISBI.2018.8363564}.
\bibitem[{Fintelmann et~al.(2018)Fintelmann, Troschel, Mario, Chretien, Knoll,
  Muniappan and Gaissert}]{Fintelmann2018}
\bibinfo{author}{Fintelmann, F.J.}, \bibinfo{author}{Troschel, F.M.},
  \bibinfo{author}{Mario, J.}, \bibinfo{author}{Chretien, Y.R.},
  \bibinfo{author}{Knoll, S.J.}, \bibinfo{author}{Muniappan, A.},
  \bibinfo{author}{Gaissert, H.A.}, \bibinfo{year}{2018}.
\newblock \bibinfo{title}{{Thoracic Skeletal Muscle Is Associated With Adverse
  Outcomes After Lobectomy for Lung Cancer}}.
\newblock \bibinfo{journal}{The Annals of Thoracic Surgery}
  \bibinfo{volume}{105}, \bibinfo{pages}{1507--1515}.
\newblock \DOIprefix\doi{10.1016/j.athoracsur.2018.01.013}.
\bibitem[{Fourni{\'{e}} et~al.(2019)Fourni{\'{e}}, Baer-Beck and
  Stierstorfer}]{Fournie2019}
\bibinfo{author}{Fourni{\'{e}}, {\'{E}}.}, \bibinfo{author}{Baer-Beck, M.},
  \bibinfo{author}{Stierstorfer, K.}, \bibinfo{year}{2019}.
\newblock \bibinfo{title}{{CT Field of View Extension Using Combined Channels
  Extension and Deep Learning Methods}}, in: \bibinfo{booktitle}{Medical
  Imaging with Deep Learning (MIDL)}, pp. \bibinfo{pages}{1--4}.
\bibitem[{Gazourian et~al.(2020)Gazourian, Durgana, Huntley, Rizzo, Thedinger,
  Regis, Price, Pagura, Lamb, Rieger-Christ, Thomson, Stefanescu, Sanayei,
  Long, McKee, Washko, Est{\'{e}}par, Wald, Liesching and
  McKee}]{Gazourian2020}
\bibinfo{author}{Gazourian, L.}, \bibinfo{author}{Durgana, C.S.},
  \bibinfo{author}{Huntley, D.}, \bibinfo{author}{Rizzo, G.S.},
  \bibinfo{author}{Thedinger, W.B.}, \bibinfo{author}{Regis, S.M.},
  \bibinfo{author}{Price, L.L.}, \bibinfo{author}{Pagura, E.J.},
  \bibinfo{author}{Lamb, C.}, \bibinfo{author}{Rieger-Christ, K.},
  \bibinfo{author}{Thomson, C.C.}, \bibinfo{author}{Stefanescu, C.F.},
  \bibinfo{author}{Sanayei, A.}, \bibinfo{author}{Long, W.P.},
  \bibinfo{author}{McKee, A.B.}, \bibinfo{author}{Washko, G.R.},
  \bibinfo{author}{Est{\'{e}}par, R.S.J.}, \bibinfo{author}{Wald, C.},
  \bibinfo{author}{Liesching, T.N.}, \bibinfo{author}{McKee, B.J.},
  \bibinfo{year}{2020}.
\newblock \bibinfo{title}{{Quantitative Pectoralis Muscle Area is Associated
  with the Development of Lung Cancer in a Large Lung Cancer Screening
  Cohort}}.
\newblock \bibinfo{journal}{Lung} \bibinfo{volume}{198},
  \bibinfo{pages}{847--853}.
\newblock \DOIprefix\doi{10.1007/s00408-020-00388-5}.
\bibitem[{Gierada et~al.(2009)Gierada, Garg, Nath, Strollo, Fagerstrom and
  Ford}]{Gierada2009}
\bibinfo{author}{Gierada, D.S.}, \bibinfo{author}{Garg, K.},
  \bibinfo{author}{Nath, H.}, \bibinfo{author}{Strollo, D.C.},
  \bibinfo{author}{Fagerstrom, R.M.}, \bibinfo{author}{Ford, M.B.},
  \bibinfo{year}{2009}.
\newblock \bibinfo{title}{{CT Quality Assurance in the Lung Screening Study
  Component of the National Lung Screening Trial: Implications for Multicenter
  Imaging Trials}}.
\newblock \bibinfo{journal}{American Journal of Roentgenology}
  \bibinfo{volume}{193}, \bibinfo{pages}{419--424}.
\newblock \DOIprefix\doi{10.2214/AJR.08.1995}.
\bibitem[{Hittner et~al.(2003)Hittner, May and Silver}]{hittner2003monte}
\bibinfo{author}{Hittner, J.B.}, \bibinfo{author}{May, K.},
  \bibinfo{author}{Silver, N.C.}, \bibinfo{year}{2003}.
\newblock \bibinfo{title}{A monte carlo evaluation of tests for comparing
  dependent correlations}.
\newblock \bibinfo{journal}{The Journal of general psychology}
  \bibinfo{volume}{130}, \bibinfo{pages}{149--168}.
\bibitem[{Hofmanninger et~al.(2020)Hofmanninger, Prayer, Pan, R{\"{o}}hrich,
  Prosch and Langs}]{Hofmanninger2020}
\bibinfo{author}{Hofmanninger, J.}, \bibinfo{author}{Prayer, F.},
  \bibinfo{author}{Pan, J.}, \bibinfo{author}{R{\"{o}}hrich, S.},
  \bibinfo{author}{Prosch, H.}, \bibinfo{author}{Langs, G.},
  \bibinfo{year}{2020}.
\newblock \bibinfo{title}{{Automatic lung segmentation in routine imaging is
  primarily a data diversity problem, not a methodology problem}}.
\newblock \bibinfo{journal}{European Radiology Experimental}
  \bibinfo{volume}{4}, \bibinfo{pages}{50}.
\newblock \DOIprefix\doi{10.1186/s41747-020-00173-2}.
\bibitem[{Hsieh et~al.(2004)Hsieh, Chao, Thibault, Grekowicz, Horst, McOlash
  and Myers}]{Hsieh2004}
\bibinfo{author}{Hsieh, J.}, \bibinfo{author}{Chao, E.},
  \bibinfo{author}{Thibault, J.}, \bibinfo{author}{Grekowicz, B.},
  \bibinfo{author}{Horst, A.}, \bibinfo{author}{McOlash, S.},
  \bibinfo{author}{Myers, T.J.}, \bibinfo{year}{2004}.
\newblock \bibinfo{title}{{A novel reconstruction algorithm to extend the CT
  scan field-of-view}}.
\newblock \bibinfo{journal}{Medical Physics} \bibinfo{volume}{31},
  \bibinfo{pages}{2385--2391}.
\newblock \DOIprefix\doi{10.1118/1.1776673}.
\bibitem[{Huang et~al.(2021)Huang, Preuhs, Manhart, Lauritsch and
  Maier}]{Huang2021}
\bibinfo{author}{Huang, Y.}, \bibinfo{author}{Preuhs, A.},
  \bibinfo{author}{Manhart, M.}, \bibinfo{author}{Lauritsch, G.},
  \bibinfo{author}{Maier, A.}, \bibinfo{year}{2021}.
\newblock \bibinfo{title}{{Data Extrapolation From Learned Prior Images for
  Truncation Correction in Computed Tomography}}.
\newblock \bibinfo{journal}{IEEE Transactions on Medical Imaging}
  \bibinfo{volume}{40}, \bibinfo{pages}{3042--3053}.
\newblock \DOIprefix\doi{10.1109/TMI.2021.3072568}.
\bibitem[{Iizuka et~al.(2017)Iizuka, Simo-Serra and Ishikawa}]{Iizuka2017}
\bibinfo{author}{Iizuka, S.}, \bibinfo{author}{Simo-Serra, E.},
  \bibinfo{author}{Ishikawa, H.}, \bibinfo{year}{2017}.
\newblock \bibinfo{title}{{Globally and locally consistent image completion}}.
\newblock \bibinfo{journal}{ACM Transactions on Graphics} \bibinfo{volume}{36},
  \bibinfo{pages}{1--14}.
\newblock \DOIprefix\doi{10.1145/3072959.3073659}.
\bibitem[{Isola et~al.(2017)Isola, Zhu, Zhou and Efros}]{Isola2017}
\bibinfo{author}{Isola, P.}, \bibinfo{author}{Zhu, J.Y.},
  \bibinfo{author}{Zhou, T.}, \bibinfo{author}{Efros, A.A.},
  \bibinfo{year}{2017}.
\newblock \bibinfo{title}{{Image-to-Image Translation with Conditional
  Adversarial Networks}}, in: \bibinfo{booktitle}{2017 IEEE Conference on
  Computer Vision and Pattern Recognition (CVPR)}, \bibinfo{publisher}{IEEE}.
  pp. \bibinfo{pages}{5967--5976}.
\newblock \DOIprefix\doi{10.1109/CVPR.2017.632}.
\bibitem[{Kang et~al.(2021)Kang, Shin, Seo, Byun, Lee, Kim, Lee and
  Lee}]{Kang2021}
\bibinfo{author}{Kang, S.K.}, \bibinfo{author}{Shin, S.A.},
  \bibinfo{author}{Seo, S.}, \bibinfo{author}{Byun, M.S.},
  \bibinfo{author}{Lee, D.Y.}, \bibinfo{author}{Kim, Y.K.},
  \bibinfo{author}{Lee, D.S.}, \bibinfo{author}{Lee, J.S.},
  \bibinfo{year}{2021}.
\newblock \bibinfo{title}{{Deep learning-Based 3D inpainting of brain MR
  images}}.
\newblock \bibinfo{journal}{Scientific Reports} \bibinfo{volume}{11},
  \bibinfo{pages}{1673}.
\newblock \DOIprefix\doi{10.1038/s41598-020-80930-w}.
\bibitem[{Kazerooni et~al.(2014)Kazerooni, Austin, Black, Dyer, Hazelton,
  Leung, McNitt-Gray, Munden and Pipavath}]{Kazerooni2014}
\bibinfo{author}{Kazerooni, E.A.}, \bibinfo{author}{Austin, J.H.},
  \bibinfo{author}{Black, W.C.}, \bibinfo{author}{Dyer, D.S.},
  \bibinfo{author}{Hazelton, T.R.}, \bibinfo{author}{Leung, A.N.},
  \bibinfo{author}{McNitt-Gray, M.F.}, \bibinfo{author}{Munden, R.F.},
  \bibinfo{author}{Pipavath, S.}, \bibinfo{year}{2014}.
\newblock \bibinfo{title}{Acr–str practice parameter for the performance and
  reporting of lung cancer screening thoracic computed tomography (ct)}.
\newblock \bibinfo{journal}{Journal of Thoracic Imaging} \bibinfo{volume}{29},
  \bibinfo{pages}{310--316}.
\newblock \URLprefix \url{https://journals.lww.com/00005382-201409000-00012},
  \DOIprefix\doi{10.1097/RTI.0000000000000097}.
\bibitem[{Ketola et~al.(2021)Ketola, Heino, Juntunen, Nieminen and
  Inkinen}]{Ketola2021}
\bibinfo{author}{Ketola, J.H.}, \bibinfo{author}{Heino, H.},
  \bibinfo{author}{Juntunen, M.A.K.}, \bibinfo{author}{Nieminen, M.T.},
  \bibinfo{author}{Inkinen, S.I.}, \bibinfo{year}{2021}.
\newblock \bibinfo{title}{{Deep learning-based sinogram extension method for
  interior computed tomography}}, in: \bibinfo{editor}{Bosmans, H.},
  \bibinfo{editor}{Zhao, W.}, \bibinfo{editor}{Yu, L.} (Eds.),
  \bibinfo{booktitle}{Medical Imaging 2021: Physics of Medical Imaging},
  \bibinfo{publisher}{SPIE}. p. \bibinfo{pages}{123}.
\newblock \DOIprefix\doi{10.1117/12.2580886}.
\bibitem[{Kim et~al.(2016)Kim, Kim, Park, Ahn, Cho, Jeong and Kim}]{Kim2016}
\bibinfo{author}{Kim, E.Y.}, \bibinfo{author}{Kim, Y.S.},
  \bibinfo{author}{Park, I.}, \bibinfo{author}{Ahn, H.K.},
  \bibinfo{author}{Cho, E.K.}, \bibinfo{author}{Jeong, Y.M.},
  \bibinfo{author}{Kim, J.H.}, \bibinfo{year}{2016}.
\newblock \bibinfo{title}{{Evaluation of sarcopenia in small-cell lung cancer
  patients by routine chest CT}}.
\newblock \bibinfo{journal}{Supportive Care in Cancer} \bibinfo{volume}{24},
  \bibinfo{pages}{4721--4726}.
\newblock \DOIprefix\doi{10.1007/s00520-016-3321-0}.
\bibitem[{Krishnan et~al.(2019)Krishnan, Teterwak, Sarna, Maschinot, Liu,
  Belanger and Freeman}]{Krishnan2019}
\bibinfo{author}{Krishnan, D.}, \bibinfo{author}{Teterwak, P.},
  \bibinfo{author}{Sarna, A.}, \bibinfo{author}{Maschinot, A.},
  \bibinfo{author}{Liu, C.}, \bibinfo{author}{Belanger, D.},
  \bibinfo{author}{Freeman, W.}, \bibinfo{year}{2019}.
\newblock \bibinfo{title}{{Boundless: Generative Adversarial Networks for Image
  Extension}}, in: \bibinfo{booktitle}{2019 IEEE/CVF International Conference
  on Computer Vision (ICCV)}, \bibinfo{publisher}{IEEE}. pp.
  \bibinfo{pages}{10520--10529}.
\newblock \DOIprefix\doi{10.1109/ICCV.2019.01062}.
\bibitem[{Krist et~al.(2021)Krist, Davidson, Mangione, Barry, Cabana, Caughey,
  Davis, Donahue, Doubeni, Kubik, Landefeld, Li, Ogedegbe, Owens, Pbert,
  Silverstein, Stevermer, Tseng and Wong}]{Krist2021}
\bibinfo{author}{Krist, A.H.}, \bibinfo{author}{Davidson, K.W.},
  \bibinfo{author}{Mangione, C.M.}, \bibinfo{author}{Barry, M.J.},
  \bibinfo{author}{Cabana, M.}, \bibinfo{author}{Caughey, A.B.},
  \bibinfo{author}{Davis, E.M.}, \bibinfo{author}{Donahue, K.E.},
  \bibinfo{author}{Doubeni, C.A.}, \bibinfo{author}{Kubik, M.},
  \bibinfo{author}{Landefeld, C.S.}, \bibinfo{author}{Li, L.},
  \bibinfo{author}{Ogedegbe, G.}, \bibinfo{author}{Owens, D.K.},
  \bibinfo{author}{Pbert, L.}, \bibinfo{author}{Silverstein, M.},
  \bibinfo{author}{Stevermer, J.}, \bibinfo{author}{Tseng, C.W.},
  \bibinfo{author}{Wong, J.B.}, \bibinfo{year}{2021}.
\newblock \bibinfo{title}{{Screening for Lung Cancer}}.
\newblock \bibinfo{journal}{JAMA} \bibinfo{volume}{325}, \bibinfo{pages}{962}.
\newblock \DOIprefix\doi{10.1001/jama.2021.1117}.
\bibitem[{Kuch et~al.(2001)Kuch, Gneiting, D{\"{o}}ring, Muscholl,
  Br{\"{o}}ckel, Schunkert and Hense}]{Kuch2001}
\bibinfo{author}{Kuch, B.}, \bibinfo{author}{Gneiting, B.},
  \bibinfo{author}{D{\"{o}}ring, A.}, \bibinfo{author}{Muscholl, M.},
  \bibinfo{author}{Br{\"{o}}ckel, U.}, \bibinfo{author}{Schunkert, H.},
  \bibinfo{author}{Hense, H.W.}, \bibinfo{year}{2001}.
\newblock \bibinfo{title}{{Indexation of left ventricular mass in adults with a
  novel approximation for fat-free mass}}.
\newblock \bibinfo{journal}{Journal of Hypertension} \bibinfo{volume}{19},
  \bibinfo{pages}{135--142}.
\newblock \DOIprefix\doi{10.1097/00004872-200101000-00018}.
\bibitem[{Lenchik et~al.(2021)Lenchik, Barnard, Boutin, Kritchevsky, Chen, Tan,
  Cawthon, Weaver and Hsu}]{Lenchik2021}
\bibinfo{author}{Lenchik, L.}, \bibinfo{author}{Barnard, R.},
  \bibinfo{author}{Boutin, R.D.}, \bibinfo{author}{Kritchevsky, S.B.},
  \bibinfo{author}{Chen, H.}, \bibinfo{author}{Tan, J.},
  \bibinfo{author}{Cawthon, P.M.}, \bibinfo{author}{Weaver, A.A.},
  \bibinfo{author}{Hsu, F.C.}, \bibinfo{year}{2021}.
\newblock \bibinfo{title}{{Automated Muscle Measurement on Chest CT Predicts
  All-Cause Mortality in Older Adults From the National Lung Screening Trial}}.
\newblock \bibinfo{journal}{The Journals of Gerontology: Series A}
  \bibinfo{volume}{76}, \bibinfo{pages}{277--285}.
\newblock \DOIprefix\doi{10.1093/gerona/glaa141}.
\bibitem[{Li et~al.(2020)Li, Wang, Zhang, Du and Tao}]{Li2020}
\bibinfo{author}{Li, J.}, \bibinfo{author}{Wang, N.}, \bibinfo{author}{Zhang,
  L.}, \bibinfo{author}{Du, B.}, \bibinfo{author}{Tao, D.},
  \bibinfo{year}{2020}.
\newblock \bibinfo{title}{{Recurrent Feature Reasoning for Image Inpainting}},
  in: \bibinfo{booktitle}{2020 IEEE/CVF Conference on Computer Vision and
  Pattern Recognition (CVPR)}, \bibinfo{publisher}{IEEE}. pp.
  \bibinfo{pages}{7757--7765}.
\newblock \DOIprefix\doi{10.1109/CVPR42600.2020.00778}.
\bibitem[{Liu et~al.(2018)Liu, Reda, Shih, Wang, Tao and Catanzaro}]{Liu2018}
\bibinfo{author}{Liu, G.}, \bibinfo{author}{Reda, F.A.}, \bibinfo{author}{Shih,
  K.J.}, \bibinfo{author}{Wang, T.C.}, \bibinfo{author}{Tao, A.},
  \bibinfo{author}{Catanzaro, B.}, \bibinfo{year}{2018}.
\newblock \bibinfo{title}{{Image Inpainting for Irregular Holes Using Partial
  Convolutions BT - Computer Vision – ECCV 2018}}, in:
  \bibinfo{editor}{Ferrari, V.}, \bibinfo{editor}{Hebert, M.},
  \bibinfo{editor}{Sminchisescu, C.}, \bibinfo{editor}{Weiss, Y.} (Eds.),
  \bibinfo{booktitle}{Computer Vision – ECCV 2018},
  \bibinfo{publisher}{Springer International Publishing},
  \bibinfo{address}{Cham}. pp. \bibinfo{pages}{89--105}.
\bibitem[{Magudia et~al.(2021)Magudia, Bridge, Bay, Babic, Fintelmann,
  Troschel, Miskin, Wrobel, Brais, Andriole, Wolpin and
  Rosenthal}]{Magudia2021}
\bibinfo{author}{Magudia, K.}, \bibinfo{author}{Bridge, C.P.},
  \bibinfo{author}{Bay, C.P.}, \bibinfo{author}{Babic, A.},
  \bibinfo{author}{Fintelmann, F.J.}, \bibinfo{author}{Troschel, F.M.},
  \bibinfo{author}{Miskin, N.}, \bibinfo{author}{Wrobel, W.C.},
  \bibinfo{author}{Brais, L.K.}, \bibinfo{author}{Andriole, K.P.},
  \bibinfo{author}{Wolpin, B.M.}, \bibinfo{author}{Rosenthal, M.H.},
  \bibinfo{year}{2021}.
\newblock \bibinfo{title}{{Population-Scale CT-based Body Composition Analysis
  of a Large Outpatient Population Using Deep Learning to Derive Age-, Sex-,
  and Race-specific Reference Curves}}.
\newblock \bibinfo{journal}{Radiology} \bibinfo{volume}{298},
  \bibinfo{pages}{319--329}.
\newblock \DOIprefix\doi{10.1148/radiol.2020201640}.
\bibitem[{Mathur et~al.(2020)Mathur, Rozenberg, Verweel, Orsso and
  Singer}]{Mathur2020}
\bibinfo{author}{Mathur, S.}, \bibinfo{author}{Rozenberg, D.},
  \bibinfo{author}{Verweel, L.}, \bibinfo{author}{Orsso, C.E.},
  \bibinfo{author}{Singer, L.G.}, \bibinfo{year}{2020}.
\newblock \bibinfo{title}{{Chest computed tomography is a valid measure of body
  composition in individuals with advanced lung disease}}.
\newblock \bibinfo{journal}{Clinical Physiology and Functional Imaging}
  \bibinfo{volume}{40}, \bibinfo{pages}{360--368}.
\newblock \DOIprefix\doi{10.1111/cpf.12652}.
\bibitem[{McDonald et~al.(2014)McDonald, Diaz, Ross, {San Jose Estepar}, Zhou,
  Regan, Eckbo, Muralidhar, Come, Cho, Hersh, Lange, Wouters, Casaburi, Coxson,
  MacNee, Rennard, Lomas, Agusti, Celli, Black-Shinn, Kinney, Lutz, Hokanson,
  Silverman and Washko}]{McDonald2014}
\bibinfo{author}{McDonald, M.L.N.}, \bibinfo{author}{Diaz, A.A.},
  \bibinfo{author}{Ross, J.C.}, \bibinfo{author}{{San Jose Estepar}, R.},
  \bibinfo{author}{Zhou, L.}, \bibinfo{author}{Regan, E.A.},
  \bibinfo{author}{Eckbo, E.}, \bibinfo{author}{Muralidhar, N.},
  \bibinfo{author}{Come, C.E.}, \bibinfo{author}{Cho, M.H.},
  \bibinfo{author}{Hersh, C.P.}, \bibinfo{author}{Lange, C.},
  \bibinfo{author}{Wouters, E.}, \bibinfo{author}{Casaburi, R.H.},
  \bibinfo{author}{Coxson, H.O.}, \bibinfo{author}{MacNee, W.},
  \bibinfo{author}{Rennard, S.I.}, \bibinfo{author}{Lomas, D.A.},
  \bibinfo{author}{Agusti, A.}, \bibinfo{author}{Celli, B.R.},
  \bibinfo{author}{Black-Shinn, J.L.}, \bibinfo{author}{Kinney, G.L.},
  \bibinfo{author}{Lutz, S.M.}, \bibinfo{author}{Hokanson, J.E.},
  \bibinfo{author}{Silverman, E.K.}, \bibinfo{author}{Washko, G.R.},
  \bibinfo{year}{2014}.
\newblock \bibinfo{title}{{Quantitative Computed Tomography Measures of
  Pectoralis Muscle Area and Disease Severity in Chronic Obstructive Pulmonary
  Disease. A Cross-Sectional Study}}.
\newblock \bibinfo{journal}{Annals of the American Thoracic Society}
  \bibinfo{volume}{11}, \bibinfo{pages}{326--334}.
\newblock \DOIprefix\doi{10.1513/AnnalsATS.201307-229OC}.
\bibitem[{Nazeri et~al.(2019)Nazeri, Ng, Joseph, Qureshi and
  Ebrahimi}]{Nazeri2019}
\bibinfo{author}{Nazeri, K.}, \bibinfo{author}{Ng, E.},
  \bibinfo{author}{Joseph, T.}, \bibinfo{author}{Qureshi, F.},
  \bibinfo{author}{Ebrahimi, M.}, \bibinfo{year}{2019}.
\newblock \bibinfo{title}{{EdgeConnect: Structure Guided Image Inpainting using
  Edge Prediction}}, in: \bibinfo{booktitle}{2019 IEEE/CVF International
  Conference on Computer Vision Workshop (ICCVW)}, \bibinfo{publisher}{IEEE}.
  pp. \bibinfo{pages}{3265--3274}.
\newblock \DOIprefix\doi{10.1109/ICCVW.2019.00408}.
\bibitem[{Ogawa et~al.(1984)Ogawa, Nakajima and Yuta}]{Ogawa1984}
\bibinfo{author}{Ogawa, K.}, \bibinfo{author}{Nakajima, M.},
  \bibinfo{author}{Yuta, S.}, \bibinfo{year}{1984}.
\newblock \bibinfo{title}{A reconstruction algorithm from truncated
  projections}.
\newblock \bibinfo{journal}{IEEE Transactions on Medical Imaging}
  \bibinfo{volume}{3}, \bibinfo{pages}{34--40}.
\newblock \DOIprefix\doi{10.1109/TMI.1984.4307648}.
\bibitem[{Ohnesorge et~al.(2000)Ohnesorge, Flohr, Schwarz, Heiken and
  Bae}]{Ohnesorge2000}
\bibinfo{author}{Ohnesorge, B.}, \bibinfo{author}{Flohr, T.},
  \bibinfo{author}{Schwarz, K.}, \bibinfo{author}{Heiken, J.P.},
  \bibinfo{author}{Bae, K.T.}, \bibinfo{year}{2000}.
\newblock \bibinfo{title}{{Efficient correction for CT image artifacts caused
  by objects extending outside the scan field of view}}.
\newblock \bibinfo{journal}{Medical Physics} \bibinfo{volume}{27},
  \bibinfo{pages}{39--46}.
\newblock \DOIprefix\doi{10.1118/1.598855}.
\bibitem[{Pathak et~al.(2016)Pathak, Krahenbuhl, Donahue, Darrell and
  Efros}]{Pathak2016}
\bibinfo{author}{Pathak, D.}, \bibinfo{author}{Krahenbuhl, P.},
  \bibinfo{author}{Donahue, J.}, \bibinfo{author}{Darrell, T.},
  \bibinfo{author}{Efros, A.A.}, \bibinfo{year}{2016}.
\newblock \bibinfo{title}{{Context Encoders: Feature Learning by Inpainting}},
  in: \bibinfo{booktitle}{2016 IEEE Conference on Computer Vision and Pattern
  Recognition (CVPR)}, \bibinfo{publisher}{IEEE}. pp.
  \bibinfo{pages}{2536--2544}.
\newblock \DOIprefix\doi{10.1109/CVPR.2016.278}.
\bibitem[{Pickhardt(2022)}]{Pickhardt2022}
\bibinfo{author}{Pickhardt, P.J.}, \bibinfo{year}{2022}.
\newblock \bibinfo{title}{{Value-added Opportunistic CT Screening: State of the
  Art}}.
\newblock \bibinfo{journal}{Radiology} \bibinfo{volume}{303},
  \bibinfo{pages}{241--254}.
\newblock \DOIprefix\doi{10.1148/radiol.211561}.
\bibitem[{Pickhardt et~al.(2021)Pickhardt, Summers and Garrett}]{Pickhardt2021}
\bibinfo{author}{Pickhardt, P.J.}, \bibinfo{author}{Summers, R.M.},
  \bibinfo{author}{Garrett, J.W.}, \bibinfo{year}{2021}.
\newblock \bibinfo{title}{{Automated CT-Based Body Composition Analysis: A
  Golden Opportunity}}.
\newblock \bibinfo{journal}{Korean Journal of Radiology} \bibinfo{volume}{22},
  \bibinfo{pages}{1934}.
\newblock \DOIprefix\doi{10.3348/kjr.2021.0775}.
\bibitem[{Pishgar et~al.(2021)Pishgar, Shabani, {Quinaglia A. C. Silva},
  Bluemke, Budoff, Barr, Allison, Post, Lima and Demehri}]{Pishgar2021}
\bibinfo{author}{Pishgar, F.}, \bibinfo{author}{Shabani, M.},
  \bibinfo{author}{{Quinaglia A. C. Silva}, T.}, \bibinfo{author}{Bluemke,
  D.A.}, \bibinfo{author}{Budoff, M.}, \bibinfo{author}{Barr, R.G.},
  \bibinfo{author}{Allison, M.A.}, \bibinfo{author}{Post, W.S.},
  \bibinfo{author}{Lima, J.A.C.}, \bibinfo{author}{Demehri, S.},
  \bibinfo{year}{2021}.
\newblock \bibinfo{title}{{Quantitative Analysis of Adipose Depots by Using
  Chest CT and Associations with All-Cause Mortality in Chronic Obstructive
  Pulmonary Disease: Longitudinal Analysis from MESArthritis Ancillary Study}}.
\newblock \bibinfo{journal}{Radiology} \bibinfo{volume}{299},
  \bibinfo{pages}{703--711}.
\newblock \DOIprefix\doi{10.1148/radiol.2021203959}.
\bibitem[{Rezatofighi et~al.(2019)Rezatofighi, Tsoi, Gwak, Sadeghian, Reid and
  Savarese}]{Rezatofighi2019}
\bibinfo{author}{Rezatofighi, H.}, \bibinfo{author}{Tsoi, N.},
  \bibinfo{author}{Gwak, J.}, \bibinfo{author}{Sadeghian, A.},
  \bibinfo{author}{Reid, I.}, \bibinfo{author}{Savarese, S.},
  \bibinfo{year}{2019}.
\newblock \bibinfo{title}{{Generalized Intersection Over Union: A Metric and a
  Loss for Bounding Box Regression}}, in: \bibinfo{booktitle}{2019 IEEE/CVF
  Conference on Computer Vision and Pattern Recognition (CVPR)},
  \bibinfo{publisher}{IEEE}. pp. \bibinfo{pages}{658--666}.
\newblock \DOIprefix\doi{10.1109/CVPR.2019.00075}.
\bibitem[{Ruchala et~al.(2002)Ruchala, Olivera, Kapatoes, Reckwerdt and
  Mackie}]{Ruchala2002}
\bibinfo{author}{Ruchala, K.J.}, \bibinfo{author}{Olivera, G.H.},
  \bibinfo{author}{Kapatoes, J.M.}, \bibinfo{author}{Reckwerdt, P.J.},
  \bibinfo{author}{Mackie, T.R.}, \bibinfo{year}{2002}.
\newblock \bibinfo{title}{Methods for improving limited field-of-view
  radiotherapy reconstructions using imperfect a priori images}.
\newblock \bibinfo{journal}{Medical Physics} \bibinfo{volume}{29},
  \bibinfo{pages}{2590--2605}.
\newblock \DOIprefix\doi{10.1118/1.1513163}.
\bibitem[{Salimova et~al.(2022)Salimova, Hinrichs, Gutberlet, Meyer, Wacker and
  von Falck}]{Salimova2022}
\bibinfo{author}{Salimova, N.}, \bibinfo{author}{Hinrichs, J.B.},
  \bibinfo{author}{Gutberlet, M.}, \bibinfo{author}{Meyer, B.C.},
  \bibinfo{author}{Wacker, F.K.}, \bibinfo{author}{von Falck, C.},
  \bibinfo{year}{2022}.
\newblock \bibinfo{title}{{The impact of the field of view (FOV) on image
  quality in MDCT angiography of the lower extremities}}.
\newblock \bibinfo{journal}{European Radiology} \bibinfo{volume}{32},
  \bibinfo{pages}{2875--2882}.
\newblock \DOIprefix\doi{10.1007/s00330-021-08391-x}.
\bibitem[{Schaapveld et~al.(2011)Schaapveld, Aleman, van Eggermond, Janus,
  Krol, van~der Maazen, Roesink, Raemaekers, de~Boer, Zijlstra, van Imhoff,
  Petersen, Poortmans, Beijert, Lybeert, Mulder, Visser, Louwman, Krul,
  Lugtenburg and van Leeuwen}]{Schaapveld2011}
\bibinfo{author}{Schaapveld, M.}, \bibinfo{author}{Aleman, B.M.},
  \bibinfo{author}{van Eggermond, A.M.}, \bibinfo{author}{Janus, C.P.},
  \bibinfo{author}{Krol, A.D.}, \bibinfo{author}{van~der Maazen, R.W.},
  \bibinfo{author}{Roesink, J.}, \bibinfo{author}{Raemaekers, J.M.},
  \bibinfo{author}{de~Boer, J.P.}, \bibinfo{author}{Zijlstra, J.M.},
  \bibinfo{author}{van Imhoff, G.W.}, \bibinfo{author}{Petersen, E.J.},
  \bibinfo{author}{Poortmans, P.M.}, \bibinfo{author}{Beijert, M.},
  \bibinfo{author}{Lybeert, M.L.}, \bibinfo{author}{Mulder, I.},
  \bibinfo{author}{Visser, O.}, \bibinfo{author}{Louwman, M.W.},
  \bibinfo{author}{Krul, I.M.}, \bibinfo{author}{Lugtenburg, P.J.},
  \bibinfo{author}{van Leeuwen, F.E.}, \bibinfo{year}{2011}.
\newblock \bibinfo{title}{Reduced lung-cancer mortality with low-dose computed
  tomographic screening}.
\newblock \bibinfo{journal}{New England Journal of Medicine}
  \bibinfo{volume}{365}, \bibinfo{pages}{395--409}.
\newblock \URLprefix \url{http://www.nejm.org/doi/10.1056/NEJMoa1505949
  http://www.nejm.org/doi/10.1056/NEJMoa1102873},
  \DOIprefix\doi{10.1056/NEJMoa1102873}.
\bibitem[{Schlegl et~al.(2019)Schlegl, Seeb{\"{o}}ck, Waldstein, Langs and
  Schmidt-Erfurth}]{Schlegl2019}
\bibinfo{author}{Schlegl, T.}, \bibinfo{author}{Seeb{\"{o}}ck, P.},
  \bibinfo{author}{Waldstein, S.M.}, \bibinfo{author}{Langs, G.},
  \bibinfo{author}{Schmidt-Erfurth, U.}, \bibinfo{year}{2019}.
\newblock \bibinfo{title}{{f-AnoGAN: Fast unsupervised anomaly detection with
  generative adversarial networks}}.
\newblock \bibinfo{journal}{Medical Image Analysis} \bibinfo{volume}{54},
  \bibinfo{pages}{30--44}.
\newblock \DOIprefix\doi{10.1016/j.media.2019.01.010}.
\bibitem[{Seeram(2015)}]{Seeram2015}
\bibinfo{author}{Seeram, E.}, \bibinfo{year}{2015}.
\newblock \bibinfo{title}{{Computed Tomography-E-Book: Physical Principles,
  Clinical Applications, and Quality Control}}.
\newblock \bibinfo{edition}{4th editio} ed., \bibinfo{publisher}{Elsevier
  Health Sciences}.
\bibitem[{Shen et~al.(2021)Shen, Zhu, Wang, Xing, Pauly, Turkbey, Harmon,
  Sanford, Mehralivand, Choyke, Wood and Xu}]{Shen2021}
\bibinfo{author}{Shen, L.}, \bibinfo{author}{Zhu, W.}, \bibinfo{author}{Wang,
  X.}, \bibinfo{author}{Xing, L.}, \bibinfo{author}{Pauly, J.M.},
  \bibinfo{author}{Turkbey, B.}, \bibinfo{author}{Harmon, S.A.},
  \bibinfo{author}{Sanford, T.H.}, \bibinfo{author}{Mehralivand, S.},
  \bibinfo{author}{Choyke, P.L.}, \bibinfo{author}{Wood, B.J.},
  \bibinfo{author}{Xu, D.}, \bibinfo{year}{2021}.
\newblock \bibinfo{title}{{Multi-Domain Image Completion for Random Missing
  Input Data}}.
\newblock \bibinfo{journal}{IEEE Transactions on Medical Imaging}
  \bibinfo{volume}{40}, \bibinfo{pages}{1113--1122}.
\newblock \DOIprefix\doi{10.1109/TMI.2020.3046444}.
\bibitem[{Shen et~al.(2004)Shen, Punyanitya, Wang, Gallagher, St.-Onge, Albu,
  Heymsfield and Heshka}]{Shen2004}
\bibinfo{author}{Shen, W.}, \bibinfo{author}{Punyanitya, M.},
  \bibinfo{author}{Wang, Z.}, \bibinfo{author}{Gallagher, D.},
  \bibinfo{author}{St.-Onge, M.P.}, \bibinfo{author}{Albu, J.},
  \bibinfo{author}{Heymsfield, S.B.}, \bibinfo{author}{Heshka, S.},
  \bibinfo{year}{2004}.
\newblock \bibinfo{title}{{Total body skeletal muscle and adipose tissue
  volumes: estimation from a single abdominal cross-sectional image}}.
\newblock \bibinfo{journal}{Journal of Applied Physiology}
  \bibinfo{volume}{97}, \bibinfo{pages}{2333--2338}.
\newblock \DOIprefix\doi{10.1152/japplphysiol.00744.2004}.
\bibitem[{Silver et~al.(2004)Silver, Hittner and May}]{silver2004testing}
\bibinfo{author}{Silver, N.C.}, \bibinfo{author}{Hittner, J.B.},
  \bibinfo{author}{May, K.}, \bibinfo{year}{2004}.
\newblock \bibinfo{title}{Testing dependent correlations with nonoverlapping
  variables: A monte carlo simulation}.
\newblock \bibinfo{journal}{The Journal of Experimental Education}
  \bibinfo{volume}{73}, \bibinfo{pages}{53--69}.
\bibitem[{Sourbelle et~al.(2005)Sourbelle, Kachelriess and
  Kalender}]{Sourbelle2005}
\bibinfo{author}{Sourbelle, K.}, \bibinfo{author}{Kachelriess, M.},
  \bibinfo{author}{Kalender, W.A.}, \bibinfo{year}{2005}.
\newblock \bibinfo{title}{{Reconstruction from truncated projections in CT
  using adaptive detruncation}}.
\newblock \bibinfo{journal}{European Radiology} \bibinfo{volume}{15},
  \bibinfo{pages}{1008--1014}.
\newblock \DOIprefix\doi{10.1007/s00330-004-2621-9}.
\bibitem[{Tang et~al.(2021a)Tang, Gao, Han, Chen, Gao, Nath, Bermudez, Savona,
  Bao, Lyu, Huo and Landman}]{Tang2021}
\bibinfo{author}{Tang, Y.}, \bibinfo{author}{Gao, R.}, \bibinfo{author}{Han,
  S.}, \bibinfo{author}{Chen, Y.}, \bibinfo{author}{Gao, D.},
  \bibinfo{author}{Nath, V.}, \bibinfo{author}{Bermudez, C.},
  \bibinfo{author}{Savona, M.R.}, \bibinfo{author}{Bao, S.},
  \bibinfo{author}{Lyu, I.}, \bibinfo{author}{Huo, Y.},
  \bibinfo{author}{Landman, B.A.}, \bibinfo{year}{2021}a.
\newblock \bibinfo{title}{{Body Part Regression With Self-Supervision}}.
\newblock \bibinfo{journal}{IEEE Transactions on Medical Imaging}
  \bibinfo{volume}{40}, \bibinfo{pages}{1499--1507}.
\newblock \DOIprefix\doi{10.1109/TMI.2021.3058281}.
\bibitem[{Tang et~al.(2021b)Tang, Tang, Zhu, Xiao and Summers}]{Tang2021a}
\bibinfo{author}{Tang, Y.}, \bibinfo{author}{Tang, Y.}, \bibinfo{author}{Zhu,
  Y.}, \bibinfo{author}{Xiao, J.}, \bibinfo{author}{Summers, R.M.},
  \bibinfo{year}{2021}b.
\newblock \bibinfo{title}{{A disentangled generative model for disease
  decomposition in chest X-rays via normal image synthesis}}.
\newblock \bibinfo{journal}{Medical Image Analysis} \bibinfo{volume}{67},
  \bibinfo{pages}{101839}.
\newblock \DOIprefix\doi{10.1016/j.media.2020.101839}.
\bibitem[{Thibault et~al.(2012)Thibault, Genton and Pichard}]{Thibault2012}
\bibinfo{author}{Thibault, R.}, \bibinfo{author}{Genton, L.},
  \bibinfo{author}{Pichard, C.}, \bibinfo{year}{2012}.
\newblock \bibinfo{title}{{Body composition: Why, when and for who?}}
\newblock \bibinfo{journal}{Clinical Nutrition} \bibinfo{volume}{31},
  \bibinfo{pages}{435--447}.
\newblock \DOIprefix\doi{10.1016/j.clnu.2011.12.011}.
\bibitem[{Troschel et~al.(2020)Troschel, Troschel, Best, Gaissert, Torriani,
  Muniappan, {Van Seventer}, Nipp, Roeland, Temel and
  Fintelmann}]{Troschel2020}
\bibinfo{author}{Troschel, A.S.}, \bibinfo{author}{Troschel, F.M.},
  \bibinfo{author}{Best, T.D.}, \bibinfo{author}{Gaissert, H.A.},
  \bibinfo{author}{Torriani, M.}, \bibinfo{author}{Muniappan, A.},
  \bibinfo{author}{{Van Seventer}, E.E.}, \bibinfo{author}{Nipp, R.D.},
  \bibinfo{author}{Roeland, E.J.}, \bibinfo{author}{Temel, J.S.},
  \bibinfo{author}{Fintelmann, F.J.}, \bibinfo{year}{2020}.
\newblock \bibinfo{title}{{Computed Tomography–based Body Composition
  Analysis and Its Role in Lung Cancer Care}}.
\newblock \bibinfo{journal}{Journal of Thoracic Imaging} \bibinfo{volume}{35},
  \bibinfo{pages}{91--100}.
\newblock \DOIprefix\doi{10.1097/RTI.0000000000000428}.
\bibitem[{Troschel et~al.(2019)Troschel, Troschel, Muniappan, Gaissert and
  Fintelmann}]{Troschel2019}
\bibinfo{author}{Troschel, A.S.}, \bibinfo{author}{Troschel, F.M.},
  \bibinfo{author}{Muniappan, A.}, \bibinfo{author}{Gaissert, H.A.},
  \bibinfo{author}{Fintelmann, F.J.}, \bibinfo{year}{2019}.
\newblock \bibinfo{title}{{Role of skeletal muscle on chest computed tomography
  for risk stratification of lung cancer patients}}.
\newblock \bibinfo{journal}{Journal of Thoracic Disease} \bibinfo{volume}{11},
  \bibinfo{pages}{S483--S484}.
\newblock \DOIprefix\doi{10.21037/jtd.2019.01.73}.
\bibitem[{Vonder et~al.(2021)Vonder, Dorrius and Vliegenthart}]{Vonder2021}
\bibinfo{author}{Vonder, M.}, \bibinfo{author}{Dorrius, M.D.},
  \bibinfo{author}{Vliegenthart, R.}, \bibinfo{year}{2021}.
\newblock \bibinfo{title}{Latest ct technologies in lung cancer screening:
  Protocols and radiation dose reduction}.
\newblock \bibinfo{journal}{Translational Lung Cancer Research}
  \bibinfo{volume}{10}, \bibinfo{pages}{1154--1164}.
\newblock \DOIprefix\doi{10.21037/tlcr-20-808}.
\bibitem[{Wang et~al.(2015)Wang, Sui, Schoepf, Song, Xue, Jin, Schmidt, Flohr,
  Canstein, Spearman, Chen and Meinel}]{Wang2015}
\bibinfo{author}{Wang, R.}, \bibinfo{author}{Sui, X.},
  \bibinfo{author}{Schoepf, U.J.}, \bibinfo{author}{Song, W.},
  \bibinfo{author}{Xue, H.}, \bibinfo{author}{Jin, Z.},
  \bibinfo{author}{Schmidt, B.}, \bibinfo{author}{Flohr, T.G.},
  \bibinfo{author}{Canstein, C.}, \bibinfo{author}{Spearman, J.V.},
  \bibinfo{author}{Chen, J.}, \bibinfo{author}{Meinel, F.G.},
  \bibinfo{year}{2015}.
\newblock \bibinfo{title}{Ultralow-radiation-dose chest ct: Accuracy for lung
  densitometry and emphysema detection}.
\newblock \bibinfo{journal}{American Journal of Roentgenology}
  \bibinfo{volume}{204}, \bibinfo{pages}{743--749}.
\newblock \DOIprefix\doi{10.2214/AJR.14.13101}.
\bibitem[{Wang et~al.(2019)Wang, Tao, Shen and Jia}]{Wang2019}
\bibinfo{author}{Wang, Y.}, \bibinfo{author}{Tao, X.}, \bibinfo{author}{Shen,
  X.}, \bibinfo{author}{Jia, J.}, \bibinfo{year}{2019}.
\newblock \bibinfo{title}{{Wide-Context Semantic Image Extrapolation}}, in:
  \bibinfo{booktitle}{2019 IEEE/CVF Conference on Computer Vision and Pattern
  Recognition (CVPR)}, \bibinfo{publisher}{IEEE}. pp.
  \bibinfo{pages}{1399--1408}.
\newblock \DOIprefix\doi{10.1109/CVPR.2019.00149}.
\bibitem[{Weston et~al.(2019)Weston, Korfiatis, Kline, Philbrick, Kostandy,
  Sakinis, Sugimoto, Takahashi and Erickson}]{Weston2019}
\bibinfo{author}{Weston, A.D.}, \bibinfo{author}{Korfiatis, P.},
  \bibinfo{author}{Kline, T.L.}, \bibinfo{author}{Philbrick, K.A.},
  \bibinfo{author}{Kostandy, P.}, \bibinfo{author}{Sakinis, T.},
  \bibinfo{author}{Sugimoto, M.}, \bibinfo{author}{Takahashi, N.},
  \bibinfo{author}{Erickson, B.J.}, \bibinfo{year}{2019}.
\newblock \bibinfo{title}{{Automated Abdominal Segmentation of CT Scans for
  Body Composition Analysis Using Deep Learning}}.
\newblock \bibinfo{journal}{Radiology} \bibinfo{volume}{290},
  \bibinfo{pages}{669--679}.
\newblock \DOIprefix\doi{10.1148/radiol.2018181432}.
\bibitem[{Xu et~al.(2022)Xu, Gao, Tang, Deppen, Sandler, Kammer, Antic,
  Maldonado, Huo, Khan and Landman}]{Xu2022}
\bibinfo{author}{Xu, K.}, \bibinfo{author}{Gao, R.}, \bibinfo{author}{Tang,
  Y.}, \bibinfo{author}{Deppen, S.}, \bibinfo{author}{Sandler, K.},
  \bibinfo{author}{Kammer, M.}, \bibinfo{author}{Antic, S.},
  \bibinfo{author}{Maldonado, F.}, \bibinfo{author}{Huo, Y.},
  \bibinfo{author}{Khan, M.}, \bibinfo{author}{Landman, B.A.},
  \bibinfo{year}{2022}.
\newblock \bibinfo{title}{{Extending the value of routine lung screening CT
  with quantitative body composition assessment}}, in:
  \bibinfo{editor}{I{\v{s}}gum, I.}, \bibinfo{editor}{Colliot, O.} (Eds.),
  \bibinfo{booktitle}{Medical Imaging 2022: Image Processing},
  \bibinfo{publisher}{SPIE}. p.~\bibinfo{pages}{54}.
\newblock \DOIprefix\doi{10.1117/12.2611784}.
\bibitem[{Yu et~al.(2019)Yu, Lin, Yang, Shen, Lu and Huang}]{Yu2019}
\bibinfo{author}{Yu, J.}, \bibinfo{author}{Lin, Z.}, \bibinfo{author}{Yang,
  J.}, \bibinfo{author}{Shen, X.}, \bibinfo{author}{Lu, X.},
  \bibinfo{author}{Huang, T.}, \bibinfo{year}{2019}.
\newblock \bibinfo{title}{{Free-Form Image Inpainting With Gated Convolution}},
  in: \bibinfo{booktitle}{2019 IEEE/CVF International Conference on Computer
  Vision (ICCV)}, \bibinfo{publisher}{IEEE}. pp. \bibinfo{pages}{4470--4479}.
\newblock \DOIprefix\doi{10.1109/ICCV.2019.00457}.
\bibitem[{Zhou et~al.(2018)Zhou, {Rahman Siddiquee}, Tajbakhsh and
  Liang}]{Zhou2018}
\bibinfo{author}{Zhou, Z.}, \bibinfo{author}{{Rahman Siddiquee}, M.M.},
  \bibinfo{author}{Tajbakhsh, N.}, \bibinfo{author}{Liang, J.},
  \bibinfo{year}{2018}.
\newblock \bibinfo{title}{{UNet++: A Nested U-Net Architecture for Medical
  Image Segmentation BT - Deep Learning in Medical Image Analysis and
  Multimodal Learning for Clinical Decision Support}}, in:
  \bibinfo{editor}{Stoyanov, D.}, \bibinfo{editor}{Taylor, Z.},
  \bibinfo{editor}{Carneiro, G.}, \bibinfo{editor}{Syeda-Mahmood, T.},
  \bibinfo{editor}{Martel, A.}, \bibinfo{editor}{Maier-Hein, L.},
  \bibinfo{editor}{Tavares, J.M.R.S.}, \bibinfo{editor}{Bradley, A.},
  \bibinfo{editor}{Papa, J.P.}, \bibinfo{editor}{Belagiannis, V.},
  \bibinfo{editor}{Nascimento, J.C.}, \bibinfo{editor}{Lu, Z.},
  \bibinfo{editor}{Conjeti, S.}, \bibinfo{editor}{Moradi, M.},
  \bibinfo{editor}{Greenspan, H.}, \bibinfo{editor}{Madabhushi, A.} (Eds.),
  \bibinfo{booktitle}{Deep Learning in Medical Image Analysis and Multimodal
  Learning for Clinical Decision Support. DLMIA ML-CDS 2018},
  \bibinfo{publisher}{Springer International Publishing},
  \bibinfo{address}{Cham}. pp. \bibinfo{pages}{3--11}.

\end{thebibliography}
